\documentclass[11pt]{article}
\pdfoutput=1 
\usepackage{shorthand}
\usepackage{mathtools}
\usepackage{booktabs}
\usepackage[english]{babel}
\usepackage{amsmath,amssymb,amsbsy,amstext, amsthm, simplewick, amsfonts,braket}
\usepackage{graphicx}
\usepackage[small]{caption}
\usepackage{siunitx}
\usepackage{upgreek}
\usepackage{framed}
\usepackage{wrapfig}
\usepackage{multirow}
\usepackage{bbm}
\usepackage[numbers,sort&compress]{natbib}
\usepackage[svgnames,dvipsnames,x11names]{xcolor}
\usepackage[utf8x]{inputenc}
\usepackage{selinput}
\usepackage{bm}
\usepackage{float}
\usepackage{dsfont}
\usepackage{caption}
\usepackage{subcaption}
\usepackage{sidecap}
\usepackage{longtable}
\usepackage{anyfontsize}

\usepackage{epstopdf}
\usepackage{cancel}
\usepackage{tcolorbox}
\usepackage{latexsym,amsmath,amssymb,epsfig}
\usepackage{braket}
\usepackage{tensor}
\usepackage{tocloft}

\usepackage{tikz-cd}

\usepackage{ytableau}
\ytableausetup{centertableaux,boxsize=.8em}
\usepackage{genyoungtabtikz}
\Yboxdim{14pt} 
\Ylinethick{.9pt}

\newcommand*{\medoplus}{\raisebox{-0.05ex}{\scalebox{1.25}{$\oplus$}}}

\usepackage{tcolorbox}
\definecolor{greyish2}{rgb}{.96,.96,.96}

\def\xyma{\xymatrix@M.7em}
\def\xymas{\xymatrix@M.1em}

\newcommand{\Comment}[1]{{}}
\definecolor{darkblue}{rgb}{0.15,0.35,0.55}
\definecolor{reddish}{rgb}{0.65, 0.2, 0.2}
\definecolor{darkgreen}{RGB}{50,150,0}
\definecolor{greyish2}{rgb}{.96,.96,.96}
\usepackage[linktocpage=true]{hyperref}
\hypersetup{
colorlinks=true,
citecolor=darkblue,
linkcolor=reddish,
urlcolor=darkblue,
pdfauthor={},
pdftitle={},
pdfsubject={}
}

\DeclareFontFamily{OT1}{rsfs10}{}
\DeclareFontShape{OT1}{rsfs10}{m}{n}{ <-> rsfs10 }{}
\DeclareMathAlphabet{\mathscript}{OT1}{rsfs10}{m}{n}

\def\gsim{ \lower .75ex \hbox{$\sim$} \llap{\raise .27ex \hbox{$>$}} }
\def\lsim{ \lower .75ex \hbox{$\sim$} \llap{\raise .27ex \hbox{$<$}} }
\def\be{\begin{equation}}
\def\ee{\end{equation}}
\def\bea{\begin{eqnarray}}
\def\eea{\end{eqnarray}}

\newcommand{\baaa}{\begin{eqnarray}}
\newcommand{\eaaa}{\end{eqnarray}}

\newcommand{\rd}{{\rm d}}

\newcommand{\tr}{\text{tr}}

\DeclareMathOperator{\E}{e}

\renewcommand{\Im}{\operatorname{Im}}

\newcommand{\ord}[1]{\mathcal{O}({#1})}
\usepackage[letterpaper,margin=1in]{geometry}

\usepackage{tikz}
\usetikzlibrary{decorations}
\pgfdeclaredecoration{complete sines}{initial}
{
    \state{initial}[
        width=+0pt,
        next state=upsine,
        persistent precomputation={\pgfmathsetmacro\matchinglength{
            \pgfdecoratedinputsegmentlength / int(\pgfdecoratedinputsegmentlength/\pgfdecorationsegmentlength)}
            \setlength{\pgfdecorationsegmentlength}{\matchinglength pt}
        }] {}
    \state{upsine}[width=\pgfdecorationsegmentlength,next state=downsine]{
        \pgfpathsine{\pgfpoint{0.25\pgfdecorationsegmentlength}{0.5\pgfdecorationsegmentamplitude}}
        \pgfpathcosine{\pgfpoint{0.25\pgfdecorationsegmentlength}{-0.5\pgfdecorationsegmentamplitude}}
    }
    \state{downsine}[width=\pgfdecorationsegmentlength,next state=upsine]{
        \pgfpathsine{\pgfpoint{0.25\pgfdecorationsegmentlength}{-0.5\pgfdecorationsegmentamplitude}}
        \pgfpathcosine{\pgfpoint{0.25\pgfdecorationsegmentlength}{0.5\pgfdecorationsegmentamplitude}}
}
    \state{final}{}
}

\definecolor{greyish}{rgb}{.90,.90,.90}
\definecolor{greyish2}{rgb}{.96,.96,.96}
\usepackage{xcolor,colortbl}
\usepackage{tcolorbox}

\usepackage[all]{xy}


\numberwithin{equation}{section}

\begin{document}

%
\renewcommand{\thefootnote}{\fnsymbol{footnote}}
\vspace{0truecm}
\thispagestyle{empty}

\begin{center}
{\fontsize{20}{24} \bf
Higher-Dimensional Black Holes\\[8pt]
and Effective Field Theory}
\end{center}

\vspace{-.2truecm}

\begin{center}
{\fontsize{13.25}{18}\selectfont
Daniel Glazer,${}^{\rm a}$
Austin Joyce,${}^{\rm a}$
Maria J. Rodriguez,${}^{\rm b,c,d}$
Luca Santoni,${}^{\rm e}$\\[4.5pt]
Adam R. Solomon,${}^{\rm f,g}$
and Luis Fernando Temoche${}^{\rm b}$
}
\end{center}

\vspace{.1truecm}

\begin{small}
  
\centerline{{\it ${}^{\rm a}$Kavli Institute for Cosmological Physics, Department of Astronomy and Astrophysics}}
\centerline{{\it University of Chicago, Chicago, IL 60637, USA} } 
 
\vspace{.3cm}

\centerline{{\it ${}^{\rm b}$Department of Physics, Utah State University,}}
\centerline{{\it 4415 Old Main Hill Road, UT 84322, USA} } 

 \vspace{.3cm}
 
 \centerline{{\it ${}^{\rm c}$Black Hole Initiative, Harvard University, Cambridge MA 02138, USA.}}
 
  \vspace{.3cm}

 \centerline{{\it ${}^{\rm d}$Instituto de Fisica Teorica UAM/CSIC, Universidad Autonoma de Madrid,}}
\centerline{{\it 13-15 Calle Nicolas Cabrera, 28049 Madrid, Spain} }

 \vspace{.3cm}

\centerline{{\it ${}^{\rm e}$Universit\'e Paris Cit\'e, CNRS, Astroparticule et Cosmologie,}}
\centerline{{\it 10 Rue Alice Domon et L\'eonie Duquet, F-75013 Paris, France}} 

 \vspace{.3cm}

\centerline{{\it ${}^{\rm f}$Department of Physics and Astronomy, McMaster University,}}
\centerline{{\it 1280 Main Street West, Hamilton ON, Canada}} 

 \vspace{.3cm}

\centerline{{\it ${}^{\rm g}$Perimeter Institute for Theoretical Physics,}}
\centerline{{\it 31 Caroline Street North, Waterloo ON, Canada}} 

\end{small}
  
\vspace{.3cm}
\begin{abstract}
\noindent
We study the scalar tidal responses of spinning higher-dimensional black holes, and their effective field theory description. After constructing the effective field theory of a spinning point particle in general dimension, we apply this theory to match the scalar responses of a variety of black hole solutions. In addition to the five-dimensional Myers--Perry black hole, we derive the scalar responses of spinning black holes in the large $D$ limit, and also study the responses of black holes in the ultra-spinning regime. We find that in the most generic case, the static responses of higher-dimensional spinning black holes do not vanish, but for special cases we find a pattern of zeroes in the responses, similar to other known examples. Further, we observe various interesting relations between the responses.
\end{abstract}

\newpage

\setcounter{page}{2}
\setcounter{tocdepth}{2}
\tableofcontents
\newpage
\renewcommand*{\thefootnote}{\arabic{footnote}}
\setcounter{footnote}{0}



\section{Introduction}
\label{sec:intro}

Higher-dimensional general relativity (GR) has a vast zoo of exact black hole solutions, each of which has its own features and peculiarities~\cite{Horowitz:2012nnc}. Though we are ultimately interested in black holes in the four dimensions that we inhabit, an effective strategy to understand physical systems is often to deform and generalize them, and observe how their properties change. As a result, in order to better understand black holes in nature, it is important and useful to abstract general lessons from the study of the more exotic objects in higher dimensions.

Four-dimensional black holes have a number of intriguing properties that we would like to understand by generalizing to higher dimensions. Perhaps the most striking feature is their simplicity. Despite being assembled from unknown microscopic constituents, and behaving like maximally chaotic quantum systems~\cite{Maldacena:2015waa} with a rich entanglement structure~\cite{Almheiri:2020cfm}, the classical features of black holes are surprisingly simple.
In fact, in some respects four-dimensional black holes behave like elementary particles: they couple minimally to gravity~\cite{Vines:2017hyw,Guevara:2018wpp,Chung:2018kqs,Arkani-Hamed:2019ymq}, and their (static) responses to external fields vanish~\cite{1972ApJ...175..243P,Martel:2005ir,Fang:2005qq,Damour:2009va,Damour:2009vw,Binnington:2009bb,Kol:2011vg,Landry:2015cva,Landry:2015zfa,Gurlebeck:2015xpa,Porto:2016pyg,Poisson:2020mdi,LeTiec:2020spy,LeTiec:2020bos,Chia:2020yla,Goldberger:2020fot,Hui:2020xxx,Charalambous:2021mea,Ivanov:2022qqt,Pereniguez:2021xcj,Rai:2024lho,1972ApJ175243P,Landry:2014jka,Charalambous:2021kcz,Hui:2021vcv,Hui:2022vbh,Charalambous:2022rre}, even nonlinearly~\cite{Poisson:2020vap,Poisson:2021yau,DeLuca:2023mio,Riva:2023rcm,Hadad:2024lsf,Iteanu:2024dvx,Combaluzier-Szteinsznaider:2024sgb,Kehagias:2024rtz,Gounis:2024hcm}, indicating that in some sense they do not have internal structure.\footnote{This vanishing of tidal responses is of course true for four-dimensional GR black holes. Considering higher-curvature corrections~\cite{Cardoso:2017cfl,Cardoso:2018ptl,DeLuca:2022tkm}, more exotic compact objects~\cite{Cardoso:2017cfl}, or higher-dimensional black holes will generically lead to nonzero tidal responses. This latter deformation will be the one of interest in the following.} It is important to understand this apparent simplicity of black holes, not least because we now have observational access to their properties via gravitational wave astronomy.\footnote{Aspects of black hole responses can be related to symmetries that emerge at low frequencies in black hole perturbation theory~\cite{Charalambous:2021kcz,Hui:2021vcv,Hui:2022vbh,Charalambous:2022rre,Combaluzier-Szteinsznaider:2024sgb,Kehagias:2024rtz,Gounis:2024hcm}. One way of framing the challenge is to understand whether these symmetries have some manifestation at arbitrary frequency and fully nonlinearly.}

In order to elucidate the properties of the tidal responses of black holes, it is helpful to have additional examples to study. Famously, black hole solutions in four spacetime dimensions are rather rigid, being uniquely determined by their mass, charge, and spin. In contrast, higher-dimensional black holes have a rich phase diagram~\cite{Emparan:2003sy,Emparan:2007wm}, and correspondingly  many different solutions exist (for example with different horizon topologies) for a given assignment of conserved charges.
In order to  abstract general lessons and principles from this rich set of examples, we need to be able to describe them in a uniform way.
Such a unified language is provided by effective field theory (EFT), which allows us to describe disparate physical objects in a way that both organizes their universal features, and corrects this universal description to capture intrinsic properties of the object.
In the present context, the relevant effective description is to imagine a black hole viewed from sufficiently far away as a point particle.
Indeed this is familiar from everyday life: as we move away from things, they begin to resemble points in the distance.
The power of EFT is that it allows us to systematically correct this approximation, the parameters of the EFT encode information about the microscopic makeup of the object. 

In this paper, we study the linear responses of higher-dimensional black holes to external fields (often called Love numbers) by 
developing aspects of the EFT description of spinning point particles in generic dimension, and then applying this effective theory to a number of black hole solutions. 
An interesting feature of these black holes is that generically both their conservative and dissipative responses are nonzero.
These two physically distinct phenomena are matched by different sectors of the EFT. While conservative responses can be matched by operators involving only the external fields on the worldline, in order to match dissipative responses we must either allow for complex couplings by considering an influence functional, or explicitly account for the worldline degrees of freedom into which energy and other conserved quantities can dissipate.

Various corners of the black hole zoo have been studied before from related perspectives~\cite{Chu:2006ce,Kol:2007rx,Kol:2011vg,Hui:2020xxx,Pereniguez:2021xcj,Charalambous:2023jgq,Rodriguez:2023xjd,Charalambous:2024tdj,Gray:2024qys}. A technical complication that precludes a completely systematic study is that the wave equations relevant for extracting linear tidal responses in black hole perturbation theory are generically Fuchsian, but with more than three singular points, having solutions involving special functions whose connection formulas are not known in closed form. Since the relations between solutions near the black hole horizon (where boundary conditions are imposed) and spatial infinity (where responses are read off) are needed in order to analytically compute tidal responses, this complicates the analytic study of generic black hole solutions in general dimension.
As a result, we 
do not attempt to be complete in our study. Rather, we study certain corners of the parameter space that are analytically tractable, and from these examples attempt to infer more general lessons. Relatedly, we restrict ourselves to the study of black hole responses to external scalar fields.
This is also for a technical reason: wave equations for electromagnetic and gravitational perturbations in higher-dimensional spinning black hole spacetimes have not been cast in a separable form, complicating their study.
In all known cases, the responses of black holes to linear scalar perturbations share the qualitative features of gravitational and electromagnetic responses, and so we expect that the insights from the study of scalar tidal responses are more generally applicable.
After constructing the EFT of interest---that of a spinning point particle---we first revisit the scalar tidal responses of Schwarzschild and Kerr black holes. Aside from casting all cases in a uniform language, these examples are useful to compare to the results obtained in more exotic situations. We then turn to consider the five-dimensional Myers--Perry black hole, cohomogeneity-1 black holes in the large-dimension limit, and black holes in the ultra-spinning limit.

What we find is that the tidal responses of  black holes in diverse dimensions and physical situations display a rich structure. Perhaps the most interesting feature is that for generic spin parameters, black hole solutions in general dimension have nonzero tidal responses, independent of the angular momentum of the applied field, $L$. This is in contrast to four dimensions, where the static Love numbers of black holes vanish for all choices of angular momenta, and to special cases in higher dimensions, where for example non-spinning black holes have Love numbers that vanish when $L/(D-3)$ is an integer. Perhaps this is unsurprising, black holes in higher dimension are more like generic objects, with different possible topologies and no unique assignment of angular momenta. Nevertheless, we do find some corners of parameter space where interesting zeroes appear. For example, $5D$ Myers--Perry black holes have vanishing static scalar Love numbers for integer $L/2$ if their two spin parameters are tuned to be equal~\cite{Charalambous:2023jgq,Rodriguez:2023xjd}, and we also find that ultra-spinning black holes with a single spin parameter have vanishing responses for certain external field profiles. In addition to this, we find a number of interesting inter-relations between the black hole responses. While some of these zeroes and facts can be explained either by symmetries or by other physical arguments, others remain somewhat mysterious. Aside from the static sector, we compute the subleading, frequency-dependent conservative tidal responses (dynamical Love numbers), and leading and subleading dissipative responses in all of these cases, some for the first time.

The richer set of black hole solutions in higher dimensions provides an interesting arena in which to study the classical properties of black holes. We anticipate that the theoretical data generated by this study will provide insights into their structure, and eventually shed light on the black holes that populate our universe.

\noindent
{\bf Outline:} In Section~\ref{sec:EFT} we describe the point particle EFT formalism that we will use to describe black holes at large distances, and clarify some aspects of spin and rotating frame effects. We further describe how to compute scalar responses in this EFT, in order to enable matching to black hole perturbation theory calculations. In Section~\ref{sec:DdimSch} we consider the Schwarzschild black hole in generic dimension. Though this case has been studied before, it serves as a useful illustration of the formalism, and a cross-check of results in the spinning cases treated in later sections. In Section~\ref{sec:Kerr} we match the scalar responses of the Kerr black hole to point particle EFT. In this case we reproduce results at subleading order in frequency, and show how they can be obtained by a suitable near-zone approximation of the wave equation (we further validate this approximation by comparing to a scattering calculation in Appendix~\ref{app:NZ}). In Section~\ref{sec:MP} we consider the five-dimensional Myers--Perry black hole with generic spin parameters. A novelty of this situation is that it requires both conservative and dissipative couplings in the EFT to match its static responses. In Section~\ref{sec:InfD}, we derive scalar responses for spinning black holes in the large dimension limit, where all of the spin parameters are turned to be equal. In this situation the problem develops an enhanced symmetry, making the problem analytically tractable. In Section~\ref{sec:US} we study spinning black holes in $D \geq 6$ in the ultra-spinning regime, where the spin parameter is scaled to be very large. In this limit the problem is again analytically tractable, and we both solve the black hole wave equation and match to point particle EFT. Finally in Section~\ref{sec:Concl} we summarize and draw some conclusions.

A number of appendices collect technical results that are important, but somewhat outside the main line of development of the text. In Appendix~\ref{app:SK}, we review aspects of the Schwinger--Keldysh formalism, which is needed to describe dissipative effects in EFT. In Appendix~\ref{appendix:Thorne} we describe the construction and properties of traceless symmetric tensors that are related to spherical harmonics (``Thorne tensors"). In Appendix~\ref{app:CPNeigen} we review some features of eigenfunctions of the charged laplacian on complex projective space, which appear in the study of large-dimension spinning black holes. In Appendix~\ref{appendix:Formulas} we list some mathematical identities that are used in many of the calculations we perform. In Appendix~\ref{app:scattering} we review the scattering of waves in point particle EFT in order to facilitate a comparison between matching computations performed using EFT one-point functions and scattering amplitudes. We also review the systematics of wave scattering in black hole perturbation theory. In 
Appendix~\ref{app:NZ} we discuss some of the systematics of so called ``near zone" approximations of the wave equation in black hole perturbation theory.

\noindent
{\bf Conventions:} We use the mostly-plus metric convention  $(-,+,\cdots,+)$, and denote the spacetime dimension by $D=d+1$, with $d$ the spatial dimension. Spacetime indices are denoted by Greek letters $\mu,\nu,\cdots$, spatial indices are denoted by Roman letters from the middle of the alphabet, $i,j,\cdots$, and spacetime particle frame indices are denoted using Roman letters from the beginning of the alphabet $A,B,\cdots$, while spatial frame indices are denoted by Roman letters from the beginning of the alphabet $a,b,\cdots$. The notation $(\cdots )_T$ indicates the trace-subtracted symmetrization of the enclosed indices. In many cases we decompose fields using spherical harmonics, where we denote the angular momentum by $L$, and the magnetic quantum numbers by (the multi-index) $M$.

\newpage
\section{Point Particle Effective Field Theory}
\label{sec:EFT}

In this Section, we construct the worldline EFT of a spinning point particle, which serves as a description of  black holes at large distances (for more details, see also~\cite{Goldberger:2004jt,Goldberger:2005cd,Hanson:1974qy,Bailey:1975fe,Porto:2005ac,Steinhoff:2011sya,Delacretaz:2014oxa,Goldberger:2020fot,Porto:2016pyg,Levi:2018nxp,Kol:2011vg,Hui:2020xxx,Charalambous:2021mea}).
We use this formalism to compute the off-shell one point function of external fields, which captures the response of the object to external perturbations. This quantity can also be computed directly in black hole spacetimes, which allows us to extract the Wilson coefficients of the EFT via matching. The details of the matching procedure are described in the subsequent sections.

\subsection{Kinematics}

As a particle moves through spacetime, it traces out a one-dimensional worldline. The action that governs its classical dynamics is simply the proper length of the worldline---the Nambu--Goto action. In the cases of interest, the particle is timelike, so the proper length of its worldline is its proper time
\begin{equation}
    S_{\rm pp}= -m\int\rd \tau=-m \int \rd\lambda \sqrt{-g_{\mu \nu}\frac{\rd x^\mu}{\rd\lambda}\frac{\rd x^\nu}{\rd\lambda}}\,,
    \label{eq:NGaction}
\end{equation}
where $x^\mu(\lambda)$ is an embedding of the particle's worldline (parameterized by $\lambda$) into spacetime. The action~\eqref{eq:NGaction} is the lowest derivative term that is both covariant in spacetime and invariant under worldline reparameterizations  $\lambda \mapsto \lambda'(\lambda)$. The particle's (timelike) velocity is given by 
\be
v^\mu = \frac{\dot x^\mu}{\sqrt{-g_{\mu \nu}\dot x^\mu\dot x^\nu}} = \frac{\rd x^\mu}{\rd\tau}\,,
\ee
where a dot denotes a derivative with respect to $\lambda$, so that $\dot x^\mu\equiv \rd x^\mu/\rd\lambda$.

In what follows, it will be more convenient to describe particle motion in Polyakov form by introducing an additional dynamical variable---a vielbein defined on the worldline which we denote as $E$. This vielbein can be written in terms of the worldline metric, $\gamma_{\lambda\lambda}$, as 
\be
\rd \tau^2 = -\gamma_{\lambda\lambda}\, \rd\lambda^2 \equiv (mE)^2\,\rd\lambda^2\,.
\ee
Including the vielbein $E$, the point particle action takes the form
\begin{equation}
    S_{\rm pp}=\frac{1}{2}\int \rd\lambda\, E\left(E^{-2}\frac{\rd x^\mu}{\rd \lambda}\frac{\rd x^\nu}{\rd \lambda}g_{\mu \nu} - m^2\right)\,.
    \label{eq:pointparticlepolyakov}
\end{equation}
In this formalism, the variable $E$ is an auxiliary field and can be integrated out using its equation of motion, 
returning the original form of the action~\eqref{eq:NGaction}. 

\vspace{-12pt}
\paragraph{Spinning particles:} In order to describe spinning objects, the point particle description must be imbued with the additional structure of a local frame $e_\mu^A(\lambda)$, which encodes the orientation of the object~\cite{Hanson:1974qy,Bailey:1975fe,Porto:2005ac,Steinhoff:2011sya,Delacretaz:2014oxa,Goldberger:2020fot}. The vielbein $e_\mu^A(\lambda)$
carries both a spacetime index, $\mu$, and internal SO$(d,1)$ particle frame index $A = 0,1,2,\cdots, d$, which we can think of as a mapping between the fixed background spacetime and the instantaneous orientation of the object.
The vielbein is related to the spacetime metric $g_{\mu \nu}$ and the internal metric $\eta_{AB}$ through
\be
\label{eq:restframevielbein}
g_{\mu\nu}e_A^\mu e_B^\nu = \eta_{AB}\,,\hspace{3cm}
\eta_{AB}e^A_\mu e^B_\nu = g_{\mu\nu}\,.
\ee
In the particular case where the spacetime is flat---so that $g_{\mu\nu} = \eta_{\mu\nu}$---the vielbein at any given instant is simply the Lorentz transformation into the particle's rest frame.

The angular velocity of the object is encoded in the antisymmetric tensor $\Omega^{AB}$ that measures the twisting of the local frame defined by $e_\mu^A$ as we move along the worldline:\footnote{The antisymmetry of $\Omega^{AB}$ follows from taking a derivative of~\eqref{eq:restframevielbein} with respect to $\lambda$.
}
\be
\Omega^{AB} =g^{\mu \nu}e^A_\mu \frac{D}{D\lambda}e^B_\nu\,,
\ee
where the total derivative with respect to the affine parameter can be written as
\be
\frac{D}{D\lambda}e^A_\mu \equiv \dot{x}^{\rho}\nabla_\rho e^A_\mu \,,
\ee
and where again an overdot denotes a derivative with respect to $\lambda$.

In order to describe the dynamics of a spinning point particle, it is convenient to work in a first-order form, where the action\footnote{Note that there is a different way of proceeding by employing the Routhian formalism, see e.g.,~\cite{Porto:2006bt,Porto:2016pyg}.} is given by~\cite{Steinhoff:2011sya,Goldberger:2020fot}
\begin{equation}
    S_{\rm pp}= \int\rd \lambda \left( p_\mu \dot x^\mu+S^{AB}\Omega_{AB} -\frac{1}{2}E\Big[p_\mu p^\mu + m^2(p,S)\Big]+E \,\xi_A S^{AB}p_B
    +\cdots
    \right)\,.
    \label{eq:pointparticleaction}
\end{equation}
In this formalism the momentum $p_\mu$ is the variable conjugate to the particle's position, the particle's angular momentum $S^{AB}$ is conjugate to its angular velocity $\Omega^{AB}$ and, in addition to $E$, the variable $\xi_A$ acts as a lagrange multiplier which enforces the constraint that the object's spin be transverse to its direction of motion. The function $m^2(p,S)$ appearing in~\eqref{eq:pointparticleaction} can have arbitrary dependence on $p^\mu$ and $S^{AB}$. This function captures some of the microscopic properties of the object (for example the relation between angular velocity and angular momentum), and in general should be determined by matching to some ultraviolet calculation.\footnote{For example, the leading-order in derivatives expression for $m$ is~\cite{Steinhoff:2011sya}
\be
m(p,S) = m_0+\frac{1}{2I}S^{AB}S_{AB} +\cdots\,, \nonumber
\ee
which, if we integrate out $p_\mu$ and $S^{AB}$ from~\eqref{eq:pointparticleaction}, leads to the action
\be
 S_{\rm pp}=  \int \rd\lambda\left(-m_0 \sqrt{-g_{\mu \nu}\dot x^\mu \dot x^\nu}+\frac{I}{2\sqrt{-g_{\mu \nu}\dot x^\mu \dot x^\nu}}\Omega_{AB}\Omega^{AB} +\cdots\right)\,, \nonumber
\ee
so that the corrections to $m_0$ in this case encode the moment of inertia of the object.
}
In addition, further higher-derivative terms that we have suppressed are denoted by $\cdots$, these terms capture finite-size effects.

The benefit of this first-order approach is that it makes it easier to both handle the kinematic constraint that the particle's spin directions should be defined transverse to its velocity, and to couple the system to gravity.
Explicitly, $p_\mu$ and $S^{AB}$ are the conjugate variables to  $x^\mu$ and $\Omega_{AB}$:
\begin{align}
p_\mu &\equiv \frac{\delta S_{\rm pp}}{\delta \dot x^\mu}\,,\\
S^{AB} &\equiv \frac{\delta S_{\rm pp}}{\delta \Omega_{AB}}\,.
\end{align}
The worldline vielbein $E$ enforces that the particle propagates on shell in spacetime, so that $p^2+m^2 = 0$, while the auxiliary field $\xi_A$ enforces a constraint called the spin-supplementary condition
\be
S^{AB}p_B = 0\,.
\label{eq:ssc}
\ee
This reduces the $D(D-1)/2$ degrees of freedom of a general, antisymmetric $S_{AB}$ down to those associated to the physical $(D-1)(D-2)/2$ spin degrees of freedom.\footnote{The choice~\eqref{eq:ssc} is not unique, but is convenient because it is covariant. See, e.g.,~\cite{Kim:2023drc} for a more complete discussion of the various possible spin-supplementary conditions.}
This particular choice also enforces that $\Omega^{AB}$ will be a pure rotation matrix when measured in the rest frame of the particle~\cite{Hanson:1974qy}.

We can express the angular momentum of the object in terms of $p_\mu$ and $S^{\mu\nu} = e_A^\mu e_B^\nu S^{AB}$ as~\cite{Hanson:1974qy}
\be
J^{\mu\nu} = x^\mu p^\nu-x^\nu p^\mu +S^{\mu\nu}\,.
\ee
In the rest frame of the particle, where $v_i = 0$, the angular momentum is therefore parameterized by the antisymmetric tensor $J^{ij} = S^{ij}$.\footnote{Classical objects' angular momentum transforms in the adjoint representation of the rotation group SO$(D-1)$, given by the tensor $J_{ij}$, with $(D-1)(D-2)/2$ independent components. By choosing spatial coordinates adapted to the orientation of the object, the antisymmetric tensor $J_{ij}$ can be brought to block-diagonal form and written as a sum over the generators of the Cartan subalgebra $T^a_{ij}$ as $J_{ij} = J_a T^a_{ij}$. Here $a$ is a Lie algebra index running over the generators in the Cartan. Recall that SO$(2r)$ and SO$(2r+1)$ have rank $r$, so the Cartan has $r$ elements. Each element of the Cartan generates a rotation of some 2-plane and the $r$ different values of $J_a$ then have the interpretation as the angular momenta associated to these spin planes of the object.}

\vspace{-12pt}
\paragraph{Rest frame and corotating frame:} 
Though the construction of the EFT is covariant, it will be convenient in what follows to 
define two distinguished reference frames (or choices of gauge): the frame where the point particle is at rest (has no linear velocity) but could appear to rotate, and the frame which both is at rest with respect to the object and which co-rotates with the object. These frames can be naturally defined using $v^\mu$ and $e_\mu^A$.

We first note that we can boost and/or rotate our coordinates to set~\cite{Porto:2016pyg}
\begin{equation}
    e^\mu_0=v^\mu \,.
\end{equation}
Using~\eqref{eq:restframevielbein}, we see that the velocity in the worldline-adapted coordinates has no spatial components
\be
\label{eq:restframev}
v^A = e^A_\mu v^\mu = e^A_\mu e^{\mu}_0= \delta_0^A = (1,\vec 0)\,.
\ee
The particle's spacetime velocity defines a distinguished time direction, which it is natural to then use to split quantities into temporal and spatial components. We can phrase this split covariantly by defining the projector
\be
\label{eq:projector}
P^\mu_\nu \equiv \delta^\mu_\nu + v^\mu v_\nu\,,
\ee
which projects onto the subspace transverse to the particle's motion. Similarly, we can isolate the time direction by contracting with the particle's velocity $v^\mu$. This spacetime velocity can also be taken to be purely in the time direction like~\eqref{eq:restframev} by aligning the worldline time direction with that of spacetime by choosing coordinates where $e_\mu^A$ is block diagonal. We will refer to the coordinates where the spacetime velocity of the particle is $v^\mu = (1,\vec 0)$,
as the particle's {\it rest frame} and will denote spatial indices in the rest frame by Latin indices $i,j,k,\cdots$. Expressions involving rest frame spatial indices can be covariantized using the projector~\eqref{eq:projector} as  $T^{i_1\cdots i_n} \to P^{\mu_1\cdots \mu_n}_{\nu_1\cdots \nu_n}T^{\nu_1\cdots \nu_n}$.

We can also use the projector $P^A_B = e^A_\mu e_B^\nu P^\mu_\nu= \delta^A_B +v^A v_B$ to isolate spatial frame indices, which we denote by $a,b,\cdots$.
The spatial part of the vielbein, $e^i_a$, which we can write covariantly as $P^B_A\, P^\mu_\nu e^\nu_B$, is a pure rotation matrix, and we can think of it as the (time-dependent) rotation that transforms to a frame which corotates with the particle. 
A further simplification that we can make is to choose coordinates so that the spatial vielbein is diagonal $e_i^a = \delta_i^a$. In this (generically non-inertial) frame, the coordinates corotate with the object, so that it does not seem to be rotating. We will therefore call this frame the {\it corotating frame}. 

Depending on the situation, it can be convenient to define quantities either in the rest frame of the particle, or in its corotating frame. However, one must be careful because there are nontrivial rotating frame effects, which we discuss in Section~\ref{sec:rotframes}.

\subsection{Finite-size Effects}

The description of an object as a point particle is an approximation, one which becomes more and more precise at larger and larger distances. 
As a practical matter, the way that we learn about the properties of objects is by probing them with external sources, like electromagnetic, gravitational, or scalar fields. 
The microphysical makeup of the object is then encoded in its response to these external stimuli. (For example, different objects will have different electric polarizabilities or magnetic susceptibilities, and measuring these features gives us insights into their properties.)
The way that this internal structure of objects is encoded in EFT is by higher-derivative operators that couple the worldline of the particle to external fields, which lead to responses in the presence of background fields.
These operators are higher dimensional, so they require a length scale to define them. This length scale is typically the characteristic size of the object, $R$, so that the EFT is an expansion in the characteristic size of the object divided by our distance from it $R/r\ll 1$.\footnote{More precisely, the typical scale suppressing higher-derivative terms is the characteristic splitting of the microscopic energy levels of the object.}

In the following, we will imagine using a massless scalar field as an external probe and characterize the response of various black holes in EFT.
This obviously represents an idealization---not least because there are no known fundamental massless scalar fields in nature---but this simplification will allow us to treat a variety of objects in the same language. Moreover, the responses of black holes to external scalar fields retain many of the conceptually interesting properties of their responses to (more realistic) gravitational or electromagnetic perturbations~\cite{1972ApJ...175..243P,Teukolsky:1973ha,Detweiler:1980uk,Kol:2011vg,Hui:2020xxx,Charalambous:2021mea,Ivanov:2022hlo,Charalambous:2023jgq}.
We therefore expect the insights obtained by studying scalar responses apply to more realistic settings.

We therefore want to consider the combined action $S_{\rm pp}+S_{\phi}+S_{\rm int}$. The action $S_{\rm pp}$~\eqref{eq:pointparticleaction} describes the dynamics of the point particle, while $S_\phi$ is the action for a massless scalar field, which propagates in  spacetime according to:
\be
  S_{\phi}=\int \rd^Dx \sqrt{-g}\left(-\frac{1}{2}(\partial\phi)^2\right) \,.
\ee
Finite size effects are encoded in the interactions between the worldline and $\phi$, denoted by $S_{\rm int}$. Different physical features of the object correspond to different classes of couplings.

\vspace{-16pt}
\paragraph{Charge multipoles:} The simplest terms that we can include on the worldline involve couplings to a single $\phi$ field. These capture the permanent charge multipole moments of the object of interest via couplings of the form 
\be
    \label{eq:Scharges}
S = \int\rd\lambda\, E\,\bigg( g\phi+\sum_{L=1}^\infty\frac{1}{L!}\lambda_{(L)}^{i_1\cdots i_L} \partial_{(i_1}\cdots\partial_{i_L)_T}\phi
\bigg)\,,
\ee
where $(\ldots)_T$ means we symmetrize the enclosed indices and remove the traces. Here we have written the coupling directly in the rest frame of the object, but these operators can be made covariant using the projector~\eqref{eq:projector}.
 Here we are focusing on external scalar couplings, but
more generally we could parameterize the electromagnetic or gravitational multipole moments of object by including couplings to (derivatives of) the Maxwell and Weyl tensors.\footnote{An intriguing feature of $D=4$ Kerr black holes is that they have the gravitational multipole moments of an elementary spinning particle~\cite{Vines:2017hyw,Guevara:2018wpp,Chung:2018kqs,Arkani-Hamed:2019ymq}.}

\vspace{-16pt}
\paragraph{Conservative tidal effects:}  The response of the particle to external fields can be split into two distinct categories: conservative and dissipative. Conservative interactions may be modeled straightforwardly with the ingredients we have already introduced;  they are captured by terms of the form~\cite{Goldberger:2004jt,Goldberger:2005cd,Kol:2011vg,Rothstein:2014sra,Porto:2016zng,Haddad:2020que,Aoude:2020ygw,Hui:2020xxx,Charalambous:2021mea}
\begin{equation}
    \label{eq:Scons}
    S_{\rm cons.} \supset \int \rd\tau \, \lambda_{\rm cons.}(\tau)^{i_1\cdots i_L\,j_1\cdots j_{L'}}C^{(2)}_{i_1\cdots i_L}(\tau)\,C^{(1)}_{j_1\cdots j_{L'}}(\tau)\,.
\end{equation}
where $C^{(n)}_{i_1\cdots i_L}$ stands schematically for composite operators built from external probe fields. In general the coupling can be between two completely different operators, and parameterizes the induced response in the operator $C^{(2)}$ in the presence of a nontrivial profile for $C^{(1)}$ (or vice versa).
The case  of interest in the following is where the external field is a scalar, which for example can couple to the worldline through the operator 
\begin{equation}
    C_{i_1\cdots i_L} = \partial_{(i_1}\cdots\partial_{i_L)_T}\phi \,. 
    \label{eq:externaloperator|}
\end{equation}
With this choice of $C$ in~\eqref{eq:Scons} for both operators, the coupling takes the form
\begin{equation}
    \label{eq:Scons2}
    S_{\rm cons.} \supset \int \rd\tau \, \lambda_{\rm cons.}(\tau)^{i_1\cdots i_L\,j_1\cdots j_{L'}}\partial_{(i_1}\cdots\partial_{i_L)_T}\phi\,\partial_{(j_1}\cdots\partial_{j_{L'})_T}\phi\,.
\end{equation}
Conceptually, this interaction
captures the (linear) response of the particle to an externally applied $\phi$ field, either as a tidal field, or in a scattering process.
More generally these operators could be built from photons or gravitons if we want to probe the system in other ways. We have only written the simplest conservative interaction, which captures the linear response of the object, but one can also contemplate nonlinear responses in the presence of stronger background fields, which would be captured by operators with more insertions of the relevant field~\cite{Bern:2020uwk,DeLuca:2023mio,Riva:2023rcm,Hadad:2024lsf,Iteanu:2024dvx,Combaluzier-Szteinsznaider:2024sgb}.
Note that in a unitary theory, $\lambda_{\rm cons.}$ will be purely real, and so couplings of the form~\eqref{eq:Scons2}
can only model conservative effects. To capture dissipative effects in the effective description, we must enlarge the EFT.

\paragraph{Dissipative effects:} 
An action like~\eqref{eq:Scons2} can only model conservative physics. Since all the interactions conserve energy and momentum, there is
nowhere for energy to escape. However, we know that black holes {\it do} dissipate energy (e.g., if you shine a light on a black hole, the light can fall in). Thus, we must adjust our EFT framework to include these dissipative interactions.

To model dissipation, we follow~\cite{Goldberger:2004jt,Goldberger:2005cd} and introduce some degrees of freedom localized on the worldline of the object, which we denote collectively by $X$.
These additional degrees of freedom can absorb energy, and provide an effective description of dissipative processes.
For example, if the particle of interest were a hydrogen atom, the various states of the $X$ would correspond to the energy levels of the hydrogen atom.
These additional $X$ degrees of freedom enter the free particle action~\eqref{eq:pointparticleaction} as~\cite{Goldberger:2005cd,Goldberger:2020fot}
\begin{equation}
    S_{\rm pp}= \int\rd \lambda \left( p_\mu(X) \dot x^\mu+S^{AB}(X)\Omega_{AB} -\frac{E}{2}\Big[p_\mu(X) p^\mu(X) - L_X(X)\Big]+E \,\xi_A S^{AB}(X)p_B(X)
    \right).
 \label{eq:disspointparticleaction}
\end{equation}
Here the momentum $p_\mu(X)$ and spin $S^{AB}(X)$ are constructed as composites of the internal degrees of freedom. The dynamics of these $X$ degrees of freedom are governed by the Lagrangian $L_X$. Since $p_\mu$ and $S^{AB}$ are not really independent variables---because they are built from the fundamental $X$ degrees of freedom---we do not vary them in the action. Instead, we infer the relation between them and $\dot x^\mu,\, \Omega^{AB}$ by imposing the spin supplementary constraint~\eqref{eq:ssc}. 

We now consider how these worldline degrees of freedom couple to external fields. Beyond the conservative charge/multipole terms written in~\eqref{eq:Scharges}, we may add interactions of the form 
\be
S_{\rm int} \supset \int\rd\tau E\, Q^{i_1\cdots i_L}(X,E)\,C_{i_1\cdots i_L}\,,
\label{eq:sint}
\ee
where $Q^{i_1\cdots i_L}(X,E)$ are composite operators built from  $X$ and the vielbein $E$. 
We want to understand how the coupling to the $X$ degrees of freedom affects the response to external fields. However, if we perform some experiment at long distances from the object of interest, we typically cannot track the detailed microstate of the degrees of freedom that comprise the object. We therefore need to average over the possible configurations that they could be in. From the field theory point of view, this amounts to integrating out these degrees of freedom. 

Since the $X$ degrees of freedom can be gapless, the remaining fields that we do track (the point particle itself and external probes) are effectively an open quantum system and can exhibit dissipation, among other phenomena, because arbitrarily small amounts of energy can excite the $X$ sector. Since we cannot specify the final state of this generically dissipative system, the appropriate framework in which to integrate out the $X$ degrees of freedom is the Schwinger--Keldysh formalism. Some general details of this formalism are summarized in Appendix~\ref{app:SK}. The key feature of Schwinger--Keldysh from the path integral point of view is that we double the fields involved in order to capture the evolution of the system along a closed time contour that runs from $t=-\infty$ to the time of interest and then back to $t = -\infty$. We are therefore interested in performing the following in-in path integral to obtain an effective action with dissipative couplings:
\begin{equation}
    \exp\Big(i \Gamma^{\text{in-in}}(F_1,F_2)\Big)=\int \mathcal{D}X_1 \mathcal{D}X_2 \,\E^{i S[X_1,F_1,\cdots]-i S[X_2,F_2,\cdots]}\,,
    \label{eq:ininint}
\end{equation}
where $ F_{1,2}= \big\{x_{1,2}^\mu, (e_{1,2})_\mu^A, E_{1,2},\phi_{1,2}\big\}$ are the long distance degrees of freedom along the two branches of the in-in contour. The $X$ degrees of freedom induce effective interactions between the fields defined on the two branches of the contour, which lead to dissipative effects in the effective action $\Gamma^\text{in-in}$. Concretely, the effective theory will include couplings of the form 
\be
\label{eq:gammaintgeneric}
\Gamma_{\rm int}^\text{in-in}\supset \int\rd\tau_1\rd\tau_2\, \lambda_{IJ}(\tau_1,\tau_2)^{i_1\cdots i_L\,j_1\cdots j_{L'}}C^{I}_{i_1\cdots i_L}(\tau_1)\,C^{J}_{j_1\cdots j_{L'}}(\tau_2)\,,
\ee
where the indices $I,J$ run over the two branches of the in-in contour (or equivalently over the doubled field content).
The effective coupling $\lambda_{IJ}(\tau_1,\tau_2)$ is essentially the Green's function between the $Q_I$ operators in the appropriate basis. If there is a hierarchy of scales between the characteristic timescale of the $X$ degrees of freedom---which we assume to be fast---and the timescale on which we want to probe the system, the dynamics of the $X$ degrees of freedom can be modeled as instantaneous. In this approximation, the Green's function associated to the $Q$ operators can be expanded as a sum of (derivatives of) delta functions\footnote{The existence of such an expansion follows from the expected rapid decay of the two-point function of operators in a thermalized system. In particular, it is expected that operators (that are not conserved charges) will have correlation functions that decay faster than any power at late times. This implies that the Fourier-space two-point function of the $Q$ operators is analytic around $\omega = 0$, which in real space corresponds to the expansion~\eqref{eq:Qgreen}. See, e.g.,~\cite{Endlich:2012vt} for more details.}
\be
G^{(Q)}(t-t') \sim \lambda^{(0)} \delta(t-t')+\lambda^{(1)}\delta'(t-t')+\cdots\,,
\label{eq:Qgreen}
\ee
where we have suppressed indices for clarity. In this regime where we can model the interactions along the worldline as instantaneous, the effective couplings in~\eqref{eq:gammaintgeneric} will be {\it local}, leading to an ordinary EFT describing the response to external fields. 

Importantly, the interactions involving $\lambda_{IJ}$ in~\eqref{eq:gammaintgeneric} need not be strictly real. We can understand this by transforming the Green's function~\eqref{eq:Qgreen} to frequency space: 
\be
G^{(Q)}(\omega) \sim \lambda^{(0)} +\lambda^{(1)} \,i\omega+\cdots\,.
\ee
Notice that this contains terms which are time reversal odd (those odd in frequency), which will lead to imaginary effective couplings/dissipative effects.\footnote{More generally, the characteristics of the object may introduce other time-reversal odd quantities---for example the spin of an object---and so we may assemble time reversal even terms from odd powers of frequency and odd powers of the spin. However, the separation into even and odd under time reversal will continue to correspond to conservative versus dissipative effects.}
These odd in $\omega$ terms cannot be mimicked by any conservative local couplings between the worldline and external fields. From a symmetry point of view, we can understand the presence of these dissipative terms as a consequence of the fact that integrating out the $X$ variables breaks the independent time translation symmetries of the in-in contour to the diagonal~\cite{Akyuz:2023lsm,Haehl:2015foa,Crossley:2015evo,Liu:2018kfw}.\footnote{Instead of introducing the auxiliary worldline degrees of freedom $X$, it is possible to directly write down couplings between the fields on the two branches of the in-in contour which lead to dissipation~\cite{Feynman:1963fq,Caldeira:1982iu,Calzetta:1986cq,Kamenev:2009jj}. The benefit of the approach that we have followed is that the couplings between external fields and the $X$ variables satisfy the usual rules of EFT, and so it is straightforward to parameterize the possible interactions.}

\vspace{-12pt}
\paragraph{Expansion in angular harmonics:} Finite size effects are captured by operators that are suppressed by some characteristic energy scale, which therefore become more and more irrelevant at large distances. Consequently, it will often be convenient to organize calculations using the rotational symmetries of spacetime at infinity. In general the object itself will not be rotationally invariant, but we can nevertheless characterize the properties of objects according to their responses to specific angular partial waves.

It is therefore convenient to write our EFT in a suitably adapted basis. This can be achieved by utilizing intertwining tensors that convert between symmetric traceless spatial tensors and spherical harmonics. These so-called Thorne tensors ${\cal Y}_{LM}^{i_1\cdots i_L}$ are traceless symmetric tensors that have the property that their contraction with the unit vector $\hat x^i \equiv x^i/\lvert \vec x\rvert$ is a hyperspherical harmonic with quantum numbers $L,M$~\cite{trautman1965lectures,Thorne:1980ru,Charalambous:2021mea} 
\be
{\cal Y}_{LM}^{i_1\cdots i_L} \hat x_{i_1}\cdots \hat x_{i_L} = Y_{LM}(\theta)\,,
\ee
where $L$ labels the angular momentum representation, and $M$ is a multi-index labeling the corresponding magnetic quantum numbers 
(see Appendix~\ref{appendix:Thorne} for more details).
It is possible to invert this relationship to write a general traceless, symmetric tensor as a linear combination of these basis tensors:
\be
T^{i_1\cdots i_L} = \sum_M T_{LM} \,{\cal Y}_{LM}^{i_1\cdots i_L}\,.
\ee
We can in particular expand the effective couplings appearing in~\eqref{eq:gammaintgeneric} as
\begin{equation}
    \lambda_{IJ}^{i_1\cdots i_L\, j_1\cdots j_{L'}}= \sum_{M,M'} \lambda_{IJ}^{ LL'MM'} \,\mathcal{Y}^{i_1\cdots i_L}_{LM} \mathcal{Y}_{L'M'}^*{}^{j_1\cdots j_{L'}}\,,
    \label{eq:thorneexpan}
\end{equation}
The coefficients $\lambda_{IJ}^{LL'MM'}$ describe how the particle couples the $(L,M)$ and $(L',M')$ modes of external probes. Notice that the definition of the Thorne tensors (or more generally spherical harmonics) requires a choice of basis, corresponding to the $r = \left \lfloor{\tfrac{d}{2}}\right\rfloor$ independent magnetic quantum numbers of the rotation group. 
For a generic spinning object, there is a natural choice of these spin planes, which are precisely the planes in which the object is spinning (see Appendix~\ref{appendix:Thorne}). Consequently, for objects that only break rotational symmetry by spinning, the decomposition~\eqref{eq:thorneexpan} is particularly natural. The manifestation of the residual SO$(2)^r$ symmetry will be that these objects' couplings will be diagonal in both $L$ and $M$, though the responses can of course depend on direction, or equivalently on the magnetic quantum numbers $M$. (The special case of a spherically symmetric non-rotating object will have the additional feature that responses for all $M$s will be the same.) In this case, it might be desirable to directly parameterize the relevant operators that couple the particle to external fields using the spin of the object (as in~\cite{Charalambous:2021mea,Saketh:2023bul}). It is straightforward to translate between the Thorne tensor basis and this spin basis, see Appendix~\ref{app:Ttensorharmonics} for details.

In Sections~\ref{sec:InfD} and~\ref{sec:US} we will see examples of objects with enhanced symmetries compared to a generic rotating object, and consequently the couplings depend on a smaller set of eigenvalues than the full set $(L,M)$, which causes many of the Wilson coefficients $\lambda_{IJ}^{LM}$ to be equal. 
More general objects with no internal symmetries will potentially couple all possible $L$ and $M$ modes together, so that the matrix $\lambda_{IJ}^{LL'MM'}$ will be totally generic.

\subsection{One-Point Function}
\label{sec:1ptfunction}

We have now assembled the necessary ingredients to calculate the response of an object when placed in an external (possibly time-dependent) field. These responses are the physical quantities that we would like to match to an ultraviolet black hole perturbation theory calculation in order to infer the properties of various black hole solutions.

Conceptually, the experiment that we imagine performing is to expose the object of interest to an external field, and to record the response of the object, which will appear as an induced field.
From the EFT point of view, these responses are a consequence of the non-minimal couplings between the external fields and the worldline of the particle. From the ultraviolet point of view, these responses reflect the internal structure of the object, which responds to the external stimulus. A paradigmatic example is the electric polarizability of an object. If we immerse an object in an external electric field, $\vec E$, the microscopic charge carriers rearrange themselves in response, leading to an induced electric field. The strength of this response relative to the applied field is captured in EFT by a coupling between $\vec E^2$ and the worldline of the object. The coefficient of this coupling is the polarizability. In the following we want to perform a similar matching for black holes.

Practically, we will match these responses between the ultraviolet theory and EFT by computing the one-point function for the field $\phi$ in the presence of some external source. For weak external fields, we expect the induced one-point function to be linear in the amplitude of the applied field, which is the essence of linear response theory.
In the EFT, the response is due to couplings between the external field and the microscopic degrees of freedom of the object of the form
\be
S_{\rm int} \supset \int\rd\tau E\, Q^{i_1\cdots i_L}(X,E)\,C_{i_1\cdots i_L}\,,
\label{eq:worldlinecoupling}
\ee
where we will be interested in the case where the operator $C_{i_1\cdots i_L}$ is linear in the external field~\eqref{eq:externaloperator|}, and $Q^{i_1\cdots i_L}$ is a composite operator built from the worldline degrees of freedom $X$.
The object's linear response then arises via two copies of this interaction, diagrammatically of the form
\begin{equation*}
\includegraphics[scale=1.65]{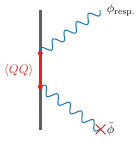}
\end{equation*}
where $\bar\phi$ represents the background source field that induces a response $\phi_{\rm resp.}$ by exciting the worldline degrees of freedom.
The properties of these degrees of freedom are captured by the two-point function of the operators $Q$. Note that the operator $Q^{i_1\cdots i_L}$ can be taken to be a traceless symmetric tensor because it couples to a traceless and symmetric combination of the external field.

We can compute this response using the Schwinger--Keldysh formalism and the coupling~\eqref{eq:worldlinecoupling}. We imagine doubling the fields as in~\eqref{eq:ininint} to capture the two branches of the in-in contour. We are interested in the classical field profile generated by the coupling to the object. To compute this response, it is convenient to rotate to the Keldysh basis 
\begin{align}
    \phi_- &\equiv \phi_1-\phi_2\,,  &\phi_+ &\equiv \tfrac{1}{2}\left(\phi_1+\phi_2\right)\,,\\
    X_- &\equiv X_1-X_2\,, &X_+ &\equiv \tfrac{1}{2}\left(X_1+X_2\right)\,,
\end{align}
where $\phi_+$ can be interpreted as  the classical field (in contrast $\phi_-$ corresponds to stochastic noise).
We then imagine splitting the classical field as
\be
\phi_+ = \bar\phi_+ + \varphi_+\,, 
\ee
where $\bar\phi$ is the applied background and $\varphi$ is the response of the object in this background. This response is given by the in-in path integral
\be
\langle \varphi_+(\vec x, t)\rangle = \int{\cal D}\varphi_+{\cal D}\varphi_-\,\varphi_+(\vec x, t) \,\E^{i\Gamma^\text{in-in}[\varphi]}\,,
\ee
where the effective action $\Gamma^\text{in-in}$ is obtained by integrating out $Q$ as in~\eqref{eq:ininint}. At leading order, we can integrate out $Q$ via its linear response
\be
\left\langle Q^{i_1\cdots i_L }(\tau) \right\rangle = \sum_{L'} \int\rd \tau' G^{(Q)}{}^{i_1\cdots i_L \,j_1\cdots j_{L'}}(\tau - \tau^\prime)C_{j_1\cdots j_{L'}}(\tau^\prime)\,,
\ee
where $G^{(Q)}{}^{i_1\cdots i_L \,j_1\cdots j_{L'}}$ is the Green's function for the $Q$ operators.
This leads to the effective action 
\be
i\Gamma_{\rm int}^\text{in-in} = - \sum_{L,L'}\int\rd\tau_1\rd\tau_2 \langle Q^{i_1\cdots i_L }_I(\tau_1)Q^{j_1\cdots j_{L'} }_J(\tau_2)\rangle\, C^I_{i_1\cdots i_{L}}(\tau_1)C^J_{j_1\cdots j_{L'}}(\tau_2)\,,
\label{eq:gammaint}
\ee
where $I,J$ run over $+,-$.
In order to compute the classical field profile, the relevant two-point function is that between the $+$ and $-$ fields (see Appendix~\ref{app:SK})
so that~\cite{Goldberger:2020fot}
\be
\langle \varphi_+(\vec x, t)\rangle = -\int\rd\tau_1\rd\tau_2 \,\langle \varphi_+(\vec x,t) C^-_{i_1\cdots i_L}(\tau_1)\rangle\langle Q^{i_1\cdots i_L }_+(\tau_1)Q^{j_1\cdots j_{L'} }_-(\tau_2)\rangle  C^+_{j_1\cdots j_{L'}}(\tau_2)\,.
\label{eq:phi1ptexp1}
\ee
The $\langle Q_+Q_-\rangle$ two-point function is just the causal Green's function
\be
G_{R}^{(Q)}{}^{i_1\cdots i_L \, j_1\cdots j_{L'}}(\tau_1-\tau_2) =i\langle Q_+^{i_1\cdots i_L}(\tau_1) Q_-^{j_1\cdots j_{L'}}(\tau_2)\rangle= i\theta(\tau_1-\tau_2)\langle [Q^{i_1\cdots i_L}(\tau_1),Q^{j_1\cdots j_{L'}}(\tau_2)]\rangle\,,
\label{eq:QQ2pt}
\ee
and similarly for the $\phi$ field, so that we can rewrite~\eqref{eq:phi1ptexp1} as 
\begin{tcolorbox}[colframe=white,arc=0pt,colback=greyish2]
\be
\label{eq:1ptfunctionexp}
\!\!\!\langle \varphi_+(\vec x, t)\rangle=\! \sum_{L,L'}\int\!\rd\tau_1\rd\tau_2 \, (-1)^L \, \partial_{i_1}\cdots \partial_{i_L}G_{R}^{(\varphi)}(\vec x, t-\tau_1)\,G_{R}^{(Q)}{}^{i_1\cdots i_L\, j_1\cdots j_{L'} }(\tau_1-\tau_2)
\partial_{j_1}\cdots \partial_{j_{L'}}\bar \phi(\tau_2)\,,
\ee
\end{tcolorbox}
\noindent
which is the response of the object to some applied external field.
Here we have written everything in terms of the causal Green's functions for the $Q$ variables and for the field $\varphi$.
(Note that we have dropped the explicit trace subtraction from the derivatives in~\eqref{eq:1ptfunctionexp} because they are contracted with the traceless Green's function of the $Q$ operators.)
In the following we will use this general expression to compute linear responses in the EFT.

\subsubsection{Rotating frames}
\label{sec:rotframes}
 
 Since many of the objects we will study will be rotating, it is worth discussing some subtleties related to rotating frame effects (see also~\cite{Goldberger:2020fot}).
For rotating objects, there is a mismatch between the (inertial) laboratory frame that is stationary at infinity and the (non-inertial) frame that corotates with the object. Similarly, there is a mismatch between frequencies defined in the two frames. (For example, a time-independent external field with  nontrivial angular dependence in the laboratory frame appears time dependent from the perspective of an observer corotating with the object.)

In equation~\eqref{eq:1ptfunctionexp}, the Green's function $G_{R}^{(Q)}{}^{i_1\cdots i_L \, j_1\cdots j_{L'} }$ is implicitly defined in the ``lab frame," which is the rest frame at infinity.\footnote{As a practical matter we will typically consider objects at rest so that the lab frame coincides with the rest frame of the object.}
However, in some cases we might be interested in defining various objects in a frame that corotates with the object.
The shift of the apparent frequency of the external field in different frames will cause 
observers in the lab and  corotating frame to disagree on the precise functional form of the frequency space Green's function. 
However, we can explicitly calculate this shift 
using the vielbein~\eqref{eq:restframevielbein} 
to change coordinates between the lab frame---here denoted with $i,j$ indices---and the black hole's corotating frame, denoted with $a,b$ indices.

The worldline Green's function, as defined in the lab frame, can be written in Fourier space in terms of Thorne tensors as~\eqref{eq:freqspacegreensf}
\be
    \label{eq:labframeGFfourierexpansion}
    G_{R}^{(Q)}{}^{i_1\cdots i_L \,j_1\cdots j_{L'} }(\omega) = \sum_{M,M'} \lambda^{({\rm lab})}_{LL'MM'}(\omega) \,\mathcal{Y}^{i_1\cdots i_L}_{LM} \mathcal{Y}_{L'M'}^*{}^{j_1\cdots j_{L'}} \,,
\ee
where we have denoted that the coefficients $\lambda^{({\rm lab})}(\omega)$ are defined in the {\it lab} frame.
As we have discussed, when the dynamics of the worldline degrees of freedom are fast compared to the external modes of interest, the function $\lambda(\omega)$ can be reliably Taylor expanded.

The Fourier transform of the Green's function defined in the {\it corotating} frame would define a different set of coefficients $\lambda^{({\rm rot.})}_{LL'MM'}(T)$, which is the Green's function measured in the corotating frame
\begin{equation}
    \label{eq:corotatingframeGFfourierexpansion}
    G_{R}^{(Q)}{}^{a_1\cdots a_L \,b_1\cdots b_{L'} }(\tau-\tau') = \sum_{M,M'} \lambda^{({\rm rot})}_{LL'MM'}(\tau-\tau') \,\mathcal{Y}^{a_1\cdots a_L}_{LM} \mathcal{Y}_{L'M'}^*{}^{b_1\cdots b_{L'}} \,.
\end{equation}
We can use the vielbein to relate the lab and corotating Green's functions  to determine their relationship
\be
G_{R}^{(Q)}{}^{i_1\cdots i_L \, j_1\cdots j_{L'}}(\tau-\tau') 
= e_{a_1}^{i_1}(\tau)\cdots e_{a_L}^{i_L}(\tau)e_{b_1}^{j_1}(\tau')\cdots e_{b_L}^{j_{L'}}(\tau') \,
G_{R}^{(Q)}{}^{a_1\cdots a_L \,b_1\cdots b_{L'} }(\tau-\tau')\,.
\ee
It is convenient to use the time-translation invariance of the system to shift $\tau' \to 0$, relabel $\tau=T$, and
utilize the expansion in Thorne tensors so that
\be
G_{R}^{(Q)}{}^{i_1\cdots i_L \, j_1\cdots j_{L'}}(T) = \sum_{M,M'}\lambda^{({\rm rot.})}_{LL'MM'}(T)\,e_{a_1}^{i_1}(T)\cdots e_{a_L}^{i_L}(T)\,e_{b_1}^{j_1}(0)\cdots e_{b_L}^{j_{L'}}(0) \mathcal{Y}^{a_1\cdots a_L}_{LM} \mathcal{Y}_{L'M'}^*{}^{b_1\cdots b_{L'}}.
\ee
A benefit of expanding in terms of Thorne tensors is that we can use the following fact to simplify the rotating-frame effects (see Appendix~\ref{appendix:Thorne})
\be
e_{a_1}^{i_1}(T)\cdots e_{a_L}^{i_L}(T)\,\mathcal{Y}^{a_1\cdots a_L}_{LM}= \E^{-i M_n \Omega_n T}\, \mathcal{Y}^{i_1\cdots i_L}_{LM}\,,
\label{eq:rotviel}
\ee
where $M_n\Omega_n \equiv m_1\Omega_2+m_2\Omega_2+\cdots$ is the sum of the magnetic quantum numbers times the corresponding angular frequencies.
Together, this means that we can write the lab-frame Green's function abstractly as 
\be
    G_{R}^{(Q)}{}^{i_1\cdots i_L \,j_1\cdots j_{L'} }(T)=\sum_{M,M'} \E^{-iM_n \Omega_n T} \lambda^{({\rm rot.})}_{LL'MM'}(T)\,\mathcal{Y}^{i_1\cdots i_L}_{LM} \mathcal{Y}_{L'M'}^*{}^{j_1\cdots j_{L'}}\,.
\ee
Now if we Fourier transform this expression, we get
\be
    G_{R}^{(Q)}{}^{i_1\cdots i_L \,j_1\cdots j_{L'} }(\omega)=\sum_{M,M'}\lambda^{({\rm rot.})}_{LL'MM'}(\omega-M_n \Omega_n)\,\mathcal{Y}^{i_1\cdots i_L}_{LM} \mathcal{Y}_{L'M'}^*{}^{j_1\cdots j_{L'}}\,.
\label{eq:rotatingframegreens}
\ee
Comparing this to~\eqref{eq:labframeGFfourierexpansion}, we see that $\lambda^{({\rm lab})}_{LL'MM'}(\omega)=\lambda^{({\rm rot.})}_{LL'MM'}(\omega- M_n \Omega_n)$---the effective frequency in the corotating frame is shifted to $(\omega- M_n \Omega_n)$. This yields the expected result: if we apply a static external field, it will have an apparent time dependence in the non-inertial, corotating frame of the object.

It is worth emphasizing that the function $\lambda^{({\rm rot.})}_{LL',MM'}$ itself may depend explicitly on the spin of the object, so that it is not possible to fully determine the properties of a rotating object from measurements of a non-rotating one.\footnote{A simple example of this phenomenon is that one cannot a priori predict the spin-induced multipole moments of an object from knowledge of the object's static multipoles.}
However, we can make one inference from the non-rotating Green's function:
the leading $\ord{\omega}$ Taylor coefficient of the Green's function~\eqref{eq:labframeGFfourierexpansion} in the laboratory frame captures the leading dissipative response of the object. From the relation between the Green's functions in the lab and corotating frame, we see that the {\it leading order in spin} dissipative response at zero frequency will take the same numerical value for a spinning object as it does at order $\omega$ for a non-spinning object.
This can be made explicit by Taylor expanding the relation  $\lambda^{({\rm lab})}_{LL'MM'}(\omega)=\lambda^{({\rm rot.})}_{LL'MM'}(\omega- M_n \Omega_n)$ to obtain
\be
 \lambda^{({\rm lab})}_{LL'MM'}(\Omega^2)+\omega \lambda'{}^{({\rm lab})}_{LL'MM'}(\Omega^2)+\cdots =\lambda^{({\rm rot.})}_{LL'MM'}(\Omega^2)+\lambda'{}^{({\rm rot.})}_{LL'MM'}(\Omega^2) \,(\omega-M_n\Omega_n)+\cdots\,,
 \label{eq:disspiativeexp}
\ee
where we have kept explicit that the various Taylor coefficients can depend on the spin of the black hole squared $\Omega^2$. As $\Omega\to 0$, we must have that 
\begin{align}
 \lambda^{({\rm lab})}_{LL'MM'}(0) &= \lambda^{({\rm rot.})}_{LL'MM'}(0)\,,\\
 \lambda'{}^{({\rm lab})}_{LL'MM'}(0) &= \lambda'{}^{({\rm rot.})}_{LL'MM'}(0) \,,
 \label{eq:disspiativeisrotating}
\end{align}
which just expresses the fact that quantities measured in the lab or rotating frame must be the same in the limit that the object is not spinning. Then, we see that in the leading $\Omega \to 0$ limit,~\eqref{eq:disspiativeexp} and~\eqref{eq:disspiativeisrotating} together imply that the $\ord{\Omega}$ dissipative response in the rotating frame is the same as the $\ord{\omega}$ dissipative response of a non-rotating object. This does not survive at higher order because different orders in the Taylor expansion~\eqref{eq:disspiativeexp} can mix when expanded in $\Omega$.

\subsubsection{Small frequency expansion}

We will be interested in the responses of black holes to relatively slowly varying external fields.
So, we would like to use the formula~\eqref{eq:1ptfunctionexp} to compute these responses order-by-order in a small frequency expansion. Concretely, we consider an external field of the form
\begin{equation}
    \bar{\phi}=\E^{-i \omega t}\mathcal{E}_{j_1 \cdots j_{\ell}} x^{j_1} \cdots x^{j_{\ell}},
\end{equation}
where importantly both $\omega$ and $x^j$ are defined with respect to the {\it lab} frame. In this frame, the particle has spacetime coordinates $x^\mu(\tau)=\tau\delta^{\mu}_0$ (i.e., it sits at the origin of spatial coordinates). It is therefore useful to first notice that
\be
\partial_{(i_1}\cdots \partial_{i_{L'})_T} x^{j_1} \cdots x^{j_{\ell}}  =\delta_{\ell L'} L'!\,\delta^{j_1\cdots j_{L'}}_{(i_1\cdots i_{L'})_T}\,,
\ee
so that we can set $\ell = L'$ in the following.

In order to use the formula~\eqref{eq:1ptfunctionexp}, it is convenient to solve for the field in momentum space in an expansion in the $\omega\to 0$ limit.
At leading order, we can use the static propagator for $\varphi$, which in momentum space is: 
\be
G^{(\varphi)}(\vec p, t-t')= -\frac{1}{p^2} \delta(t -t')\,.
\label{eq:staticphiprop}
\ee
This approximation is accurate to order $\ord{\omega}$ when solving for the spatial profile of $\varphi$. Using this expression in~\eqref{eq:1ptfunctionexp} and evaluating the expression in Fourier space yields 
\be
\label{eq:1ptgreens2}
\langle \varphi_+(\vec p, t)\rangle 
=\sum_L(-i)^L\E^{-i \omega t}\frac{p_{(i_1}\cdots p_{i_L)_T}}{p^2}\,L'!\,\mathcal{E}_{j_1 \cdots j_{L'}}\int\rd T\E^{i \omega T} \,G_{R}^{(Q)}{}^{i_1\cdots i_L |j_1\cdots j_{L'} }(T)
\,,
\ee
where we have defined $T\equiv t-\tau_2$. This formula involves the frequency Fourier transform of the lab frame Green's function, which can be written as
\be
    G_{R}^{(Q)}{}^{i_1\cdots i_L |j_1\cdots j_{L'} }(\omega) = \sum_{M,M'} \lambda_{LL'MM'}(\omega) \,\mathcal{Y}^{i_1\cdots i_L}_{LM} \mathcal{Y}_{L'M'}^*{}^{j_1\cdots j_{L'}} \,.
    \label{eq:freqspacegreensf}
\ee
Using this expression in~\eqref{eq:1ptgreens2}, we obtain 
\be
\langle \varphi_+(\vec p, t)\rangle 
=\sum_L(-i)^L\E^{-i \omega t}\frac{p_{(i_1}\cdots p_{i_L)_T}}{p^2}L'!\,\mathcal{E}_{j_1 \cdots j_{L'}} \sum_{M,M'} \,\lambda_{LL'\,MM'}(\omega)\,\mathcal{Y}^{i_1\cdots i_L}_{LM} \mathcal{Y}_{L'M'}^*{}^{j_1\cdots j_{L'}}
\,.
\ee
We can then
go back to real space using the following formula\footnote{This formula can be obtained by taking $x_i$ derivatives of the standard Fourier integral 
\begin{equation}\label{propagatorFouriertoreal}
    \int \frac{\rd^d p}{(2 \pi)^d} \E^{i\vec p\cdot \vec x} \frac{1}{p^2}= \frac{\Gamma(\tfrac{d}{2}-1)}{(4\pi)^{d/2}}\left(\frac{\vec x^2}{4}\right)^{1-\tfrac{d}{2}}\,.
\end{equation}
}
\begin{equation} 
    i^L \int \frac{\rd^d p}{(2 \pi)^d} \E^{i\vec p\cdot \vec x}\, \frac{p^{(i_1} \cdots p^{i_L)_T}}{p^2}=  \frac{\Gamma(\tfrac{d}{2}-1)\Gamma(2-\tfrac{d}{2})}{2^L(4\pi)^{d/2} \Gamma(2-\tfrac{d}{2}-L)}  x^{(i_1} \cdots x^{i_L)_T}\left(\frac{x^2}{4}\right)^{1-\tfrac{d}{2}-L}\,.
    \label{eq:fourierintegral}
\end{equation}
Putting this all together, we find that the real space response is given by
\begin{align}
    \langle \varphi_+(\vec x, t)\rangle 
    &= \E^{-i \omega t}\sum_{L,M,M'} \lambda_{LL'\,MM'}(\omega)\, c_{LL'}\, r^{-2L-d+2}\Big( r^L\mathcal{E}_{L'M'} Y_{LM}\Big)\,,\\
    c_{LL'}&\equiv  
    \frac{\Gamma(L+\tfrac{d}{2}-1)\Gamma(L'+\tfrac{d}{2})}{2^{3-L-L'}\pi^d}\,,
\end{align}
where we have used~\eqref{eq:ThorneTensorcomponentNormgeneralD}.
Note that $\sum_Mr^L\mathcal{E}_{LM} Y_{LM}=\bar{\phi}$, so $\langle \varphi \rangle \propto \bar{\phi}$ as expected. 
We can expand this expression in powers of $r_s \omega$ as\footnote{Note that the formula~\eqref{eq:smallfreq1ptfinal} is only really reliable to order $\omega$, which will be sufficient for our purposes. If we wanted to match to subleading order in $\omega$, we would have to go beyond the approximation~\eqref{eq:staticphiprop}.}
\begin{tcolorbox}[colframe=white,arc=0pt,colback=greyish2]
\be
    \label{eq:smallfreq1ptfinal}
    \langle \varphi_+(\vec x, t)\rangle =\E^{-i \omega t}\sum_{L,M,M',n} \lambda^{(n)}_{LL'\,MM'}  (r_s  \omega)^n c_{LL'}\, r^{-2L-d+2}\Big( r^L\mathcal{E}_{L'M'} Y_{LM}\Big)\,.
\ee
\end{tcolorbox}
\noindent
At this level, $r_s$ is an arbitrary dimensionful scale, which will be determined through matching. This organizes all the unknown information into the Taylor coefficient matrices ${\lambda}^{(n)}_{LL'\,MM'}$, which we determine by matching to an explicit UV calculation.

To isolate the conservative and dissipative contributions to the one point function, we further expand each Wilson coefficient into its real and imaginary components as 
\begin{equation}
    \label{eq:knuexpansionoflambda}
    \lambda^{(n)}_{LL'\,MM'}= k^{(n)}_{LL'\,MM'} +i\, \nu^{(n)}_{LL'\,MM'}\,.
\end{equation}
The real part of these coefficients, $k$, encode information about the conservative response of the object, while the imaginary part, $\nu$, contains information about dissipative physics. In the following we will determine these Wilson coefficients by matching to various general relativity solutions.

Here we have worked directly in terms of lab-frame quantities. This is natural from the point of view of performing a thought experiment outside a black hole.
These quantities can be straightforwardly related to coefficients measured in the co-rotating frame using the formulas in Section~\ref{sec:rotframes}.

\subsection{Matching}
The formula~\eqref{eq:smallfreq1ptfinal} encapsulates the response in the EFT of the particle to an external field that couples through an interaction of the form~\eqref{eq:worldlinecoupling}. At this level, the formula still depends on a number of unknown parameters---the Taylor coefficients of the frequency-space Green's function~\eqref{eq:freqspacegreensf}. There is also scheme dependence in the fact that we can---in addition to couplings like~\eqref{eq:worldlinecoupling} to worldline $X$ variables---include conservative contact interactions involving the external fields on the worldline. The effects of these contact interactions are degenerate with time reversal even terms in the Green's function of the $X$ (or $Q$) operators. In order to fix these unknowns, we {\it match} the physical response computed in the EFT to a microphysical calculation. The utility of this approach is that once we have fixed the couplings in the EFT by matching, we can then use the effective description to compute any further quantity of interest.

In the case of interest, the ultraviolet theory will be Einstein gravity. More specifically, we will consider small perturbations around various black hole solutions in diverse dimensions. The EFT one-point function captures the linear response of an object to external fields. From the ultraviolet point of view, these responses are captured by solutions to the wave equation in black hole geometries---subject to certain boundary conditions. Therefore, it is essentially the ultraviolet information about boundary conditions that ends up being encoded in the Wilson coefficients in the EFT. In the following we will carry out this matching procedure in a number of examples.

\newpage
\section{Schwarzschild}
\label{sec:DdimSch}

As a first example, we consider the Schwarzschild black hole. The response coefficients in this case are well studied~\cite{Damour:2009vw,Binnington:2009bb,Kol:2011vg,Hui:2020xxx,Ivanov:2022hlo}, but it will provide a simple illustrative arena. There are two aspects to the computation: we must first consider solutions of the wave equation in a Schwarzschild background with the right boundary conditions in order to extract the induced response to a tidal field. We then match these responses to extract the Wilson coefficients in the EFT.

\vspace{-4pt}
\subsection{General Relativity Calculation}
\label{sec:DdimSchGRcalc}

We first consider the Einstein gravity calculation that will be used to infer the properties of the black hole.
 At the microscopic level, a black hole's linear tidal response is captured by linearized fluctuations in the background geometry sourced by the black hole. We can expand the resulting Klein--Gordon equation in a small-frequency expansion. The particular solution of interest involves fixing boundary conditions at infinity corresponding to a tidal external field, and demanding ingoing boundary conditions at the horizon of the black hole.
These solutions can be used to identify the black hole's conservative and dissipative responses to being immersed in a scalar field background by  matching the field profiles to the EFT one-point function~\eqref{eq:smallfreq1ptfinal}.

We start by
considering the Schwarzschild--Tangherlini metric in $D$ spacetime dimensions,
\begin{equation}
    \rd s^2= -f(r) \rd t^2 +\frac{1}{f(r)}\rd r^2 + r^2 \rd\Omega^2_{S^{D-2}},~~\text{where} \quad f(r)= 1-\left(\frac{r_s}{r}\right)^{D-3}\,,
    \label{eq:schmetric}
\end{equation}
where $r_s$ parametrizes the event horizon and $\rd\Omega^2_{S^{D-2}}$ is the line element on the $(D-2)$-sphere.\footnote{Recall that the line element of the generic $(k+1)$-sphere can be defined recursively as
$
\rd \Omega_{S^{k+1}}^2= \rd\theta_{k+1}^2 + \sin^2\theta_{k+1} \rd \Omega_{S^{k}}^2\,,$
where the $1$-sphere line element is simply $\rd \Omega_{S^1}^2 = \rd\theta^2$.} We want to study the dynamics of a massless scalar in this background, which satisfies the Klein--Gordon equation $\square\Phi = 0$.
Since the background is spherically symmetric and time-translation invariant, it is convenient to decompose the field as 
\begin{equation}
    \Phi(x)= \sum_{L,M} \E^{-i\omega t}R_L(r) Y_{LM}(\theta) \,,
\end{equation}
where $Y_{LM}(\theta)$ are spherical harmonics on the $(D-2)$-sphere.
Using the eigenvalue equation $\Delta_{S^{D-2}}Y_{LM} =-L(L+D-3)Y_{LM}$, the Klein--Gordon equation implies $R_L(r)$ satisfies
\begin{equation}
    \partial_r\Big(r^{D-2} f(r) \partial_r R_L\Big)+ \left(\frac{\omega ^2 r^{D-2}}{f(r)}-L (L+D-3) r^{D-4}\right)R_L=0\,.
    \label{eq:Requationsch}
\end{equation}
In order to study this equation, it is convenient to make the following variable definitions 
\begin{equation}
   P\equiv \frac{-r_s\omega}{D-3},\qquad\quad \hat{L}\equiv \frac{L}{D-3},\qquad\quad \rho \equiv \left(\frac{r}{r_s}\right)^{D-3}\,,
\end{equation}
in terms of which~\eqref{eq:Requationsch} takes the form
\begin{equation}
    \rho(\rho-1)\partial_\rho \Big(\rho(\rho-1)\partial_\rho R_L \Big)+\left(P^2 \rho^{2+\tfrac{2}{D-3}}-\hat{L}(\hat{L}+1)\rho(\rho-1)\right)R_L=0\,.
    \label{eq:ddimscheq}
\end{equation}

For generic $P$, this differential equation is of the confluent Heun type, with $\rho = \infty$ an irregular singularity. In contrast, when $P=0$, the equation reduces to a Legendre equation. We want to solve this equation for nonzero $P$; to do so, we will need to make
a {\it near zone} approximation, which is effectively a small-frequency approximation (in these variables a small $P$ approximation).\footnote{While it is possible to solve~\eqref{eq:ddimscheq} for arbitrary values of the frequency in terms of (confluent) Heun functions, the connection formulas required  to extract responses are not known in complete generality (though see~\cite{Aminov:2020yma,Bonelli:2021uvf,Consoli:2022eey,Bautista:2023sdf} for recent progress). Consequently we approximate this equation to cast it in hypergeometric form.}

\vspace{-12pt}
\paragraph{Near zone approximation:}To cast~\eqref{eq:ddimscheq} in hypergeometric form, we approximate
\be
P^2 \rho^{2+\tfrac{2}{D-3}} \to P^2\,,
\ee
in~\eqref{eq:ddimscheq}. This approximation is reliable when $r\omega \ll 1$, and for this reason we refer to it as a near zone approximation. That is, solutions to the approximated equation will agree with solutions to the full equation to high accuracy at distances nearer to the black hole than the inverse frequency. The guiding principles behind making this particular replacement are that
\vspace{-4pt}
\begin{itemize}
\item Near the horizon ($\rho = 1$), we want the solution of the approximated equation to have the same fall-off as solutions to the full equation, which scale as $\sim\left[(\rho-1)/\rho\right]^{\pm iP}$.

\vspace{-2pt}
\item The equation of motion is not altered at the order in which we would like to trust our final result, here $\ord{P}$. (Note that it is not possible to leave all terms at $\ord{P^2}$ unaltered and cast the resulting equation in hypergeometric form.)
\end{itemize}
\vspace{-4pt}
Note that
these two requirements would allow the inclusion of terms of the form $ P^2 \rho$ or $P^2 \rho^2$ while maintaining hypergeometric form, however we exclude them here so that the near zone equation has a deformation of the symmetries identified in~\cite{Hui:2021vcv}. Since we have altered~\eqref{eq:ddimscheq} at $\ord{P^2}$, we should not expect the solutions to be reliable at this order, but lower orders in $P$ capture the correct behavior.\footnote{This is somewhat subtle. In principle if we only want to be able to trust the solution at $\ord P$, it would seem that we could just drop the $P^2$ term in~\eqref{eq:ddimscheq}. While this would capture the static limit correctly, it would not produce the right fall-off matching the full equation near the horizon at finite frequency. This requires keeping some terms at $\ord{P^2}$, but since we have changed the equation at this order, the part of the solution at $\ord{P^2}$ is not reliable. In contrast, the $\ord P$ terms come from boundary conditions, and so can be trusted.} (See Appendix~\ref{app:NZ} for a more general discussion on near-zone approximations.) 
After making the near zone approximation, the equation that we want to solve is
\begin{tcolorbox}[colframe=white,arc=0pt,colback=greyish2]
\begin{equation}
    \label{eq:DdimSchzEoM}
    \rho(\rho-1)\partial_\rho \Big(\rho(\rho-1)\partial_\rho R_L \Big)+\left(P^2 -\hat{L}(\hat{L}+1)\rho(\rho-1)\right)R_L=0\,.
\end{equation}
\end{tcolorbox}
\noindent
In order to cast this equation in hypergeometric form, it is convenient to define a new coordinate $z\equiv \rho -1$, which shifts the horizon to $z=0$, and extract a factor by defining
\begin{equation}
    R_L(z)=\left(\frac{z}{z+1}\right)^{i P} u_L(z)\,,
\end{equation}
after which $u(z)$ satisfies
\begin{equation}
    z(z+1) u_L^{\prime \prime}(z) + (1+2iP +2z) u_L^\prime(z) -\hat L(\hat L+1) u_L(z)=0\,.
    \label{eq:standardhypersch}
\end{equation}
which is a hypergeometric equation in standard form.

\paragraph{Solutions and boundary conditions:} We now want to solve~\eqref{eq:standardhypersch}. To do so, we have to impose boundary conditions. We first choose the solution to have a  leading field profile which grows like $R\sim r^L$ as $r\to\infty$. This boundary condition captures the physics of immersing the black hole in an external scalar field profile of some given amplitude. The second boundary condition is that we require that the solution is purely ingoing at the black hole horizon (which is the fall-off that scales like $\left[z/(z+1)\right]^{i P}$ as $z\to 0$).
The solution that satisfies these two boundary conditions is~\cite{1972ApJ175243P,Detweiler:1980uk,Charalambous:2021mea} 
\begin{equation}
u_L(z) \propto\,{}_2F_1\left[\begin{array}{c}
-\hat L\,,\,\hat L+1\\[-3pt]
1+2iP
\end{array}\Big\rvert \,-z\,\right]\,.
\label{eq:regularsolution}
\end{equation}
The quantity of interest is the ratio of fall-offs of this solution at spatial infinity ($z\to\infty$). We can extract these by expanding the solution~\eqref{eq:regularsolution} using the connection formula (see e.g., the~\href{https://dlmf.nist.gov/15.10\#ii}{\tt DLMF})
\be
\label{eq:2F1connectionformula}
\begin{aligned}
{}_2F_1\left[\begin{array}{c}
-\hat L\,,\,\hat L+1\\[-3pt]
1+2iP
\end{array}\Big\rvert \,-z\,\right] =~& \frac{\Gamma(1+2iP) \Gamma(2\hat L+1)}{\Gamma(\hat L+1) \Gamma(1+\hat L+2iP)} z^{\hat L}{}_2F_1\left[\begin{array}{c}
-\hat L\,,\,\hat L-2iP\\[-3pt]
-2\hat L
\end{array}\Big\rvert \,-\frac{1}{z}\,\right] \\
&
+\frac{\Gamma(1+2iP) \Gamma(-2\hat L-1)}{\Gamma(-\hat L)\Gamma(-\hat L+2iP)} z^{-\hat L-1}
{}_2F_1\left[\begin{array}{c}
\hat L+1\,,\,\hat L+1-2iP\\[-3pt]
2\hat L+2
\end{array}\Big\rvert \,-\frac{1}{z}\,\right] \,.
\end{aligned}
\ee
From this, we find that the expansion of $R_L(r)$ as $r\to \infty$ is (recalling $z= (r/r_s)^{D-3} -1$)
\begin{tcolorbox}[colframe=white,arc=0pt,colback=greyish2]
\begin{equation}\label{SWUVLovenumber}
    R_L(r) \xrightarrow[r \rightarrow \infty]{} \left(\frac{r}{r_s}\right)^L  \left( 1+ \ldots + \left(\frac{r}{r_s}\right)^{-2L-D+3} \frac{\Gamma(-2 \hat{L} -1)\Gamma(\hat{L}+1) \Gamma(1+\hat{L}-2i \frac{r_s\omega}{D-3})}{\Gamma(2\hat{L}+1)\Gamma(-\hat{L})\Gamma(-\hat{L}-2i \frac{r_s\omega}{D-3})}\right)\,.
\end{equation}
\end{tcolorbox}
\noindent
Using this solution, we can extract the conservative and dissipative responses of a Schwarzschild black hole up to $\ord{\omega}$, as we will see in the next section.  

\vspace{-4pt}
\subsection{EFT Matching}

We now want to match the solution~\eqref{SWUVLovenumber} to the EFT one-point function~\eqref{eq:smallfreq1ptfinal} in order to extract the conservative and dissipative responses of general dimension Schwarzschild black holes.

We first expand the result of the microscopic calculation~\eqref{SWUVLovenumber} in the small $\omega$ limit:
\be
    R_L(r)= \left(\frac{r}{r_s}\right)^L \Bigg( 1+ \cdots + \left(\frac{r}{r_s}\right)^{2-2L-d} \frac{ \Gamma (\hat{L} +1)^4}{2\pi \Gamma(2 \hat L+1)\Gamma(2\hat L +2)}\left[ \tan(\pi \hat L)+  \frac{2 i \pi r_s\omega}{d-2} +\ord{\omega^2} \right]
    \Bigg).
    \label{eq:schmicro}
\ee
Here we have dropped terms that are subleading because only terms up to $\ord{\omega}$ are 
reliable, given the approximations that we have made.
We have factored out an overall factor of  $r^L$, so that the induced response explicitly is proportional to the applied field.
Notice that the leading static response is real (indicating it is a conservative effect), while the first subleading response at $\ord{\omega}$ is purely imaginary (indicating it is a dissipative effect).

\vspace{-12pt}
\paragraph{EFT calculation:} We can adapt the EFT calculation to this situation. 
Note first that for a spherically symmetric black hole, rotational invariance guarantees that we will be able to diagonalize the two-point function~\eqref{eq:freqspacegreensf} so that $L' = L$ and $M' = M$.  We can then write~\eqref{eq:smallfreq1ptfinal} as
\begin{equation}
    \label{eq:smallfreq1ptfinalschwarz}
    \langle \varphi_+(\vec x, t)\rangle = \E^{-i \omega t}\sum_Mr^L\mathcal{E}_{LM}\, Y_{LM}\left( \lambda^{(0)}_{LM}+(r_s\omega)\,
    \lambda^{(1)}_{LM}+\cdots \right)\frac{\Gamma(L+\tfrac{d}{2})\Gamma(L+\tfrac{d}{2}-1)}{2^{3-2L}\pi^d}r^{2-2L-d}\,,
\end{equation}
where we may further expand $\lambda^{(n)}_{LM}= k^{(n)}_{LM} +i\, \nu^{(n)}_{LM}$. 
We can now compare~\eqref{eq:smallfreq1ptfinalschwarz} and~\eqref{eq:schmicro} to read off the Wilson coefficients.

\vspace{-4pt}
\subsubsection{Conservative response}

We first match the conservative response, which is given by the real part of the expansion~\eqref{eq:schmicro}. 
All of the $k,\nu$ coefficients will be proportional to the factor
\be
    \lambda_{L}^c \equiv r_s^{2L+d-2}\frac{2^{-2L-d+3}(2\pi)^{d} \Gamma (\hat L +1)^4}{\Gamma(L+\tfrac{d}{2})\Gamma(L+\tfrac{d}{2}-1)\Gamma(2\hat L+1)\Gamma(2\hat L +2)}\,,
\ee
which never vanishes or diverges for physical values of the angular momentum.
Then, comparing to~\eqref{eq:smallfreq1ptfinalschwarz}, and using the fact that $\bar\phi = r^L\mathcal{E}_{LM}\, Y_{LM}$, we deduce
\begin{tcolorbox}[colframe=white,arc=0pt,colback=greyish2]
\vspace{-12pt}
    \begin{align}
        \label{eq:Schwconservativeresp}
        k_{LM}^{(0)} &=  \lambda_{L}^c \frac{\tan \big(\pi \hat L\big)}{2\pi}\,, \\ 
        k_{LM}^{(1)} &= 0 \,,
    \end{align}
\end{tcolorbox}
\noindent
which matches~\cite{Kol:2011vg,Hui:2020xxx}. Note that the response does not depend on the magnetic quantum numbers, which is to be expected because the Schwarzschild black hole is rotationally symmetric. As is by now well known, these responses have some interesting features. Most notably, when $\hat L = L/(D-3)$ is an integer, the response vanishes. Additionally, when $\hat L$ is half-integer, the response formally diverges, but a careful treatment reveals that it scales logarithmically with distance, indicating a classical running~\cite{Kol:2011vg,Hui:2020xxx}. As there are no other time reversal odd quantities in the Schwarzschild problem aside from $\omega$, the term linear in $\omega$ must break time reversal symmetry and contribute a dissipative response. Thus, the vanishing of $k_{LM}^{(1)}$ is expected.

\vspace{-4pt}
\subsubsection{Dissipative response}
We can also read off the dissipative response by comparing~\eqref{eq:schmicro} with~\eqref{eq:smallfreq1ptfinalschwarz} at
$\ord{\omega}$, to find
\begin{tcolorbox}[colframe=white,arc=0pt,colback=greyish2]
\vspace{-12pt}
    \begin{align}
        \nu_{LM}^{(0)} &=  0 \,,\\ 
        \nu_{LM}^{(1)} &=  \frac{\lambda_{L}^c}{D-3} \,,
    \end{align}
\end{tcolorbox}
\noindent
We see that again the response does not depend on $M$, as expected. (These results are also consistent with those of~\cite{Ivanov:2022hlo,Charalambous:2024tdj}.) We again expect $\nu_{LM}^{(0)}$ to vanish because the only time reversal odd quantity is the frequency, so that the dissipative response scales with $\omega$. The dissipative response is a simple multiple of the overall prefactor that never vanishes or diverges.

\newpage
\section{Kerr}
\label{sec:Kerr}

We next consider the four-dimensional Kerr black hole. Similar to the Schwarzschild case, we first solve the wave equation in this geometry in order to extract the tidal response. These coefficients have been computed before~\cite{1972ApJ175243P,LeTiec:2020spy,Chia:2020yla,Goldberger:2020fot,LeTiec:2020bos,Charalambous:2021mea,Charalambous:2021kcz,Hui:2021vcv}, but it is worthwhile in the present context to review the calculation because it nicely illustrates subtleties associated to rotating frame effects. 
By choosing a suitable near-zone truncation, we are further able to compute the Kerr responses to all orders in the spin at leading order in the external field's frequency.

\subsection{General Relativity Calculation}
\label{sec:KerrGRcalc}
Here we compute the response of a Kerr black hole to being immersed in a time dependent scalar field in Einstein gravity.
We start with the Kerr metric in Boyer--Lindquist coordinates
\be
    \label{eq:BLmetric}
    \rd s^2=\!-\left(1-\frac{r_sr}{\Sigma}\right)\rd t^2 -\frac{2ar_sr\sin^2\theta}{\Sigma}\rd t \rd \phi + \frac{\Sigma}{\Delta}\rd r^2 +\Sigma \rd \theta^2+
    \left(r^2+a^2+\frac{r_sra^2 \sin^2\theta}{\Sigma}\right)\sin^2\theta \rd \phi^2,    
\ee
where we have defined the quantities
\begin{equation}
    \Sigma \equiv r^2 +a^2 \cos^2\theta,\qquad \Delta \equiv r^2 -r_sr +a^2 = (r-r_-)(r-r_+), \qquad r_\pm \equiv \tfrac{r_s}{2}\pm\sqrt{\tfrac{r_s^2}{4}-a^2} \,.
\end{equation}
Here $r_\pm$ are the locations of the inner and outer horizons (with $r_s = 2GM$).

The Kerr spacetime is both axially symmetric and stationary, so it is natural to make the ansatz
\begin{equation}
    \Phi=\E^{-i \omega t}\E^{i m \phi} S_{\omega\lambda m}(\theta)R_{\omega\lambda m}(r)\,,
\end{equation}
with this parameterization, the Klein--Gordon equation ($\square \Phi=0$) separates into two equations that govern the angular and radial components of the field:
\begin{align}
\label{eq:spheroidaleq}
    \frac{1}{\sin{\theta}}\frac{\rd}{\rd \theta}\left(\sin{\theta}\frac{\rd S_{\omega\lambda m}(\theta)}{\rd\theta}\right)+\left[a^2 \omega^2 \cos^2{\theta}-\frac{m^2}{\sin^2{\theta}}+\lambda\right]S_{\omega\lambda m}(\theta)&=0\,,\\
    \label{eq:KerrradialKGequation}
    \Delta \frac{\rd}{\rd r}\left(\Delta \frac{\rd R_{\omega\lambda m}(r)}{\rd r}\right)+\Big[ \omega ^2\left(a^2+r^2\right)^2-2 a m  r_s \omega \,r+a^2 m^2-\Delta  \left(a^2 \omega ^2+\lambda \right)\Big]R_{\omega\lambda m}(r)&=0\,.
\end{align}
where $\lambda$ is a separation constant. In general the functions $S_{\omega\lambda m}(\theta)$ are so-called spheroidal harmonics~\cite{Berti:2005gp}, and the separation constants $\lambda$ are not known in closed form. However, when $\omega r_s \ll 1$ (and consequently $\omega a \ll 1$) the equation~\eqref{eq:spheroidaleq} reduces to the associated Legendre equation and the spheroidal harmonics reduce to associated Legendre polynomials. Thus, at small frequency, the angular sector of the wave equation in a Kerr background is solved by spherical harmonics, $Y_{L m}(\theta,\phi)$, and we may set $\lambda \rightarrow L(L+1) +\ord{\omega^2}$.\footnote{Of course, this approximation fails at higher orders in $\omega$ and we must confront the fact that the geometry is only axially symmetric. We will only be concerned with leading order behavior at small frequency, and so we will not need to deal with this complication here.}

 As before, we must make some approximations in order to be able to solve~\eqref{eq:KerrradialKGequation} analytically.

\paragraph{Near zone:}
The near zone approximation for Kerr is chosen with similar guiding principles as in the Schwarzschild case considered in Section~\ref{sec:DdimSchGRcalc}. That is, we want to approximate the radial KG equation so that it takes a hypergeometric form. 

As before, one requirement is that the approximate solution matches the full solution at the black hole horizon (i.e. that it scales as $\sim(r-r_+)^{iP}$ as $r\to r_+$, for an appropriately defined $P$).

In the Kerr case, the second requirement that we imposed in Section~\ref{sec:DdimSchGRcalc} (leaving the equation unchanged at ${\cal O}(\omega)$)
 is more restrictive. In the Schwarzschild case, we could additionally require that the symmetries of the $\omega = 0$ equation continue to be symmetries of the finite-frequency equation. Here, this would require modifying the equation at $\ord{\omega}$. (Indeed, there are various truncations of the equation that lead to different symmetries~\cite{Castro:2010fd,Lowe:2011aa,Charalambous:2021kcz,Hui:2022vbh} depending on which terms are kept.) In the following we will instead be maximally conservative and only modify the equation at $\ord{\omega^2}$ in order to make it hypergeometric. The resulting equation will lack the full set of symmetries of the static equation, but the $\ord{\omega}$ part of its solutions will be reliable.

In order to make the approximation explicit and put~\eqref{eq:KerrradialKGequation} in standard hypergeometric form, it is useful to redefine the radial coordinate and make the following parameter definitions 
\begin{align}
\label{eq:kerrvars1}
  z&\equiv\frac{r-r_+}{r_+ - r_-}\,, &
    \Omega_H &\equiv \frac{a}{r_+ r_s}\,, &  \kappa &\equiv\frac{r_+ - r_-}{2 r_+ r_s}\,, \\
    U &\equiv \frac{m \Omega_H}{2\kappa}\,, &  W&\equiv\frac{\omega}{2\kappa}, \quad &
    P&\equiv U-W= \frac{am- r_s r_+ \omega}{r_+ - r_-} \,.
\label{eq:kerrvars2}
\end{align}
The parameter $\Omega_H$ is the angular velocity of the black hole at the horizon, $\kappa$ is the surface gravity, and $P$ is the superradiance factor. The parameters $U$ and $W$ are dimensionless versions of the angular frequency and ordinary frequency, respectively. With these redefinitions~\eqref{eq:KerrradialKGequation} becomes 
\begin{align}
    \label{eq:KerrradialKGinz}
    z(z+1)\frac{\rd}{\rd z}\left(z(z+1)\frac{\rd R_{\omega\lambda m}}{\rd z}\right)&+ \Bigg[P^2 (z+1)-(P+\omega r_s)^2z -\lambda\, z(z+1)\\
    &+z (z+1)\,\omega^2 \bigg( r_s^2+\left(r_+ +2r_+ \kappa \, z\right)\left(r_s +r_+ +2r_+ \kappa\, z\right)\bigg)\Bigg]R_{\omega\lambda m}=0\,.\nonumber
\end{align}
The near zone approximation of interest can be summarized as sending
\begin{equation}
    r_s^2\omega^2 z^n \to 0\,,
    \label{eq:kerrNZapprox}
\end{equation}
in the second line of~\eqref{eq:KerrradialKGinz}, for any exponent $n\ge 1$. (This amounts to neglecting the second line of~\eqref{eq:KerrradialKGinz}.)
This approximation is reliable so long as
\be
\omega (r-r_+) \ll 1\,.
\ee
Since the coefficients of the $(z+1)$ and $z$ terms in~\eqref{eq:KerrradialKGinz} control the behavior of the solution near the singular points $z=0,-1$, we have left them unapproximated. Importantly, in the small frequency limit we can also replace the spheroidal harmonic separation constant as $\lambda\to L(L+1)+\ord{\omega^2 r_s^2}$. Similar to the Schwarzschild case, there are terms of the form $(r_s\omega)^2 z$, $(r_s\omega)^2 z^2$ which we could have included while keeping~\eqref{eq:KerrradialKGinz} in hypergeometric form, though they do not change the final result at $\ord{r_s \omega}$. (See Appendix~\ref{app:NZ} for a solution including these terms as well as more details about our choice of near zone.)

With this approximation, the radial equation becomes
\begin{tcolorbox}[colframe=white,arc=0pt,colback=greyish2]
\vspace{-12pt}
\be
z(z+1)\frac{\rd}{\rd z} \left(z(z+1)\frac{\rd}{\rd z}R_{LM}\right)+\Big[P^2 (z+1)-(P+\omega r_s)^2z-z(z+1)L(L+1)\Big]R_{LM}=0\,.
\label{eq:KerrzEoM}
\ee
\end{tcolorbox}
\noindent
Note that this equation is essentially the same as~\eqref{eq:DdimSchzEoM}, with the addition of a term linear in $z$.\footnote{Note that this term breaks the translation symmetry of the $L=0$ equation of motion that leads to the potential being transparent~\cite{Hui:2021vcv,Hui:2022vbh}. We should therefore expect Love numbers to be nonzero at $\ord{a\omega}$ (which indeed is the case~\cite{Charalambous:2021mea,Saketh:2023bul,Perry:2023wmm,Ivanov:2024sds}).
}
If we rescale $R$ as
\begin{equation}
    R_{LM}(z)=\left(\frac{z}{z+1}\right)^{iP}\left(z+1\right)^{-i\omega r_s} u(z) \,,
\end{equation}
then~\eqref{eq:KerrzEoM} takes the hypergeometric form
\begin{equation}
    z(z+1) u^{\prime \prime}(z) + \Big[1+2iP +2(1-i \omega r_s )z\Big] u^\prime(z) -\Big[L(L+1)+i\, \omega r_s(1-\, i\omega r_s) \Big]u(z)=0\,.
    \label{eq:standardhyperkerr}
\end{equation}
Next, we want to solve this equation and impose physical boundary conditions.

\vspace{-12pt}
\paragraph{Solutions and boundary conditions:}
In order to solve~\eqref{eq:standardhyperkerr}, we impose the same boundary conditions as in the Schwarzschild problem considered in Section~\ref{sec:DdimSchGRcalc}. That is we insist $R\sim z^L$ as $z\to\infty$ and $R \sim \left[z/(z+1)\right]^{i P}$ as $z\to 0$. The solution satisfying these boundary conditions is 
\begin{equation}
    u(z) \propto\,{}_2F_1\left[\begin{array}{c}
    -L-i\omega r_s\,,\,L+1-i\omega r_s\\[-3pt]
    1+2iP
    \end{array}\Big\rvert \,-z\,\right]\,.
\end{equation}
We would again like to extract the ratio of the fall-offs of this solution at spatial infinity ($z\to \infty$). With the help of a similar connection formula to~\eqref{eq:2F1connectionformula}, we find
\begin{equation}
    \label{eq:KerrUVLovenumber}
    R(z)\xrightarrow{z \to \infty}  z^{L}\left( 1+ \cdots + z^{-2L-1} \frac{\Gamma(-2L-1)\Gamma(L+1-i\omega r_s) \Gamma(L+1+2iP+i\omega r_s)}{\Gamma(2L+1)\Gamma(-L -i\omega r_s)\Gamma(-L+2iP+i\omega r_s)}\right)\,.
\end{equation}
As a consistency check, we note that in the zero spin limit ($a \to 0$) we have $P\to -r_s\omega$. Hence, this expression agrees with the Schwarzschild response~\eqref{SWUVLovenumber} in the zero spin, $D=4$ limit to $\ord{\omega}$.  
\subsection{EFT Matching}
\label{sec:KerrEFT}

We now want to match the Kerr solution~\eqref{eq:KerrUVLovenumber} to the EFT one-point function~\eqref{eq:smallfreq1ptfinal}  to extract the conservative and dissipative responses of Kerr black holes to external scalar perturbations. 
In order to do this, 
we first expand~\eqref{eq:KerrUVLovenumber} to first order in the small $\omega$ limit:
\be
    R(r)\xrightarrow{r \to \infty}\left(\frac{r}{r_+-r_-}\right)^L \Bigg(1+\cdots +A_L\, \frac{\Gamma(-2L-1)\Gamma(L+1) \Gamma(L+1+2iU)}{\Gamma(2L+1)\Gamma(-L)\Gamma(-L+2iU)} \left(\frac{r}{r_+-r_-}\right)^{-2L-1}\Bigg)\,,
\ee
where we have defined the following prefactor of the subleading fall-off:
\be
A_L \equiv 1-r_s\omega\, {\rm Im} \Delta\psi + i r_s \omega\left(\pi \cot(L \pi)+ \frac{1-r_s\kappa}{r_s \kappa}\Big[\pi \cot(\pi (L-2iU))\Big]\right)\,.
\ee
Recall $U = \frac{m \Omega_H}{2\kappa}$ and the difference 
\be
\Delta\psi\equiv \psi(L+1-2i U)-\psi(L+1+2i U)\,,
\ee
is a purely imaginary linear combination of the digamma function (See Appendix~\ref{appendix:Formulas} for details).

In order to identify the conservative and dissipative responses associated to this fall-off, it is convenient to use the identities in Appendix~\ref{appendix:Formulas} to write (for $L\in {\mathbb Z}$)
\be
\label{eq:KerrUVrealandimsplit}
\begin{aligned}
A_L\, \frac{\Gamma(-2L-1)\Gamma(L+1) \Gamma(L+1+2iU)}{\Gamma(2L+1)\Gamma(-L)\Gamma(-L+2iU)} = &~
U\frac{(L!)^2}{(2L)!(2L+1)!}\prod_{k=1}^{L} (k^2+4U^2) \bigg[ B_L+i C_L\bigg]\,,
\end{aligned}
\ee
where we have defined the parameters
\begin{align}
 B_L &=  \pi r_s \omega \cot(\pi L)\,, \\
 C_L &= -1+r_s\omega\left[\pi \coth(2\pi U)+\frac{1-r_s\kappa}{r_s\kappa}\,\left(\sum_{k=0}^{L} \frac{4U}{k^2+4 U^2}-\frac{1}{2 U}\right)\,\right]\,.
\end{align}
Notice that $C_L$ starts at $\ord{1/U}$, so in the $a\to0$ limit, a nontrivial dissipative response survives at $\ord{\omega}$, as expected. Note also that $B_L$ is divergent for integer $L$, which indicates a nontrivial running of the conservative response at $\ord{\omega}$.

\vspace{-12pt}
\paragraph{EFT calculation:}
Now we wish to match the EFT using the above solution. At the order in $\omega$ we are working,  spheroidal harmonics reduce to spherical harmonics and so the responses will be diagonal in $L, m$ space. That is, $L'=L$ and $m'=m$ in the two-point function~\eqref{eq:freqspacegreensf}. 
Adapting the response to $D=4$, we can write the one-point function~\eqref{eq:smallfreq1ptfinal} as
\begin{equation}
    \label{eq:smallfreq1ptfinalkerr}
    \langle \varphi_+(\vec x, t)\rangle =\E^{-i \omega t}r^L\mathcal{E}_{Lm}\, Y_{Lm}\left( \lambda^{(0)}_{Lm}+(r_s\omega)\,
    \lambda^{(1)}_{Lm}+\cdots \right)2^{2 L-3} \pi^{-3} \Gamma \left(L+\tfrac{3}{2}\right) \Gamma\left( L+\tfrac{1}{2}\right) r^{-2L-1}\,,
\end{equation}
and then expand the coefficients in their real and imaginary parts
\begin{equation}
    \lambda^{(0)}_{Lm}+(r_s\omega)\,
    \lambda^{(1)}_{Lm}= k^{(0)}_{Lm}+i\nu^{(0)}_{Lm} +(r_s\omega)\left(k^{(1)}_{Lm}+i\nu^{(1)}_{Lm}\right)\,.
\end{equation}
It is now possible to compare~\eqref{eq:smallfreq1ptfinalkerr} and~\eqref{eq:KerrUVrealandimsplit} to read off the response coefficients. For each value of $L$, the quantum number $m$ can run over $2L+1$ different values. Since the angular sectors of the EFT and microscopic calculations are the same, we can associate the $L$ in the EFT with the $L$ in the microscopic calculation. It is convenient to note the responses share an overall, real prefactor
\begin{equation}
    \lambda_{Lm}^{\rm Kerr}\equiv (r_+-r_-)^{2L+1} 2^{2L+2} (2\pi)^2 \left(\frac{(L!)^2}{(2L)!(2L+1)!}\right)^2U \prod_{k=1}^{L} (k^2+4U^2)\,.
\end{equation}
It is perhaps useful to note that in the extremal limit, $a\to r_s/2$, this prefactor reduces to
\begin{equation}
    \lambda_{Lm}^{\rm Kerr} \xrightarrow{a \to \frac{r_s}{2}}  2^{2L+1} (2\pi)^2(m r_s)^{2L+1}\left(\frac{(L!)^2 }{(2L)!(2L+1)!}\right)^2 \,,
    \label{eq:extremalprefactor}
\end{equation}
which is finite.\footnote{To take the extremal limit, we introduce a small parameter, $\epsilon\equiv (r_+-r_-)/r_s= \sqrt{1-4a^2/r_s^2}$. When taking the extremal and small frequency limit simultaneously, we need to decide how to scale the ratio $r_s \omega / \epsilon=2W-\omega r_s$. In~\eqref{eq:extremalprefactor}, we have scaled $r_s \omega / \epsilon\to\infty$ in the extremal limit, in which case the $m=0$ response vanishes at $\ord{\omega}$.} We now turn to matching the EFT to read off response coefficients.
\subsubsection{Conservative response}

We first consider the conservative response, which is given by
 the real part of~\eqref{eq:KerrUVrealandimsplit}. By comparing with~\eqref{eq:smallfreq1ptfinalschwarz}, we find
\begin{tcolorbox}[colframe=white,arc=0pt,colback=greyish2]
\vspace{-12pt}
\begin{align}
    k^{(0)}_{Lm}&=0\,,\\
    k^{(1)}_{Lm} &=  \lambda_{Lm}^{\rm Kerr}\, \pi \cot(\pi L)\,.
    \label{eq:DLN}
\end{align}
\end{tcolorbox}
\noindent
In contrast to Schwarzschild, we can see this response does depend on the magnetic quantum number $m$ (implicit in $U=\frac{m \Omega_H}{2\kappa}$). This is expected because the black hole's spin picks a preferred direction.
The fact that $k_{Lm}^{(0)} = 0$, is the well-known result that static Love numbers of the Kerr black hole vanish~\cite{1972ApJ175243P,Landry:2014jka,LeTiec:2020spy,Chia:2020yla,Goldberger:2020fot,LeTiec:2020bos,Charalambous:2021mea,Charalambous:2021kcz,Hui:2021vcv}.
Since Kerr contains a second time scale, $\Omega_H$, it is possible to have a conservative response linear in the frequency, which we observe as a non-vanishing ``dynamical Love number" $k^{(1)}_{LM}$. For physical angular momentum values $L\in{\mathbb Z}$ this coefficient is superficially divergent, but as in the Schwarzschild case, careful treatment of this divergence reveals a logarithmic dependence on radius---a classical running. 
Notice that this expression is reliable to all orders in the spin at this order in frequency, and so we can consider taking the extremal limit where $\lambda_{Lm}^{\rm Kerr}$ is given by~\eqref{eq:extremalprefactor}.
We have checked that the scheme-independent coefficient of this logarithmic piece agrees with~\cite{Charalambous:2021mea,Saketh:2023bul,Perry:2023wmm,Ivanov:2024sds}. Further, in order to validate the near zone approximation that we have made in deriving~\eqref{eq:DLN}, we have compared with the result of a scattering calculation (see Appendix~\ref{app:scattering}), finding agreement.

\subsubsection{Dissipative response}

The dissipative response is given by the imaginary part of~\eqref{eq:KerrUVrealandimsplit}. By comparing with~\eqref{eq:smallfreq1ptfinalschwarz}, we can match 
\begin{tcolorbox}[colframe=white,arc=0pt,colback=greyish2]
\vspace{-8pt}
    \begin{align}
        \nu^{(0)}_{Lm}&=-\lambda_{Lm}^{\rm Kerr}\\
        \nu^{(1)}_{Lm} &= \lambda_{Lm}^{\rm Kerr} \left[\pi  \coth(2\pi U)+\frac{1-r_s\kappa}{r_s\kappa}\left(\sum_{k=0}^{L} \frac{4U}{k^2+4 U^2}-\frac{1}{2 U}\right)\right]\,.
    \end{align}
\end{tcolorbox}
\noindent
At leading order in the spin $a$, we find $\nu^{(0)}_{Lm,\rm Kerr}=-U \nu^{(1)}_{Lm,\rm Schw}+\ord{a^2}$.
This is a consequence of rotating frame kinematics, as described in Section~\ref{sec:rotframes}. (We also note that $\nu^{(1)}_{Lm,\rm Kerr}\to \nu^{(1)}_{Lm,\rm Schw}$ in the $a\to0$ limit, as expected.) 
Like the Schwarzschild case, neither of these responses diverge for physical angular momentum values $L \in \mathbb{Z}$. In the extremal limit, $\lambda^{\rm Kerr}_{Lm}$ remains finite, and $\nu^{(1)}_{Lm}$ becomes 
\begin{equation}
    \nu^{(1)}_{Lm}\to \lambda_{Lm}^{\rm Kerr,\, ext} \left(\pi +\frac{2L+1}{m}\right)\,,
\end{equation}
where the extremal limit of $\lambda_{Lm}^{\rm Kerr}$ is given by~\eqref{eq:extremalprefactor}.
\newpage
\section{Myers--Perry}
\label{sec:MP}

In order to gain further insight into the properties of spinning black holes, we would now like to consider a higher-dimensional example. There is some sense in which the physics of spinning objects becomes more generic in higher dimensions, as there are now multiple different planes in which an object can be spinning. We would like to investigate this phenomenon in EFT. As a natural next example, we consider the five-dimensional Myers--Perry black hole. (See also~\cite{Charalambous:2023jgq,Rodriguez:2023xjd}.) We will see that this example displays a number of interesting phenomena. In particular, for generic values of the spins, this object displays both conservative and dissipative static responses, which must be matched by different sectors in the EFT.
However, when the two spin parameters of the black hole are equal, the problem simplifies and we obtain results analogous to the $4D$ Kerr problem.

\subsection{General Relativity Calculation}
\label{sec:MPGRcalc}
We begin with the line element for a five-dimensional spinning black hole. In the analogue of Boyer--Lindquist coordinates, this is given by~\cite{Myers:1986un}
\be
\begin{aligned}
    \label{eq:MPmetric}
    \rd s^2=-\rd t^2&+\frac{r_s^2}{\Sigma}\Big(\!-\rd t +a \sin^2{\theta} \rd\phi +b \cos^2{\theta}\rd\psi\Big)^2+\frac{ \Sigma}{\Delta}r^2 \rd r^2  \\[2pt]
    & +\Sigma \rd\theta^2+(r^2+a^2)\sin^2{\theta}\rd\phi^2+(r^2+b^2)\cos^2{\theta}\rd\psi^2\,,
\end{aligned}
\ee
where the various functions appearing in this line element are defined by
\begin{align}
 \Sigma &=r^2+a^2\cos^2{\theta}+b^2\sin^2{\theta}\,,\\
 \Delta &=(r^2+a^2)(r^2+b^2)-r_s^2r^2=(r^2-r_+^2)(r^2-r_-^2)\,,\\
 r_{\pm}^2 &=\frac{1}{2}\left(r_s^2-a^2-b^2 \pm \sqrt{(r_s^2-a^2-b^2)^2-4a^2b^2}\right). 
\end{align}
It is worth noting that the angular coordinates in~\eqref{eq:MPmetric} agree with the Hopf parametrization of the three-sphere $S^3$ as $r \rightarrow \infty$. We can therefore think of $a$ as the angular momentum in the plane rotated by changing the coordinate $\phi$ and similarly for $b$ and the coordinate $\psi$. More explicitly, the mass and angular momentum parameters associated to the metric~\eqref{eq:MPmetric} are~\cite{Myers:1986un}
\begin{equation}
M_{\rm ADM} = \frac{3\pi}{8 G} r_s^{2} \,
\qquad\quad
J_\phi = \frac{2M_{\rm ADM}}{3} a \, , \quad\qquad J_\psi = \frac{2M_{\rm ADM}}{3} b\,,
\end{equation}
where $M_{\rm ADM}$ is the ADM mass.
The background is axially symmetric in the $\phi$ and $\psi$ directions,\footnote{The vector fields $\partial_\phi$ and $\partial_\psi$ are Killing vectors of the spacetime.} so it is natural to make the  ansatz
\begin{equation}
    \Phi=\E^{-i\omega t}\E^{i m_{\phi} \phi}\E^{i m_{\psi}\psi}S(\theta)R(r)\,.
\end{equation}
With this splitting, the massless Klein--Gordon equation separates into 
\begin{align}
    \label{eq:AngularLaplacian5D}
    &\frac{1}{\cos{\theta}\sin{\theta}}\frac{\rd}{\rd\theta}\left(\cos{\theta}\sin{\theta}\frac{\rd S}{\rd\theta}\right)+\bigg[\omega ^2 \left(a^2 \cos ^2\theta +b^2 \sin ^2\theta \right)-m_{\psi}^2 \sec ^2\theta  -m_{\phi}^2 \csc ^2\theta
   +\lambda\bigg]S =0\,,\\[-2pt]
    &\frac{\Delta}{r}\frac{\rd}{\rd r}\left(\frac{\Delta}{r}\frac{\rd R}{\rd r}\right)+(K(r)-\Delta \lambda)R=0\,,
    \label{eq:MPradial}
\end{align}
where in this second equation~\eqref{eq:MPradial}, we have defined the function
\be
\begin{aligned}
    K(r)\!\equiv \Delta \Bigg[\frac{m_{\phi}^2 \left(a^2-b^2\right)}{a^2+r^2}\!+\!\frac{m_{\psi}^2 \left(b^2-a^2\right)}{b^2+r^2}+r^2 \omega ^2\Bigg]
    \!
    +
    \! 
    \left(a^2+r^2\right)
   \left(b^2+r^2\right) \left[\frac{r_sa m_\phi}{a^2+r^2}+\frac{r_sb m_{\psi}}{b^2+r^2}-r_s\omega \right]^2\!\!.
\end{aligned}
\ee
In general, the solutions to~\eqref{eq:AngularLaplacian5D} are spheroidal harmonics, but they reduce to Jacobi polynomials 
in the small frequency limit so that the angular solutions are given by spherical harmonics, as before.

We can simplify the appearance of~\eqref{eq:MPradial} by making a number of variable redefinitions. First, we define the new radial coordinate
\be
z=\frac{r^2-r_+^2}{r_+^2-r_-^2}\,,
\ee 
and define the angular velocities, surface gravity, and rescaled frequency:
\be
    \Omega_a=\frac{a}{r_+^2 + a^2}\,, \qquad\quad \Omega_b=\frac{b}{r_+^2 + b^2}\,, \qquad\quad \kappa=\frac{r_+^2 - r_-^2}{r_+ r_s^2} \qquad\quad
    W\equiv \omega \frac{r_s^2 r_+}{2(r_+^2-r_-^2)}=\frac{\omega}{2\kappa}\,.
\ee
In five dimensions, the azimuthal quantum numbers in each angular direction, $m_\phi$, $m_\psi$, couple to both of the spin magnitudes. It is thus more natural to express the solution in terms of the following linear combinations of the angular velocities
\begin{equation}
    m_R\equiv m_\phi +m_\psi \,, \qquad\quad m_L \equiv m_\phi-m_\psi \,, \qquad\quad\Omega_R=\frac{\Omega_a +\Omega_b}{2}\,,  \qquad\quad\Omega_L=\frac{\Omega_a -\Omega_b}{2}\,,
\end{equation}
so that the spin parameters are
\be
    U\equiv\frac{(a+b)(m_\phi+m_\psi)}{4(r_+ - r_-)} =\frac{\Omega_R\, m_R}{2 \kappa}\,,\qquad\qquad V\equiv\frac{(a-b)(m_\phi-m_\psi)}{4(r_+ + r_-)}=\frac{\Omega_L \, m_L}{2 \kappa} \,.
\ee
It will also be useful to define the dimensionless sum and difference of the horizon locations
\begin{equation}
    z_\pm\equiv\frac{r_+ \pm r_-}{r_+}\,.
\end{equation}
Note that these are {\it not} the locations of the inner and outer horizons (which sit at $z=-1$ and $z=0$, respectively).
Finally, we define the superradiance factor $P$ 
and rescaled angular momentum 
\be
    P\equiv U+V-W=\frac{1}{2\kappa}\left( m_\phi \Omega_a + m_\psi \Omega_b -\omega\right)\,,\qquad\qquad \hat{L}\equiv\frac{L}{2}\,.
\ee
After all of these redefinitions, the radial equation~\eqref{eq:MPradial} becomes 
\begin{align}
    \label{eq:reqMP}
    z(z+1)\frac{\rd}{\rd z}\left(z(z+1)\frac{\rd R}{\rd z}\right)+ \bigg[&-\hat{L} (\hat{L}+1) z (z+1)+ P^2 +4 U V z-\left(2 z_-\, U+2 z_+\, V\right)W z  \\
    &+z_- z_+\, W^2 z+ W^2\kappa^2 z(z+1)\left(r_+^2 +r_s^2+(r_+^2-r_-^2)z\right)\bigg]R=0 \,.
    \nonumber
\end{align}
As before, to solve this equation analytically we approximate it to cast it in hypergeometric form.

\paragraph{Near Zone:}
The near zone for Myers--Perry is chosen with the same principles as for Schwarzschild and Kerr. As before we keep all terms in the equation of motion that are $\ord{r_s\omega}$.
However, in this case it is useful to also keep a single term that is $\ord{r_s^2\omega^2 z}$. Since we never trust the solution to this order, it has absolutely no effect on the physics, but including this term allows us to complete the square in various places in order to simplify the representation of the solution. In order to implement the near zone approximation, we approximate
\begin{equation}
    r_s^2\omega^2 z \approx 0\,,
\end{equation} 
in the last term of~\eqref{eq:reqMP}. We also make the approximation that $\omega a \ll 1$ in~\eqref{eq:AngularLaplacian5D}, which is then solved by higher-dimensional spherical harmonics. 
The radial equation~\eqref{eq:MPradial} then is
\begin{tcolorbox}[colframe=white,arc=0pt,colback=greyish2]
\be
\begin{aligned}
    \label{eq:MPnearzone}
    z(z+1)\frac{\rd}{\rd z}\left(z(z+1)\frac{\rd R}{\rd z}\right)+ \bigg[&-\hat{L} (\hat{L}+1) z (z+1)+ P^2 +4 U V z \\
    &-\left(2 z_-\, U+2 z_+\, V\right)W z +z_- z_+\, W^2 z\bigg]R=0 \,.
\end{aligned}
\ee
\end{tcolorbox}
\noindent
Note that this equation has terms linear in $z$ multiplying both spin and frequency $\sim \Omega \,\omega\, z$, which are analogous to a similar term in the Kerr equation of motion~\eqref{eq:KerrzEoM}. However, there is also the term $4UVz\propto \Omega^2\, z$, which remains in the static limit of the problem.\footnote{As we will see, this will lead to nonzero static Love numbers for a generic spin configuration. Much as in the Kerr case at $\ord{\omega}$, the presence of this term breaks the translation symmetry present at $\omega = L = 0$ for zero spin.}

We can put~\eqref{eq:MPnearzone} in explicit hypergeometric form by defining
\begin{equation}
    R(z)= \left(\frac{z}{z+1}\right)^{i P} (z+1)^{2iV- iz_- W} u(z)\,.
\end{equation}
The resulting equation is somewhat unwieldy, so we will not display it, but it is solved by~\eqref{eq:MPsolution}.

\vspace{-12pt}
\paragraph{Solutions and boundary conditions}
We solve~\eqref{eq:MPnearzone} using the same boundary conditions as Section~\ref{sec:DdimSchGRcalc}, that is, we require $R\sim z^L$ as $z\to\infty$ (corresponding to a tidal external field) and $R \sim \left[z/(z+1)\right]^{i P}$ as $z\to 0$ (corresponding to regularity at the horizon). The solution satisfying these boundary conditions is 
\begin{equation}
    u(z) \propto\,{}_2F_1\left[\begin{array}{c}
    -\hat L+2iV- iz_- W\,,\, \hat L+1+2iV- iz_- W\\[-3pt]
    1+2iP
    \end{array}\Big\rvert \,-z\,\right]\,.
    \label{eq:MPsolution}
\end{equation}
Expanding this at $z\to \infty$ yields 
\be
    \label{eq:MPUVLovenumber}
    R(z)\xrightarrow{z \to \infty}  z^{\hat L}\Bigg( 1+ \cdots +A_L \,z^{-2\hat L-1}  
    \Bigg)\,,
\ee
where the ratio of fall-offs is
\be
A_{L} \equiv  \frac{\Gamma (-2 \hat L-1) \Gamma (\hat L+1+2 i V-i z_- W) \Gamma (\hat L+1+2 i U-i z_+ W)}{\Gamma (2 \hat L+1) \Gamma (-\hat L+2 i V-i z_- W) \Gamma (-\hat L+2 i U-iz_+ W)}\,.
\label{eq:MPAL}
\ee
Note that $A_L$ is only reliable when expanded up to and including $\ord{\omega}$. As a check, the coefficients of this expansion match the corresponding Schwarzschild result in the limit $U,V\to 0$.\footnote{Note that~\eqref{eq:MPAL} as a full function of $\omega$ does not reduce to~\eqref{SWUVLovenumber} in the zero spin limit because we have kept different $\ord{\omega^2}$ terms in the two cases for convenience. Since we never use any part of the expansion beyond $\ord{\omega}$ (where the two agree), there is nothing inconsistent about this.}

In the limit where the spins are equal, $a=b$ (or $V=0$), the response becomes 
\be
A_{L} \xrightarrow{V\to 0}  \frac{\Gamma (-2 \hat L-1) \Gamma (\hat L+1-i z_- W) \Gamma (\hat L+1+2 i P+i z_- W)}{\Gamma (2 \hat L+1) \Gamma (-\hat L-i z_- W) \Gamma (-\hat L+2 i P+iz_- W)}\,.
\label{eq:MPequalspinresponse}
\ee
This matches the response in Kerr~\eqref{eq:KerrUVLovenumber} upon noting $r_s \omega=\frac{r_+-r_-}{r_+}W=z_{-}^{\rm Kerr} \, W$ 
and equating analogous quantities in the two calculations $(\kappa, r_\pm , U, W, \hat{L}, \ldots)$. This agreement is a reflection of the fact that the two differential equations for radial perturbations take the same form in this limit.

\subsection{EFT Matching}
\label{sec:MPEFT}

We now match the EFT one-point function~\eqref{eq:freqspacegreensf}. We first expand~\eqref{eq:MPUVLovenumber} to $\ord{\omega}$:
\be
    \label{eq:MPLNtoOw}
    R(r) \xrightarrow{r \to \infty} \left(\frac{r^2}{r_+^2 - r_-^2}\right)^{L/2} \Bigg(1+\cdots +\tilde A_L\left(\frac{r^2}{r_+^2 - r_-^2}\right)^{-L-1} 
    \Bigg)\,,
\ee
where $\tilde A_L$ is the expansion of $A_L$~\eqref{eq:MPAL} to $\ord{\omega}$ 
\begin{align}
    \tilde A_L = A_{L}^{(\omega=0)}\Bigg[1 +i\frac{\omega}{2\kappa}\bigg(&z_+ \left[\psi(-\hat L+2iU)-\psi(\hat L+1+2iU)\right] \nonumber\\
    +&z_- \left[\psi(-\hat L+2iV)-\psi(\hat L+1+2iV)\right]\bigg)\Bigg]\,,
\end{align}
and $\psi$ is the digamma function. To separate this expression into real and complex pieces, we use the identities in Appendix~\ref{appendix:Formulas}, in particular~\eqref{eq:MPresponseexpansion},~\eqref{eq:MPtrigexpansion}, and~\eqref{eq:digammaexpansion}. The result is somewhat similar to~\eqref{eq:KerrUVrealandimsplit}. We will not write the real and imaginary split explicitly, but it can be extracted implicitly from the Wilson coefficients detailed below.

We can match the EFT using the above solution. At the order in $\omega$ that we are working, the responses are diagonal in $L$ and $M$, so that the two-point function will also be diagonal in $L,M$.
 Adapting the response to $D=5$, we can write~\eqref{eq:smallfreq1ptfinal} as
\be
    \label{eq:smallfreq1ptfinalMP}
    \langle \varphi_+(\vec x, t)\rangle =\E^{-i \omega t}r^L\mathcal{E}_{LM}\, Y_{LM}\left( \lambda^{(0)}_{LM}+(r_s\omega)\,
    \lambda^{(1)}_{LM}+\cdots \right)\frac{2^{2L-3}}{\pi^4} \Gamma(L+1)\Gamma(L+2)r^{-2L-2} \,,
\ee
and split the Wilson coefficients as 
\be
    \lambda^{(0)}_{LM}+(r_s\omega)\,
    \lambda^{(1)}_{LM}= k^{(0)}_{LM}+i\nu^{(0)}_{LM} +(r_s\omega)\left(k^{(1)}_{LM}+i\nu^{(1)}_{LM}\right)\,.
\ee
Comparing~\eqref{eq:smallfreq1ptfinalMP} with~\eqref{eq:MPLNtoOw}, we see that all response coefficients have the common, real prefactor:
\begin{equation}
    \lambda_{LM}^{\rm mp}=(r_+^2-r_-^2)^{L+1}\frac{2^{-2L}(2\pi)^3 |\Gamma(\hat L+1+2iU)|^2|\Gamma(\hat L+1+2iV)|^2  }{\Gamma(L+1)
    \Gamma(L+2) \Gamma(2\hat L+1)\Gamma(2\hat L+2)}\,.
\end{equation}
We can now match the UV and EFT calculations to determine the Wilson coefficients explicitly.

\subsubsection{Conservative response}
We first match the conservative sector, to extract the real part of the Wilson coefficients. We find 
\begin{tcolorbox}[colframe=white,arc=0pt,colback=greyish2]
\vspace{-4pt}
    \begin{align}
        k^{(0)}_{LM}&=\frac{\lambda_{LM}^{\rm mp}}{2} \bigg(\csc (2 \pi  \hat L) \cosh (2 \pi  (U-V))-\cot (2 \pi  \hat L) \cosh (2 \pi  (U+V))\bigg)\,,\\[1pt]
        k^{(1)}_{LM} &= -\frac{\lambda_{LM}^{\rm mp}}{4 r_s \kappa} \Bigg(2\pi\left(\frac{ r_- }{r_+}\csc (2 \pi  \hat L) \sinh (2 \pi  (U-V))- \cot (2 \pi  \hat L) \sinh (2 \pi  (U+V))\right)\\[-2pt] \nonumber 
  & \hspace{1.75cm}+\Im(\Delta_{z}\psi)\left(\csc (2 \pi  \hat L) \cosh (2 \pi  (U-V))-\cot (2 \pi  \hat L) \cosh (2 \pi  (U+V))\right) \Bigg)\,,
    \end{align}
\end{tcolorbox}
\noindent
where we have defined the (purely imaginary) combination
\be
\Delta_{z}\psi \equiv z_+\big [\psi(L+1-2iU)-\psi(L+1+2iU)\big]+z_- \big[\psi(L+1-2iV)-\psi(L+1+2iV)\big]\,.
\ee

Both expressions involve a factor of $\csc (2 \pi  \hat L)$, which diverges for all $L \in \mathbb{Z}$. This is again associated with classical running. It is also notable that the static conservative response (Love number) {\it does not} 
vanish for generic values of $U$, $V$. 
However, when $U=0$ or $V=0$ (corresponding to $|a|=|b|$) $k^{(0)}_{LM}$ does vanish for even values of $L$. In the equal spin case this is consistent with the fact that the ratio of fall-offs takes the same form as the Kerr black hole~\eqref{eq:MPequalspinresponse}. In the case where $a=-b$, this is somewhat more mysterious and it would be interesting to understand more fully. 
The expression for the static Love number $k^{(0)}_{LM}$ agrees with~\cite{Charalambous:2023jgq,Rodriguez:2023xjd}.
Note also that the dynamical Love number at $\ord{\omega}$ is reliable given the near zone approximation that we have made.

\subsubsection{Dissipative response}
We can similarly compare the imaginary parts of the fall-offs to match the dissipative response. Using~\eqref{eq:MPLNtoOw}, we find
\begin{tcolorbox}[colframe=white,arc=0pt,colback=greyish2]
\vspace{-6pt}
    \begin{align}
        \nu^{(0)}_{LM}&=-\frac{\lambda_{LM}^{\rm mp}}{2} \sinh (2 \pi  (U+V))\,,\\[1pt]
        \nu^{(1)}_{LM}&=\frac{\lambda_{LM}^{\rm mp}}{4 r_s \kappa}\Bigg( 2 \pi  \cosh (2 \pi  (U+V))+\Im(\Delta_{z\psi})\sinh (2 \pi  (U+V))\Bigg)  \,.
    \end{align}
\end{tcolorbox}
\noindent
Like the Kerr case, these responses neither vanish nor diverge. Relative to the analogous terms in the Kerr case, there is an additional factor of $\frac{1}{2} \sinh (2 \pi  (U+V))$. We can see that, in the $a=b$ case, the responses take similar forms to those in Kerr, as expected from~\eqref{eq:MPequalspinresponse}. Additionally, we see that $\nu^{(0)}_{LM, \rm MP}=-(U+V)\nu^{(1)}_{LM, \rm Schw}+\ord{U^2,V^2,UV}$, which may be explained by the rotating frame effects described in Section~\ref{sec:rotframes}.

\newpage
 \section{Large Dimension}
\label{sec:InfD}

As another interesting example, we will consider a spinning black hole in the $1/D$ expansion. Though this regime is obviously not particularly physical, 
in our effort to explore the parameter space of higher-dimensional black holes, the arenas where things are analytically solvable serve as valuable guideposts. They display new phenomena, and allow us to make inferences about the space of possibilities beyond four dimensions. In this regard the large dimension expansion is useful. This limit in gravity has been considered in various contexts before. Concretely, many features of Einstein's equations are simpler in large dimensions~\cite{Emparan:2013moa,Emparan:2015hwa,Bhattacharyya:2015dva,Emparan:2020inr}, and even some aspects of quantum gravity simplify in this limit~\cite{Strominger:1981jg,Bjerrum-Bohr:2003veq}. We are therefore motivated to study the tidal responses of black holes in the $1/D$ expansion.
In this limit, the wave equation around a black hole becomes hypergeometric, and so can be solved exactly to all orders in frequency. Though the Klein--Gordon equation is known for a black hole in $D$ dimensions with arbitrary spin parameters~\cite{Frolov:2017kze}, for simplicity we study the case with all spin parameters set to be equal.

\subsection{General Relativity Calculation}
\label{sec:InfDGRcalc}

We begin by recalling the metric of a $D=2N+3$ dimensional Myers--Perry black hole with all its spin parameters taking the same value~\cite{Jorge:2014kra}
\begin{equation}
    \rd s^2=-f(r)^2 \rd t^2 +g(r)^2 \rd r^2 + h(r)^2 \Big(\rd\psi +B_a \rd x^a -\Omega(r)\rd t\Big)^2 +r^2 \hat{g}_{ab} \rd x^a \rd x^b\,.
    \label{eq:generalDMP}
\end{equation}
Here the functions appearing are given by
\begin{align}
    \label{equalspinsfuncts}
    g(r)^2 & = \left( 1- \frac{r_s^{2N}}{r^{2N}} + \frac{r_s^{2N} a^2}{r^{2N+2}} \right)^{-1}\, , & \Omega(r)  & =   \frac{r_s^{2N} a}{r^{2N} h(r)^2}  \, ,\\
    h(r)^2 & =  r^2 \left( 1+ \frac{r_s^{2N} a^2}{r^{2N+2}} \right)  \, , &    f(r) & = \frac{r}{g(r) h(r)} \, ,
\end{align}
and $\hat g_{ab}$, $B_a$ are respectively the Fubini--Study metric on ${\mathbb C}P^{N}$ and its K\"ahler potential (see Appendix~\ref{app:CPNeigen}). This metric reduces to~\eqref{eq:MPmetric} with $a=b$ for $N=1$.

The mass and angular momentum of the black hole relative to each spin plane are~\cite{Myers:1986un}
\be
M_{\rm ADM} = \frac{(2N+1)A_{2N+1}}{16\pi G}r_s^{2N}\,,\qquad\quad J = \frac{2}{2N+1}M_{\rm ADM}\,a\,,
\ee
where $A_{2N+1}=2\pi^{N+1}/\Gamma(N+1)$ is the area of the unit $(2N+1)$-sphere, and all the angular momenta are the same because the spin parameters are equal. 
Relatedly, the angular velocity of the black hole in any of its spin planes is given by $\Omega(r_h)=\Omega_H= a/r_h^2$.
The event horizon, $r_h$, corresponds to the largest root of the polynomial $1/g^{2}$~\cite{Jorge:2014kra}.  In the $N \rightarrow \infty$ limit, this root can be found explicitly and approaches $r_h \rightarrow r_s(1-\frac{a^2}{r_s^2})^{1/2N}$.

The benefit of this spin configuration is that the geometry of the metric simplifies for this choice and can be viewed as a warped product of a two dimensional space (coordinatized by $t, r$) with a circle bundle (with coordinate $\psi$) over complex projective space ${\mathbb C}P^N$  (which has coordinates $x^a$).
The Klein--Gordon equation separates if we make the following ansatz:
\begin{equation}
\Phi (t, r, \psi, x^a) = \E^{-i \omega t + i M \psi} R_{\omega LM}(r) \mathbb{Y}_{LM}(x^a)\, ,
\end{equation}
where $\mathbb{Y}_{L M}(x^a)$ is an  eigenfunction of the charged scalar laplacian on $\mathbb{C} P^N$, which has eigenvalues  given by~\cite{Kunduri:2006qa,Jorge:2014kra}
\be
\begin{aligned}
- \mathcal{D}^2 \mathbb{Y}_{L M} &\equiv -\hat{g}^{ab}\left(D_a - i M B_a\right) \left(D_b - i M B_b\right)\mathbb{Y}_{L M} 
\\
&= \left[L(L+2N)  - M^2\right]\mathbb{Y}_{L M} \, .
\end{aligned}
\ee
(See Appendix~\ref{app:CPNeigen} for more details about these functions.)
The equation for the radial field $R_{\omega LM}(r)$ takes the form:
\begin{align}
\label{eq:MPKGequalspins}
\partial_r \left[\frac{r^{2N+1}}{g(r)^2} \partial_r R_{\omega LM}(r) \right] +
 \Bigg[&g(r)^2\frac{r^2 \left(r_s^{2N} a^2 +r^{2N+2}\right)^2 (\omega -M \Omega (r))^2}{r^{2N+2}}
 \\\nonumber
&- r_s^{2N}a^2  \Big(L( L+2 N)-M^2\Big)-L (L+2N) r^{2 N+2}\Bigg]  \frac{r^{2 N-1}  R_{\omega LM} (r)}{r_s^{2N} a^2 +r^{2 N+2}}  =0 \, .
\end{align}
We can simplify this equation by defining the following rescaled parameters
\begin{align}
    \rho &\equiv \frac{r^{2N}}{r_s^{2N}} \, ,  & \hat a &\equiv \frac{a}{r_s} \, ,\\
    \hat{L}&\equiv \frac{L}{2N}, & \hat{M}&\equiv\frac{M}{2N}, &\hat{\omega}&\equiv\frac{\omega}{2N}\,,
\end{align}
after which the radial equation becomes
\begin{equation}
    \label{eq:LargeDinrhoEoM}
    \hat{\Delta} \partial_\rho\Big(\rho^{1-\tfrac{1}{N}}\hat{\Delta}\partial_\rho R_{\omega LM}\Big)
    +\left[\Big(\hat{\omega} r_s \rho^{\tfrac{1}{N}}-\hat{a} \hat{M}\Big)^2-\hat{\Delta}\Big(\hat{L}(\hat{L}+1)-r_s^2 \hat{\omega}^2\rho^{\tfrac{1}{N}}\Big)\right]R_{\omega LM}=0\,.
\end{equation}
In this equation, we have defined the function in analogy to the previous cases
\begin{equation}
    \hat{\Delta}(\rho)\equiv(\rho-1)\rho^{\tfrac{1}{N}}+\hat{a}^2\,.
\end{equation}
As before, the equation~\eqref{eq:LargeDinrhoEoM} is difficult to solve in complete generality, but we can simplify it by taking the limit of a large number of dimensions.

\vspace{-12pt}
\paragraph{Large dimension limit:}
The dynamics of black holes simplify greatly in large dimension~\cite{Emparan:2020inr}. In the present context, this manifests as the equation~\eqref{eq:LargeDinrhoEoM} reducing to hypergeometric form in the limit of large $N$. Specifically, we can see this by taking the limit  $N\rightarrow \infty$, keeping the parameters $\hat{L}, \hat{M}$, and $\hat{\omega}$ fixed. 
The practical consequence of this limit is to approximate
\begin{equation}
    \rho^{\tfrac{1}{N}}\to 1\,.
\end{equation}
For fixed $N$, this approximation is valid in a region $\log(\rho) \ll N$, however we will be interested only in the leading order behavior, so we will take the strict $N = \infty$ limit, where the approximation is reliable to arbitrarily large radius. It is in this sense that the equation reduces exactly to a solvable form. In the infinite $N$ limit, the equation of motion reduces to
\begin{equation}
    \Delta_0 \partial_\rho\big(\rho \Delta_0 \partial_\rho R_{\omega LM}\big)
    +\left[\big(\hat{\omega} r_s-\hat{a} \hat{M}\big)^2-\Delta_0 \big(\hat{L}(\hat{L}+1)-r_s^2 \hat{\omega}^2\big)\right]R_{\omega LM}=0\,,
\end{equation}
where we have approximated $\hat\Delta$ by
\begin{equation}
    \Delta_0\equiv  \rho-1+\hat{a}^2\,.
\end{equation}
It is possible to recast this equation in a more familiar form by again defining new parameters
\begin{align}
    \label{eq:LargeDPnu}
    \kappa &\equiv \frac{\sqrt{1-\hat{a}^2}}{2 r_s} \, , \hspace{1cm} 
    \Omega_H \equiv\frac{\hat{a}}{r_s}\,, \hspace{2.2cm} \nu(\nu+1)\equiv\hat{L}(\hat{L}+1)-\hat{\omega}^2 r_s^2\,, \\[3pt]
    U&=\frac{\hat M \Omega_H}{2\kappa}\,,   \hspace{1.45cm} 
    P^2\equiv\frac{ \big(\hat{a} \hat{M}-r_s\hat{\omega} \big){}^2}{1-\hat{a}^2}=\left(U-\frac{\hat{\omega} }{2\kappa}\right)^2 \,.
\end{align}
Physically $\kappa$ is the surface gravity and $\Omega_H$ is the angular velocity at the horizon. It is convenient to also redefine the radial variable as
\begin{equation}
    z\equiv\frac{\rho}{1-\hat{a}^2}-1\,,
\end{equation}
so that the horizon sits at $z = 0$.
Together these redefinitions recast the radial equation as
\begin{tcolorbox}[colframe=white,arc=0pt,colback=greyish2]
\begin{equation}
    \label{eq:LargeDsimpleradialeq}
    z (z+1)\partial_z\Big(z (z+1) \partial_z R_{\omega LM}\Big)+ \Big(P^2 (z+1)-\nu  (\nu +1)z (z+1)\Big)R_{\omega LM} =0\,.
\end{equation}
\end{tcolorbox}
\noindent
Note this agrees with the large $D$ limit of Schwarzschild~\eqref{eq:ddimscheq}, as we scale the spin to zero, $\hat{a}\to 0$.

An important feature of~\eqref{eq:LargeDsimpleradialeq} is that it is a hypergeometric equation for {\it any} value of the frequency. This means that in this particular case, it is possible to solve the equation exactly in frequency. 
We can put~\eqref{eq:LargeDsimpleradialeq} in the standard hypergeometric form by redefining
\begin{equation}
    R_{\omega L M}(z)= z^{i P} u(z)\,,
\end{equation}
though we will not write the equation explicitly. Given this equation, it is straightforward to find its solution that is regular at the black hole's horizon
\be
u(z) = {}_2F_1\left[\begin{array}{c}
-\nu+i P\,,\,\nu +1+i P\\[0pt]
2 i P+1
\end{array}\bigg\rvert \,-z\,\right]\,,
\ee
which we may then expand near $z\to \infty$ to find the following fall-off behavior
\begin{equation}
    R_{\omega L M}(z) \xrightarrow{z\to\infty} z^{i P}z^{\nu}\left(1+\cdots + \frac{\Gamma (-2 \nu -1) \Gamma (\nu +1+i P)^2}{\Gamma (2 \nu +1) \Gamma (-\nu+i P )^2} z^{-2\nu-1}\right)\,.
\end{equation}
Note that this result is {\it exact} in $\hat{\omega}$, we have only made the approximation of large $N$. However, for the purposes of matching, we will have to make a small-frequency approximation because we did not compute to all orders in $\omega$ in the EFT. (Though it would be interesting to do so.)
In order to match, ideally we would like to expand $R_{\omega L M}(z)$ around $z\rightarrow \infty$ and read off the ratio of the fall-offs of the form $R_{\omega L M} \propto (r^{L}+A_{\hat{L}\hat{M}}(\hat\omega)r^{-L-2N})$. However---as we can see from~\eqref{eq:LargeDPnu}---the parameter $\nu \neq \hat{L}$, which makes this somewhat tricky. 
We can overcome this difficulty by expanding in $\omega$; we know that 
$\nu=\hat{L}+\ord{\hat{\omega}^2}$, so to first order in $\omega$ we can ignore the difference between $\nu$ and $L$ and define the response coefficients in a similar manner to the above calculations. Thus we consider the ratio of the falloffs $z^\nu,\,  z^{-2\nu-1}$ as the $L^{\rm th}$ response coefficient, accurate to $\ord{\omega}$:
\begin{equation}\label{eq:LargeDResponseCoeff}
    A_{\hat{L}\hat{M}}(\hat{\omega})=\frac{\Gamma (-2 \hat{L} -1) \Gamma (\hat{L} +1+i P)^2}{\Gamma (2 \hat{L} +1) \Gamma (-\hat{L}+i P )^2} r_h^{2L+2N}\,,
\end{equation} 
which we now want to match to the EFT to determine Wilson coefficients.

\subsection{EFT Matching}
\label{sec:InfDEFT}

To extract response coefficients, it is convenient to first expand the expression~\eqref{eq:LargeDResponseCoeff} for small $\omega$. To linear order in $\omega$ we obtain
\begin{equation}
    \label{eq:LargeDfirstorderresponse}
    A_{\hat{L}\hat{M}}(\hat{\omega})=\frac{|\Gamma (\hat L+1+i U)|^4 }{\pi  \Gamma (2 \hat L+1) \Gamma (2 \hat L+2)} \csc (2 \pi \hat L) \sin ^2(\pi  [\hat L-i U])\left[1+\frac{i \hat{\omega}}{\kappa}\left(\pi \cot(\pi [\hat L -iU])+ \Delta_\psi \right)+\cdots\right]\,,
\end{equation}
where we have defined the difference of digamma functions similarly to before
\be
\Delta\psi = \psi(\hat L+1-i U)-\psi(\hat L+1+i U)\,.
\ee
We next expand~\eqref{eq:LargeDfirstorderresponse}
into real and imaginary parts using the formulas in Appendix~\ref{appendix:Formulas}, to obtain
\be
\begin{aligned}
A_{\hat{L}\hat{M}}(\hat{\omega}) =\, &\frac{\lvert\Gamma (\hat L+1+i U)\rvert^4 }{2\pi  \Gamma (2 \hat L+1) \Gamma (2 \hat L+2)}\Bigg[
\csc (2 \pi \hat L)-\cot (2 \pi  \hat L) \cosh (2 \pi  U)- i \sinh (2 \pi  U)\\
&\hspace{.75cm}+\frac{\hat\omega}{\kappa} \Big(
\pi  \cot (2 \pi  \hat L)\sinh (2 \pi  U)+ \Im(\Delta\psi)\left[\cosh(2\pi U)\cot(2 \pi \hat L)- \csc(2\pi \hat L)\right]\\
&\hspace{6.15cm}+i\left[\pi  \cosh (2 \pi  U)+ \Im(\Delta{\psi}) \sinh (2 \pi  U)\right]
\Big)
\Bigg].
\end{aligned}
\ee
We now want to use this expression to match the coefficients in the EFT.

\vspace{-12pt}
\paragraph{Manifesting Symmetries in the EFT:} Generically a spinning object---even if it is spherically symmetric at rest---will be at most invariant under $SO(2)^r$ symmetry (where $r=\lfloor\frac{D-1}{2}\rfloor$). However, the situation we are considering actually has enhanced symmetry. As a result, in order to simplify the matching procedure, it makes sense to organize the EFT in a way that respects this enhanced symmetry.
In Section~\ref{sec:EFT}, we constructed the finite-size terms present in the EFT without assuming any particular symmetry of the object in question. As a result, the response coefficients (or the Green's function of $Q$ operators) could depend arbitrarily on the angular quantum numbers $L$ and $M$ of both the source and response. In Section~\ref{sec:DdimSch} we saw a simple example of an object with enhanced symmetry. Since the Schwarzschild black hole is spherically symmetric, its responses to external fields are actually independent of the magnetic number $M$. The manifestation of this symmetry in the EFT is that the Wilson coefficients of many operators are related to each other (explicitly, the tensor $\lambda_{\rm cons.}(\tau)^{i_1\cdots i_L\,j_1\cdots j_{L'}}$ in~\eqref{eq:Scons} is constructed only from rotationally invariant $\delta_{ij}$ tensors). 
In the case of interest, the black hole is actually invariant under the SU$(N+1)$ transformations that act on $\mathbb{C} P^N$. As such, we will find it convenient to match by constructing the EFT in a basis of tensors that are invariant under SU$(N+1)$. For completeness we then work out the translation back into the basis of ordinary Thorne tensors.

We will briefly sketch the construction of the EFT, similar to the one done in Section~\ref{sec:EFT}. Here, it will be useful to write the EFT in complexified coordinates, and organize finite-size effects into interactions involving operators that transform in SU$(N+1)$ representations, denoted by $(p,q)$. 
It is convenient to phrase the construction in terms of $\mathbb{C} P^N$ embedded in $\mathbb{C}^{N+1}$.
In complex coordinates $Z^A$, the flat metric is (using the natural isomorphism $\mathbb{R}^{2N+2}\cong \mathbb{C}^{N+1}$)
\begin{equation}\label{eq:complexflatmetric}
    \rd s^2_{2N+2}= \rd Z^A \rd\bar{Z}_A\,,
\end{equation}
where $A$ runs from $0$ to $N$. To connect with the coordinates used in Section~\ref{sec:InfDGRcalc}, 
we split $A$ as $(0,a)$, with $1\leq a \leq N$, and define~\cite{Hoxha:2000jf}
\begin{equation}
    x^a = \frac{Z^a}{Z^0},\qquad Z^0=\E^{i\psi} |Z^0|, \qquad r=\sqrt{Z^A \bar{Z}_A}\,,
\end{equation}
which then correspond to the coordinates in~\eqref{eq:generalDMP}.

We now want to include finite-size interactions on the worldline in these coordinates. To do this, we build
the EFT out of interactions of the form
\be
S_{\rm int} \supset \int\rd\tau E\, Q^{a_1\cdots a_p}_{b_1 \cdots b_q} (X,E) \,C_{a_1\cdots a_p}^{b_1 \cdots b_q}\,,
\label{eq:sintLargeD}
\ee
which should be contrasted with~\eqref{eq:sint}. (Note that in this expression, the indices $a, b$ are {\it lab-frame} indices that coordinatize $\mathbb{C} P^N$, in contrast to Section~\ref{sec:EFT}, where they denoted corotating indices.) In the case of interest, the operator $C$ is built from the external scalar field as\footnote{The notation $\langle\cdots\rangle_T$ requires some explanation: it means that we symmetrize over the enclosed indices and remove traces with respect to the opposite raised/lowered indies. For example in~\eqref{eq:scalarininfD} any contraction between a lowered and raised index vanishes, but contractions between two lower or two upper indices need not vanish.}
\begin{equation}
    C_{a_1\cdots a_p}^{b_1 \cdots b_q}= \partial_{\langle a_1}\cdots\partial_{a_p\rangle_T} \bar\partial^{\langle b_1}\cdots\bar\partial^{b_q\rangle_T}\phi\,,
    \label{eq:scalarininfD}
\end{equation}
where lower indices are derivatives with respect to unbarred coordinates $Z^a$ and upper indices are derivatives with respect to barred coordinates, $\bar Z_b$. In this organization of the EFT, each of the $C$ operators transforms in a definite $(p,q)$ representation of SU$(N+1)$, which will make it easy to implement this symmetry in the EFT---essentially we will just find that the Green's function is diagonal in this space.

\noindent
{\it One-point function:}  
The calculation of the one-point function proceeds largely the same way as before. We perform a Schwinger--Keldysh path integral, integrate out the $X$ degrees of freedom, and define the Green's function
\be
G_R\, {}^{a_1\cdots a_p}_{b_1 \cdots b_q} {}^{\, c_1\cdots c_{p'}}_{\, d_1 \cdots d_{q'}}(\tau-\tau') = i\theta(\tau-\tau')\langle [Q^{a_1\cdots a_p}_{b_1 \cdots b_q}(\tau),Q^{c_1\cdots c_{p'}}_{d_1 \cdots d_{q'}}(\tau')]\rangle\,,
\label{eq:QQ2ptLargeD}
\ee
so that it has the expansion in frequency space
\be
G_R\,^{a_1\cdots a_p}_{b_1 \cdots b_q} {}^{\, c_1\cdots c_{p'}}_{\, d_1 \cdots d_{q'}}(\omega) = \lambda^{(0)}{}^{a_1\cdots a_p}_{b_1 \cdots b_q} {}^{\, c_1\cdots c_{p'}}_{\, d_1 \cdots d_{q'}}+(r_s\omega)\lambda^{(1)}{}^{a_1\cdots a_p}_{b_1 \cdots b_q} {}^{\, c_1\cdots c_{p'}}_{\, d_1 \cdots d_{q'}}+\cdots\,.
\label{eq:greensfunctionexpLargeD}
\ee
To calculate the one-point function, we start with the field profile
\begin{equation}
    \bar{\phi}=\E^{-i \omega t}\mathcal{E}_{ \langle c_1\cdots c_{p'}\rangle_T}^{\langle d_1 \cdots d_{q'}\rangle_T} Z^{ c_1 }\cdots Z^{c_{p'}}\bar{Z}_{d_1}\cdots \bar{Z}_{d_{q'}}\,,
\end{equation}
where $\mathcal{E}_{ \langle c_1\cdots c_{p'}\rangle_T}^{\langle d_1 \cdots d_{q'}\rangle_T}$ is an arbitrary tensor which is symmetric and traceless between its lower and upper indices. Following the same steps as in Section~\ref{sec:1ptfunction}, we find that
\be
\langle \varphi_+(\vec p, t)\rangle 
=(-i)^{p+q}\E^{-i \omega t}\frac{\bar{p}_{\langle a_1}\cdots \bar{p}_{a_p\rangle_T}p^{\langle b_1}\cdots p^{b_q\rangle_T}}{p^2} \,p'!\, q'!\,\mathcal{E}_{\langle c_1\cdots c_{p'}\rangle_T}^{\langle d_1 \cdots d_{q'}\rangle_T}G_{R}^{(Q)}{}^{a_1\cdots a_p}_{b_1 \cdots j_q} {}^{c_1\cdots c_{p'}}_{d_1 \cdots d_{q'}}(\omega)
\,.
\ee
Using the complex version of~\eqref{propagatorFouriertoreal}
\begin{equation}
    \int \frac{\rd^{N+1} p \, \rd^{N+1} \bar{p}}{(2 \pi)^{2N+2}} \E^{i(\bar{p}_A Z^A+ p^A \bar{Z}_A)} \frac{1}{p^2}= \frac{\Gamma(N)}{(4\pi)^{N+1}}\left(\frac{Z^A \bar{Z}_A}{4}\right)^{-N}\,,
\end{equation} 
we can take $p$ derivatives with respect to $Z^a$ and $q$ derivatives with respect to $\bar{Z}_b$ on either side, and remove the traces to yield the complex Fourier integral
\be
\begin{aligned}
    (i)^{p+q}&\int \frac{\rd^{N+1} p \, \rd^{N	+1} \bar{p}}{(2 \pi)^{2N+2}}\bar{p}_{\langle a_1}\cdots \bar{p}_{a_p\rangle_T}p^{\langle b_1}\cdots p^{b_q\rangle_T} \E^{i(\bar{p}_A Z^A+ p^A \bar{Z}_A)} \frac{1}{p^2}\\
     &\hspace{2cm}=\frac{\Gamma(N)\Gamma(1-N)}{2^{p+q}(4\pi)^{N+1}\Gamma(1-N-p-q)}\bar{Z}_{\langle a_1}\cdots \bar{Z}_{a_p\rangle_T}Z^{\langle b_1}\cdots Z^{b_q\rangle_T}\left(\frac{Z^A \bar{Z}_A}{4}\right)^{-N-p-q}\,.
\end{aligned}
\ee
Using this formula, 
the real space response can be written (for $p+q \in \mathbb{Z}$) as
\be
\label{eq:intermediatelargeDEFT}
\begin{aligned}
\langle \varphi_+(\vec{Z}, t)\rangle 
&=\E^{-i \omega t}\frac{\Gamma(N+p+q)\Gamma(1+ p')\Gamma(1+q')}{2^{2-p-q}\pi^{N+1}}\left(Z^A \bar{Z}_A\right)^{-N-p-q}\\
&\hspace{2cm}\times\bar{Z}_{\langle a_1}\cdots \bar{Z}_{a_p\rangle_T}Z^{\langle b_1}\cdots Z^{b_q\rangle_T}\mathcal{E}_{\langle c_1\cdots c_{p'}\rangle_T}^{\langle d_1 \cdots d_{q'}\rangle_T}G_{R}^{(Q)}{}^{a_1\cdots a_p}_{b_1 \cdots b_q} {}^{c_1\cdots c_{p'}}_{d_1 \cdots d_{q'}}(\omega)
\,,
\end{aligned}
\ee
which is the analogue of~\eqref{eq:1ptfunctionexp}, albeit with an explicit expression substituted in for $\bar\phi$.

\vspace{-12pt}
\paragraph{Matching to GR Calculation:}
We now want to match to the 
large $D$ microscopic calculation performed in Section~\ref{sec:InfDGRcalc}. To do this, we need to evaluate~\eqref{eq:intermediatelargeDEFT} a bit more explicitly. If desired, we could expand the Green's function in~\eqref{eq:intermediatelargeDEFT} in the basis of Thorne tensor analogues for ${\mathbb C}P^N$ and match that way. However, since many of the response coefficients are related by symmetry, it makes sense to instead expand the Green's function in terms of the invariant tensors of SU$(N+1)$.\footnote{The analogue in the Schwarzschild case is to expand the Green's function (or the worldline operators) in terms of rotationally invariant tensors built from $\delta_{ij}$, as was done in~\cite{Hui:2020xxx}.}

We first note that the parameters of the EFT and the UV calculation are related by
\begin{equation}
    L=p+q, \qquad\quad M=p-q\,.
\end{equation}
The solution obtained in the UV is diagonal in $L$ and $M$, so we may set $p=p'$, $q=q'$. 
This is a reflection of the fact that the system has SU$(N+1)$ symmetry. A further consequence is that the responses do not depend on the other magnetic quantum numbers that the states in the $(p,q)$ representation are labeled by.
We could make this more explicit by decomposing  $G_{R}^{(Q)}$ in terms of  SU$(N+1)$ invariants, which map the $(p,q)$ representation to itself. In this case, the relevant invariants are $g^{ab}$, $g_{cd}$ where $g$ is the metric~\eqref{eq:complexflatmetric}.
We can then write~\cite{Hoxha:2000jf} (See Appendix~\ref{app:CPNeigen})
\begin{equation}
    \mathcal{E}_{ \langle a_1\cdots a_{p'}\rangle_T}^{\langle b_1 \cdots b_{q'}\rangle_T} Z^{ a_1\cdots a_{p'}}\bar{Z}_{b_1 \cdots b_{q'}}=\mathcal{E}_{LM} r^L \E^{iM\psi} \mathbb{Y}_{LM}(x^a)\,.
\end{equation}
Then the fact that the response is diagonal in $L,M$, leads to the following expansion of the one-point function response
\begin{equation}
    \label{eq:EFTlargeD}
    \langle \varphi_+(\vec{x}, t)\rangle 
    =\frac{\Gamma(N+L)\Gamma(\frac{L+M+2}{2})\Gamma(\frac{L-M+2}{2})}{2^{2-L}\pi^{N+1}}r^{-2L-2N}(\lambda^{(0)}_{LM}+ r_s \omega \lambda^{(1)}_{LM}+\cdots)\,\bar\phi_{LM}\,.
\end{equation}
We again split the above Wilson coefficients into real and imaginary parts as
\begin{equation}
    \lambda^{(n)}_{LM}= k^{(n)}_{LM}+i\nu^{(n)}_{LM} \,.
\end{equation}
Comparing~\eqref{eq:EFTlargeD} with~\eqref{eq:LargeDfirstorderresponse}, we see that all response coefficients will have the common, real prefactor:
\begin{equation}
    \lambda_{LM}^{\rm LD} \equiv r_h^{2L+2N}\frac{2^{2-L}\pi^{N+1}|\Gamma (\hat L+1+i U)|^4 }{\pi\Gamma(N+L)\Gamma(\frac{L+M}{2}+1)\Gamma(\frac{L-M}{2}+1)  \Gamma (2 \hat L+1) \Gamma (2 \hat L+2)}\,,
\end{equation}
which it is therefore convenient to define as a separate parameter.

\subsubsection{Conservative response}
We begin by matching the conservative sector. Comparing~\eqref{eq:EFTlargeD} with~\eqref{eq:LargeDfirstorderresponse}, we find
\begin{tcolorbox}[colframe=white,arc=0pt,colback=greyish2]
    \begin{align}
        k^{(0)}_{LM}&=\lambda_{LM}^{\rm LD} \left(\csc (2 \pi \hat L)-\cot (2 \pi  \hat L) \cosh (2 \pi  U)\right)\,,\\
        k^{(1)}_{LM} &=\frac{\lambda_{LM}^{\rm LD}}{2N}\frac{1}{r_s \kappa } \Big(\pi  \cot (2 \pi  \hat L)\sinh (2 \pi  U) + \Im(\Delta{\psi})\left[\cosh(2\pi U)\cot(2 \pi \hat L)- \csc(2\pi \hat L)\right]\Big)\,.
    \end{align}
\end{tcolorbox}
\noindent
These responses share many of the properties of the previous cases. 
Much like the Myers--Perry case, when $U\neq 0$, the Love number $k^{(0)}_{LM}$ does not vanish for any $L$. 
The response does diverge for integer values of $\hat L$, corresponding to a classical running of the coupling. As a consistency check, we can verify that these responses agree with the Schwarzschild results~\eqref{eq:Schwconservativeresp} in the $U\to0$ limit.

The conservative responses display a somewhat curious feature: if we artificially set $N = 1$ and identify the two $5D$ Myers Perry spin parameters with $U$ (that is, set $U_{\rm MP}=V_{\rm MP}=U_{\rm Large D}$), then the responses agree exactly with the $5D$ Myers--Perry result. It would be interesting to understand if there is a physical origin for this coincidence.

\subsubsection{Dissipative response}

We can now proceed to match the dissipative sector. Comparing~\eqref{eq:EFTlargeD} with~\eqref{eq:LargeDfirstorderresponse} one again yields
\begin{tcolorbox}[colframe=white,arc=0pt,colback=greyish2]
    \begin{align}
        \nu^{(0)}&=-\lambda_{LM}^{\rm LD} \sinh (2 \pi  U) \\
        \nu^{(1)}&=\frac{\lambda_{LM}^{\rm LD} }{2 N}\frac{1}{r_s \kappa}\Big(\pi  \cosh (2 \pi  U)+ \Im(\Delta{\psi}) \sinh (2 \pi  U)\Big) \,.
    \end{align}
\end{tcolorbox}
\noindent
As in the conservative sector, we can check that these quantities agree with the Schwarzschild responses in the $U\to 0$ limit. Interestingly, the leading $1/D$ results accurately reproduce the Schwarzschild results at finite $D$ once we scale out some overall factors in defining $\hat L$.

\subsubsection{Connection with Thorne Tensor Expansion}
When matching, it was convenient to organize the EFT in terms of interactions that are invariant under SU$(N+1)$ transformations, which makes manifest the various Wilson coefficients that are related by the symmetries of the black hole geometry. 
However, we could envision a situation where we would want to compare this EFT construction to the one from Section~\ref{sec:EFT}, which is expanded in terms of Thorne tensors, which allow all Wilson coefficients to be independent. To relate these two EFT constructions, we would like a tensor that decomposes EFT interactions which transform in a fixed SO$(D-2)$ representation labeled by $L$ into terms which transform in a given SU$(N+1)$ representation, labeled by $(p,q)$. Equivalently, given a tensor $\mathcal{E}_{ \langle A_1\cdots A_{p}\rangle_T}^{\langle B_1 \cdots B_{q}\rangle_T}$ which specifies an eigenfunction of the charged laplacian on $\mathbb{C} P^N$, we would like a tensor $\mathcal{Y}_{Lpq}$ such that
\begin{equation}
    \mathcal{E}_{ \langle A_1\cdots A_{p}\rangle_T}^{\langle B_1 \cdots B_{q}\rangle_T}{\cal Y}_{Lpq}^{a_1 \cdots a_L}{}^{ \langle A_1\cdots A_{p}\rangle_T}_{\langle B_1 \cdots B_{q}\rangle_T} x_{(a_1}\cdots x_{a_L)_T}=\mathcal{E}_{ \langle A_1\cdots A_{p}\rangle_T}^{\langle B_1 \cdots B_{q}\rangle_T}Z^{A_1}\cdots Z^{A_p}\bar{Z}_{B_1}\cdots \bar{Z}_{B_q}\,,
\end{equation}
which relates the $L$ representation to the $(p,q)$ one. Here
the $a$ indices and $x$ coordinates correspond to $\mathbb{R}^{2N+2}$ and the $A, B$ indices and $Z$ coordinates correspond to $\mathbb{C}^{N+1}$.
Note that $p,q$ do not fully specify the eigenfunctions, there are additional magnetic quantum numbers implicitly contained in ${\cal E}$ that we do not write explicitly because the responses do not depend on these numbers.

To identify the tensor $\mathcal{Y}_{Lpq}$, we start by noting that we can write the isomorphism between $\mathbb{R}^{2(N+1)}$ and $\mathbb{C}^{N+1}$ as  $\hat Z^A=\hat x_{2A}+i \hat x_{2A+1}$
where $0\leq A \leq N$. This equation should be read as identifying unit vector in the complex $A$ direction in ${\mathbb C}^{N+1}$ as a complex linear combination of the unit vectors in the $2A$ and $(2A+1){}^{\rm th}$ real directions. This prescription groups together the real coordinates in pairs to form complex coordinates.
We can then define a set of helicity basis vectors in the same way as in~\eqref{eq:helicityeigenvectors}, 
\begin{equation}
    (e^A_+)^i= (\hat{x}^{2A})^i+i (\hat{x}^{2A+1})^i, \qquad\quad (e_A^-)^i= (\hat{x}_{2A})^i-i (\hat{x}_{2A+1})^i\,,
\end{equation}
where for example $(\hat x^{2A})^i$ means the components of the unit vector in the $2A{}^{\rm th}$ real direction. There are $2N+2$ such vectors, which span $\mathbb{R}^{2N+2}$, and they form complex conjugate pairs, so they also span ${\mathbb C}^{N+1}$. In particular, these objects allow us to translate between vectors in $\mathbb{R}^{2N+2}$ and ${\mathbb C}^{N+1}$ straightforwardly via the relations $ Z^A=x^i(e^A_+)_i$ and $\bar{Z}_A=x^i(e_A^-)_i$
where $Z^A$ and $x^i$ are vector components in $\mathbb{R}^{2N+2}$ and $\mathbb{C}^{N+1}$ respectively. 

From here, we see that what we want is a tensor with the right index symmetries built from $L$ basis vectors, $p$ of which are $+$ and $q = L-p$ of which are $-$:
\begin{equation}
    {\cal Y}_{Lpq}^{i_1 \cdots i_L}{}^{ \langle A_1\cdots A_{p}\rangle_T}_{\langle B_1 \cdots B_{q}\rangle_T}=(e^{\langle A_1}_+)^{(i_1}\cdots(e^{A_p \rangle_T}_+)^{i_p}(e_{\langle B_1}^-)^{i_{p+1}}\cdots(e_{B_q \rangle_T}^-)^{i_L )_T}\,.
\end{equation}
These tensors are a linear combination of all the $\mathcal{Y}_{L m_1 \cdots m_N}$ Thorne tensors which satisfy $\sum_{i=1}^{N} m_i=M=p-q$. In other words, scattering in this system only depends on the number of $+$ versus $-$ basis vectors, not which plane they came from. This is a consequence of the U$(N+1)$ symmetry of the problem which rotates these planes into each other. 
Thus, if we were to match our EFT in the Thorne tensor basis, we would find all Thorne tensors with total angular momentum $L$ and $\sum_{i=1}^{N} m_i=M$ would have the Wilson coefficients found in Section~\ref{sec:InfDEFT}. This is all consistent with the fact that the equal-spin limit is a special case of the general situation where some of the Wilson coefficients are related to each other. We will see a similar example in Section~\ref{sec:USEFT}.

\newpage
\section{Ultra-Spinning}
\label{sec:US}

A novel feature of black holes in $D\geq 6$ is that they do not necessarily have an extremality bound~\cite{Myers:1986un,Emparan:2008eg}. Consequently there are so-called ``ultra-spinning" regimes for these objects, where the spin parameter is scaled to be very large. 
The simplest example is provided by the Myers--Perry Black hole with a single spin parameter, which can be taken to infinity.
This parameter regime is interesting for two reasons: the first is that this is a new physical regime to study the properties of black holes, which does not have an analogue in four spacetime dimensions. Second, in the infinite spin limit it turns out that linear black hole tidal responses can be computed exactly in any dimension.
The ultra-spinning regime therefore represents an analytically tractable corner of parameter space of black hole properties in generic dimension with interesting physics properties.

\subsection{General Relativity Calculation}
\label{sec:USGRcalc}

We begin by considering the metric for an ultra-spinning black hole. Recall that the line element
of a Myers--Perry black hole in $D=n+4$ spacetime dimensions with a single nonzero spin parameter can be written as~\cite{Myers:1986un}
\begin{equation}
\rd s^2 = - \rd t^2 +\frac{\Sigma}{\Delta} \rd r^2 + \Sigma \rd \theta^2 + (r^2+a^2)\sin^2\theta \, \rd \varphi^2
+ \frac{r_s^{n+1}}{r^{n-1}\Sigma}\big(\rd t-a \sin^2\theta \, \rd \varphi\big)^2
+ r^2 \cos^2\theta  \, \rd \Omega_n^2\, ,
\label{MPmetric}
\end{equation}
where we have defined the functions
\begin{equation}
\Delta= r^2+a^2 - \frac{r_s^{n+1}}{r^{n-1}} \, ,
\qquad\quad
\Sigma = r^2 + a^2 \cos^2\theta \, ,
\end{equation}
and where $\rd \Omega_n^2$ is the line-element on the unit $n$-sphere. In this case the ADM mass and angular momentum are given by 
\begin{equation}
M_{\rm ADM} = \frac{(n+2)A_{n+2}}{16\pi G} r_s^{n+1} \,,
\qquad\quad
J = \frac{2M_{\rm ADM}}{n+2} a \, ,
\end{equation}
where $A_{n+2} = 2\pi^{\frac{n+3}{2}}/\Gamma\left(\frac{n+3}{2}\right)$
is the area of the $(n+2)$-dimensional unit sphere.

We can separate the Klein--Gordon equation $\square\Phi = 0$ in the background~\eqref{MPmetric} using the ansatz
\begin{equation}
\Phi(t,r,\theta,\varphi, \theta_1, \ldots, \theta_n)= \E^{-i\omega t} \E^{i m \varphi} R_{\omega L mjM}(r) S_{L m}(\theta) Y_{jM}(\theta_1, \ldots, \theta_n) \, ,
\end{equation}
where $Y_{jM}(\theta_1, \ldots, \theta_n)$ are spherical harmonics on the $n$-sphere, with angular momentum $j$, and $M$ is a multi-index denoting the remaining magnetic quantum numbers. Importantly, these magnetic quantum numbers, $m_i$, cannot be totally arbitrary, but must add up to the total angular momentum $j=\sum_{i=2}^r m_i$.
 In addition to this, one can interpret the combination
$\E^{i m \varphi}S_{L m}(\theta)Y_{jM}(\theta_1, \ldots, \theta_n)$ as an $(n+2)$-spheroidal harmonic~\cite{Teukolsky:1973ha}.

After separation of variables, the radial part of the Klein--Gordon equation takes the form~\cite{Boonserm:2014fja} (where we have dropped the labels on $R$ for simplicity):
\begin{equation}
 \left[\frac{1}{r^n}\frac{\rd}{\rd r}\left( r^n \Delta \frac{\rd}{\rd r}\right)  +
\frac{\left((r^2+a^2)\omega-ma\right)^2}{\Delta}- \frac{j(j+n-1)a^2}{r^2} - \lambda_{Ljm}
\right]R (r)=0 \, ,
\label{eq:USKGeqradial}
\end{equation}
while the angular equation is
\begin{equation}
\left[  \frac{1}{\sin\theta \, \cos^n\theta} \frac{\rd}{\rd \theta}\left(\sin\theta \, \cos^n\theta\frac{\rd}{\rd \theta}  \right)
- \left( \omega a \sin\theta -\frac{m}{\sin\theta}  \right)^2- \frac{j(j+n-1)}{\cos^2\theta} + \lambda_{L\ell m}
\right]S_{L m}(\theta)=0\, .
\label{eq:USKGangular}
\end{equation}
In these equations, $\lambda_{Lj m}$ is a separation constant.
In the static limit,\footnote{At finite frequency, the approximate separation constant depends on how we scale the product $a\omega$. If we take the limit $\omega\to0$, $a\to\infty$ such that $a \omega \ll 1$, the separation constant can be approximated as $\lambda_{Lj m} \approx L(L+n+1) -2m a \omega +\ord{(a \omega)^2}$. If instead we take the limit with $a \omega \gg 1$, the separation constant becomes $\lambda_{Lj m} \approx 2 a \omega \left(L +1 -j -m\right) +\ord{(a \omega)^0}$~\cite{Berti:2005gp}. We leave $\lambda_{Lj m}$ implicit in our microscopic calculation, but assume the former limit when matching.} the separation constant $ \lambda_{Lj m}$ takes the simple analytic  form~\cite{Berti:2005gp}
\begin{equation}
\begin{aligned}
 \lambda_{Lj m} &= L(L+n+1),\\[2pt]
 &=(2k + j +\vert m\vert)(2k + j +\vert m\vert+n+1)\,, \qquad\quad  (\text{for}~\omega=0)
 \end{aligned}
\end{equation}
where $L$ is constrained to be
\begin{equation}
L = 2k + j +\vert m\vert \, ,
\qquad\text{with}\quad k= 0,1,2,\cdots\,.
\label{defj}
\end{equation}
Note that taking the $D=4$ limit to recover the Kerr case is a little subtle, we need to set both $n=0$ and $j= L-m \mod 2 $. The latter is necessary to remove the dependence on $Y_{jM}$ in the scalar field decomposition.\footnote{We can think of $j$ as cataloging the parity transformation of the even versus odd $L$ spherical harmonics.}

When $\omega=0$, it is possible to write the angular equation~\eqref{eq:USKGangular} as
\be
\label{eq:singlespinangularLaplacianoperator}
\left[  \frac{1}{\sin\theta \, \cos^n\theta} \frac{\rd}{\rd \theta}\left(\sin\theta \, \cos^n\theta\frac{\rd}{\rd \theta}  \right)
+\frac{1}{\sin^2\theta}\frac{\partial^2}{\partial\varphi^2} + \frac{1}{\cos^2\theta}\Delta_{S^n}
\right]S_{Lm}(\theta) = - L(L+n+1)S_{Lm}(\theta)\,,
\ee
which is precisely the angular laplacian in a choice of polyspherical coordinates, and is solved by Jacobi polynomials~\cite{Berti:2005gp}. Explicitly, the relation between these coordinates and Cartesian coordinates on $\mathbb{R}^{D-1}$ is 
\begin{equation}
    \label{eq:USpolysphericalcoords}
    \hat{x}^i=
    \left(
    \begin{array}{c}
    \sin \theta \cos \varphi\\
    \sin \theta \sin \varphi\\
    \cos \theta \, \hat{y}_1\\
    \cos \theta \, \hat{y}_2\\
    \cos \theta \, \hat{y}_3\\
    \vdots
    \end{array}
    \right)\,,
\end{equation}
where $\hat{x}$ is a point on $S_{n+2}$ and $\hat{y}$ is a point on $S_n$, parametrized by $\theta_1,\ldots,\theta_n$. These relations can be seen by taking the $a, r_s \rightarrow 0$ limit of the metric. From this, we infer that $\E^{i m \varphi} S_{L m}(\theta) Y_{jM}(\theta_1, \ldots, \theta_n)$ is a spherical harmonic on $S_{n+2}$, written in these coordinates. This is the analogue of the angular sector in the Kerr case being solved by spherical harmonics in the $\omega \to 0$ limit.

\vspace{-12pt}
\paragraph{Near Zone:} 
In order to solve~\eqref{eq:USKGeqradial}, we again need to make an approximation so that it is of the hypergeometric form. In the large $a$ limit, we can approximate the horizon radius as $r_h^{n-1}= \frac{r_s^{n+1}}{a^2}$. Starting from~\eqref{eq:USKGeqradial}, if we make the redefinitions 
\be
\begin{aligned}
    \hat{a}&=\frac{a}{r_s},\qquad\qquad& \rho&=\hat{a}^2\left(\frac{r}{r_s}\right)^{n-1},\qquad& \hat{j}&=\frac{j}{n-1},\\
    \hat{m}&=\frac{m}{n-1},& \hat \omega&=\frac{\omega}{n-1},&  \hat{\lambda}_{Lj m}&=\frac{\lambda_{Lj m}}{(n-1)^2}\,,
\end{aligned}
\ee
and define the small parameter
\begin{equation}
    \delta_a\equiv \hat{a}^{-\frac{n+1}{n-1}}=\frac{r_h}{a}\,,
\end{equation}
along with the quantity
\begin{equation}
    \tilde{\Delta} \equiv \rho\left(\rho-1+\delta_a^2 \rho^{\frac{n+1}{n-1}}\right)\,,
\end{equation}
we can rewrite this equation as 
\begin{equation}
    \label{eq:USKGeqinrho}
    \tilde{\Delta}\partial_\rho \left(\tilde{\Delta} R'(\rho)\right) +\left[-\left(\hat{j}(\hat{j}+1)+ \hat{\lambda}_{Lj m} \delta_a^2 \rho^{\tfrac{2}{n-1}}\right)\tilde{\Delta}+ \delta_a^2\, \rho^{\tfrac{2n}{n-1}} \left((1+\delta_a^2 \rho^{\tfrac{2}{n-1}}) \,a \hat \omega-\hat{m}\right)^2\right]R(\rho)=0\,.
\end{equation}
The approximation that we would like to make is to set $\rho$ to its value at the horizon everywhere except in the vicinity of singularities of the differential equation.
The precise value that $\rho$ takes at the horizon is a little difficult to obtain analytically (since $\tilde{\Delta}$ is a $2n$ degree polynomial), but we can see in the large $a$ limit that $\rho_h=1-\delta_a^2+\ord{\delta_a^4}$.
Since we are making this approximation in terms that already have an overall factor of $\delta_a^2$---and we only expect the solution to be accurate to $\mathcal{O}(\delta_a)$---it suffices to set $\rho_h=1$. 
We may look at $\tilde{\Delta}$ to define the region of validity of this approximation. In particular, we see that we should expect this approximation to be reliable as long as
$\rho \gg \delta_a^2 \rho^{\frac{n+1}{n-1}}$, or when $a \gg r$. 
In sum, the near zone approximation sets
\begin{equation}
    \delta_a^2 \rho^{\frac{n+1}{n-1}}\rightarrow \delta_a^2\,.
\end{equation}
With this approximation, we can also replace 
\begin{equation}
    \tilde{\Delta}\approx \tilde{\Delta}_0\equiv\rho\left(\rho-1+\delta_a^2\right)\,.
\end{equation}
After this, the near zone equation of motion is
\begin{equation}
    \Delta_0 \partial_\rho \left(\Delta_0 R'(\rho)\right) +\left[-\left(\hat{j}(\hat{j}+1)+ \hat{\lambda}_{Lj m} \delta_a^2 \rho^{-1}\right)\Delta_0+ \delta_a^2 \rho \left((1+\delta_a^2 \rho^{-1}) \,a \hat \omega-\hat{m}\right)^2\right]R(\rho)=0\,.
\end{equation}
We can put this in hypergeometric form by first defining 
\begin{align}
    z&=\frac{\rho }{1-\delta_a^2}-1,\quad& P&=\hat{m}-a\hat{\omega},\quad &\tilde{\delta}_a^2&=\frac{\delta_a^2}{1-\delta_a^2}\,,\\
    \tilde{P}&=\tilde{\delta}_a P,\qquad &\tilde{W}&=\tilde{\delta}_a^3 a \hat{\omega},\qquad&\tilde{\lambda}&=\tilde{\delta}_a^2 \hat{\lambda}_{Lj m}\,.
\end{align}
After doing this, we find that the radial equation is
\begin{equation}
    \label{eq:USnearzoneinz}
    z(z+1)\partial_z(z(z+1)R'(z))+\left(-\hat{j}(\hat{j}+1)z(z+1)+ \tilde{P}^2 -2\tilde{P} \tilde{W} +z(\tilde{P}^2-\tilde{\lambda})+\frac{\tilde{W}^2}{z+1} \right)R(z)=0 \, .
\end{equation}
A final approximation that we can make is to set $z=z_h=0$ in the term with $\tilde W^2$ because it is $\ord{\tilde{\delta_a^6} a^2}$, and hence subleading in the large $a$ limit. The equation we want to solve is then
\begin{tcolorbox}[colframe=white,arc=0pt,colback=greyish2]
\vspace{-12pt}
\begin{equation}
    \label{eq:USnearzoneinz2}
    z(z+1)\partial_z(z(z+1)R'(z))+\left(-\hat{j}(\hat{j}+1)z(z+1)+ (\tilde{P}-\tilde{W})^2 -z(\tilde{\lambda}-\tilde{P}^2)\right)R(z)=0 \, .
\end{equation}
\end{tcolorbox}
It is convenient to define 
\begin{equation}
    P_*\equiv \tilde{P}-\tilde{W}\,.
\end{equation}
We can then cast the equation in standard hypergeometric form with the field redefinition
\begin{equation}
    R(z)=z^{i P_*}(1+z)^{i\sqrt{\tilde{\lambda}+P_*^2}}u(z)\,.
\end{equation}
The full form of the equation is rather lengthy, so we do not write it explicitly. In any case its parameters can readily be inferred from the following solution.

\vspace{-12pt}
\paragraph{Solutions and Boundary Conditions:}
The near zone equation~\eqref{eq:USnearzoneinz2} takes the same form as the Kerr equation of motion~\eqref{eq:KerrzEoM}, with some parameter redefinitions. As a result, we want to apply conceptually the same boundary conditions. The solution that is regular at the horizon is
\be
u(z) = {}_2F_1\left[\begin{array}{c}
-\hat j +i \left(P_*+\sqrt{\tilde{\lambda}+P_*^2}\right)\,,\,\hat j +1+i \left(P_*+\sqrt{\tilde{\lambda}+P_*^2}\right)\\[0pt]
1+2 i P_*
\end{array}\bigg\rvert \,-z\,\right]\,.
\ee
We want to use this solution to read off the ratio of fall-offs at large distance.
Expanding around $z \to \infty$ yields
\begin{equation}
    \label{eq:USUVLovenumberunapproximated}
    R(z)\xrightarrow{z \to \infty}  z^{\hat j}\left( 1+ \cdots + z^{-2\hat j-1} \frac{\Gamma (-2 \hat j-1) \Gamma (\hat j +1 +i P_-) \Gamma (\hat j+1 +i P_+)}{ \Gamma (2 \hat j+1) \Gamma (-\hat j+ i P_-)\Gamma (-\hat j+ iP_+)}\right)\,,
\end{equation}
where we have defined
\begin{equation}
    P_\pm=\left(P_*\pm\sqrt{\tilde{\lambda}+P_*^2}\right)\,.
\end{equation}
Since we approximated~\eqref{eq:USKGeqinrho} at $\ord{\delta_a^2}$, we should only trust this expression to $\ord{\delta_a}$. In the large $a$ limit (where $\delta_a\to 0$),~\eqref{eq:USUVLovenumberunapproximated} becomes (taking note that there is $a$ dependence in $P$ that also has to be expanded)
\be
    \label{eq:USNZresponse}
    R(r)\xrightarrow{r \to \infty}  \left(\frac{r}{r_h}\right)^{j}\Bigg( 1+ \cdots + \left(\frac{r}{r_h}\right)^{-2j-n+1} A_{\omega j}
\Big[1- 2i \hat m \delta_a \left(\pi \cot(\pi \hat j +i \pi r_h \hat \omega) - \Delta_{j\omega}\psi\right)+\cdots\Big]\Bigg)\,.
\ee
where we have defined the quantities
\begin{align}
A_{\omega j} &\equiv \frac{  |\Gamma (\hat j+1+2 i r_h \hat \omega)|^4}{ \pi  \Gamma (2 \hat j+1) \Gamma (2\hat j+2)}\csc (2\pi  \hat j)\sin (\pi  (\hat j+ i r_h \hat \omega))^2\,,\\
    \Delta_{j\omega}\psi &\equiv  \psi(\hat j+1-i r_h \hat \omega)-\psi(\hat j+1+i r_h \hat \omega)\,,
\end{align}
to simplify the expression. Note that the response has no dependence on $\lambda_{Ljm}$ at $\ord{\delta_a}$.
(Note also that the near zone response in~\eqref{eq:USNZresponse} is of the same form as the response of a large $D$ black hole in Section~\ref{sec:InfDEFT}, after shifting variables as $\{\hat L\to \hat j\,, U\to -r_h \hat \omega \,, \frac{\hat \omega}{\kappa}\to -2 \hat m \delta_a\}$.)

An important feature of the expression~\eqref{eq:USNZresponse} is that it has asymptotic fall-offs that scale as $r^j$ and $r^{-j-n+1}$. This may seem at first sight surprising, because we would expect the fall-offs of a point-like object to scale as $r^L$ and $r^{-L-n-1}$ at large distances. One way to understand this is to note that in the large-spin limit, the horizon of the black hole flattens in the plane of rotation into something resembling a membrane with topology $\mathbb{R}^2\times S^{n}$~\cite{Emparan:2003sy}. Indeed, the fall-offs appearing in~\eqref{eq:USNZresponse} are what we would expect for a $(2+1)$-dimensional membrane. In order to match to point-particle EFT, we need to be careful about the order of limits. In the EFT, we are interested in distances far from the object, $r\gg r_s$, and in the ultra-spinning case, we are interested in large spins $a\gg r_s$. In order to fully define this simultaneous limit, we have to prescribe how to scale the ratio $r/a$. In principle we can take the limit in many different ways, but the one of interest to match to point particle EFT is to take $r/a\gg1$ as we scale both of these quantities to infinity. However, this limit is not accessible to the near zone equation~\eqref{eq:USnearzoneinz2}. So, in order to match to the EFT, we must first match the GR solution to a far zone, which will then have the correct fall-offs that can be reproduced by the EFT.

\vspace{-12pt}
\paragraph{Far Zone:} In order to match to a point-particle EFT, we have to connect the solution~\eqref{eq:USNZresponse} to a far zone. In order to specify this far zone, we first define
\begin{equation}
    x\equiv \frac{r}{a},\qquad \hat a \equiv \frac{a}{r_s}=\delta_a^{-\frac{n-1}{n+1}},\qquad W\equiv a\, \omega\,, \qquad   \Delta_x\equiv1+x^2\big(1-\left(\hat{a} x\right)^{-n-1}\big)\,.
\end{equation}
With these definitions, we can then
 rewrite the full radial equation of motion~\eqref{eq:USKGeqradial} as
\begin{equation}
    x^{-n}\partial_x\left(x^n \Delta_x R'(x)\right)+ \left(-\lambda_{Ljm}-\frac{L(L-n-1)}{x^2}+\frac{\left(m-W-W x^2\right)^2}{\Delta_x}\right)R(x)=0\,.
    \label{eq:fullradialeqUS}
\end{equation}
We now take the large $x$ limit. In particular, we scale $\hat{a} x \gg 1$ (or $r\gg r_s$). In this limit, we find
\begin{equation}
    \Delta_x \xrightarrow{r \gg r_s} 1+x^2\,.
\end{equation}
In addition to the large $r$ limit, we must also make a small frequency approximation in order to transform~\eqref{eq:fullradialeqUS} into a hypergeometric equation. We scale $\omega \to 0$ sufficiently fast so that $W =\omega a \to 0$, even as $a\to\infty$. To $\ord{W}$, we can write $\lambda_{Ljm}= L(L+n+1)-2mW+\ord{W^2}$. Then, dropping terms of order $\ord{W^2}$ and $\ord{W^2x^2}$,
the far zone equation of motion becomes
\begin{tcolorbox}[colframe=white,arc=0pt,colback=greyish2]
\vspace{-10pt}
\be
    x^{-n}\partial_x\left(x^n (1+x^2)R'(x)\right)+ \left(-L(L+n+1)-\frac{L(L-n-1)}{x^2}+\frac{m^2}{1+x^2}\right)R(x)=0\,.
    \label{eq:USFZeqn}
\ee
\end{tcolorbox}
A basis of solutions to this equation (which are simple near $x=0$) is
\be
\label{eq:USFZsol}
\begin{aligned}
    R(x)= ~&c_1 x^j
    \left(x^2+1\right)^{m/2} \, 
    ~{}_2F_1\left[\!\!\begin{array}{c}
        \frac{1}{2} (j+m-L)\,,\,\frac{1}{2} (L+j+m+n+1)\\[-2pt]
        \frac{1}{2} (2j+n+1)
        \end{array}\bigg\rvert \,-x^2\,\right] \\
    &+c_2 \left(x^2+1\right)^{m/2} x^{-j-n+1} ~{}_2F_1\left[\!\!\begin{array}{c}
        \frac{1}{2} (L-j+m+2)\,,\,\frac{1}{2} (m-n+1-L-j)\\[-2pt]
        \frac{1}{2} (-2 j-n+3)
        \end{array}\bigg\rvert \,-x^2\,\right]\,.
\end{aligned}
\ee
As $x\to 0$, the solution has the expansion
\begin{equation}
    R(x)\xrightarrow{x\to0} c_1 x^j + c_2 x^{-j-n+1}\,,
\end{equation}
which we can match to~\eqref{eq:USNZresponse} in order to infer
\be
\frac{c_2}{c_1} = \delta_a^{2j+n-1}A_{\omega j}
\Big[1- 2i \hat m \delta_a \left(\pi \cot(\pi \hat j +i \pi r_h \hat \omega) - \Delta_{j\omega}\psi\right)+\ord{\delta_a^2}\Big]\,.
\label{eq:ratioc1c2}
\ee
We can then expand~\eqref{eq:USFZsol} as  $x\to \infty$ 
\begin{align}
    R(x)\xrightarrow{x\to\infty} &c_1 x^L\!\!\left(\frac{ \Gamma \left( j+\tfrac{n+1}{2}\right) \Gamma \left(L+\tfrac{n+1}{2}\right)}{\Gamma \left(\tfrac{j+L-m+n+1}{2}\right) \Gamma
    \left(\tfrac{j+L+m+n+1}{2}\right)}+\frac{c_2}{c_1}\frac{ \Gamma \left(\tfrac{3-n}{2}- j\right) \Gamma \left(L+\tfrac{n+1}{2})\right)}{\Gamma
    \left(\tfrac{L-j-m}{2}+1\right) \Gamma \left(\tfrac{L-j+m}{2}+1\right)}\right.\\
    &\!\!\!\!
    \left.+x^{-L-n-1}\left[\frac{ \Gamma \left(j+\tfrac{n+1}{2}\right) \Gamma \left(- L-\tfrac{n+1}{2}\right)}{\Gamma \left(\tfrac{j-L-m}{2}\right) \Gamma
    \left(\tfrac{j-L+m}{2}\right)}+\frac{c_2}{c_1}\frac{ \Gamma \left(\tfrac{3-n}{2}-j\right) \Gamma \left(-L-\tfrac{n+1}{2}\right)}{\Gamma
    \left(\tfrac{1-j-L-m-n}{2}\right) \Gamma \left(\tfrac{1-j-L+m-n}{2}\right)}
\right]+\cdots\right)\,.
 \nonumber
\end{align}
Substituting~\eqref{eq:ratioc1c2} into this expression, we find that the ratio of fall-offs to leading order in $\delta_a$ is
\be
R(r) \xrightarrow{r\to\infty} r^L\Big( 1+ C_{aLmjM}\, r^{-2L-n-1}\Big)\,,
\label{eq:infinityUSexpan}
\ee
where we have defined the coefficient
\begin{align}
\nonumber
 C_{aLmjM}(\omega) \equiv 
 a^{2 l+n+1}\, \delta_a^{2 j+n-1}&\frac{ \Gamma \left(\tfrac{j+L-m+n+1}{2}\right)^2 \Gamma \left(\tfrac{j+L+m+n+1}{2}\right)^2 }{\Gamma
    \left(j+\frac{n-1}{2}\right) \Gamma \left( j+\tfrac{n+1}{2}\right) \Gamma \left(L+\frac{n+3}{2}\right) \Gamma \left(
    L+\tfrac{n+1}{2}\right)} \\[2pt]
    &\hspace{1cm}\times A_{\omega j}\left[1-2 i \hat{m} \delta_a \left(H_{-\hat{j}-1-i
    r_h \hat{\omega}}-H_{\hat{j}-i r_h \hat{\omega}}\right)\right]\,.
    \label{eq:USratio}
\end{align}
Note that the near-zone ratio of fall-offs in~\eqref{eq:USNZresponse} is in principle reliable to all orders in $\omega$, since we did not have to make any small frequency approximation when we took $a\to \infty$. However, we {\it did} have to truncate the far zone at small frequency to obtain~\eqref{eq:USFZeqn}. As a result, the expression~\eqref{eq:USratio} is only reliable to $\ord{\omega}$. If we Taylor expand in frequency, we find 
\begin{align}
    \label{eq:USfarzoneapproximatedresponse}
 C&= a^{2 l+n+1}\, \delta_a^{2 j+n-1}\frac{\Gamma \left(\tfrac{j+L-m+n+1}{2}\right)^2 \Gamma \left(\tfrac{j+L+m+n+1}{2}\right)^2\Gamma(\hat{j}+1)^4}{\Gamma
    \left(j+\frac{n-1}{2}\right) \Gamma \left(j+\tfrac{n+1}{2}\right) \Gamma \left(L+\frac{n+3}{2}\right) \Gamma \left(
    L+\tfrac{n+1}{2}\right)\Gamma(2\hat j +1)\Gamma(2\hat j +2)}\nonumber\\
    &\hspace{.6cm} \times
    \left[\frac{\tan (\pi  \hat{j})}{2 \pi }-i \delta_a \hat{m}+r_h \hat{\omega}\left(i+
     \delta_a \hat{m} \Big[2 \pi \cot (\pi \hat{j})+\frac{ \tan (\pi \hat{j}) }{\pi }\big(\psi^{(1)}(\hat{j}+1)-\psi^{(1)}(-\hat{j})\big)\Big]\right)
\right] \,,
\end{align}
where we have dropped the labels on $C$ for simplicity, and
where $\psi^{(1)}(x) = \partial_x\psi(x)$ is the trigamma function. We can use polygamma identities to write
\begin{equation}
    2 \pi   \cot (\pi  \hat{j})+\frac{ \tan (\pi 
    \hat{j}) }{\pi }\left(\psi^{(1)}(\hat{j}+1)-\psi^{(1)}(-\hat{j})\right)=2\pi \cot(2\pi\hat j )+\frac{ 2\tan (\pi 
    \hat{j}) }{\pi }\sum_{k=1}^{\infty}\frac{1}{(k+\hat{j})^2}\,,
\end{equation}
which is real. This formula also allows us to read off the locations where this combination of polygammas vanish or diverge from the trigonometric components.

Notice that~\eqref{eq:infinityUSexpan} now has the correct fall-offs appropriate for a point particle, so we can match these responses with point-particle EFT.

\subsection{EFT Matching}
\label{sec:USEFT}

Here we match the EFT Wilson coefficients to reproduce the general relativity result~\eqref{eq:USfarzoneapproximatedresponse}.
 
The angular sector of the EFT one-point function~\eqref{eq:smallfreq1ptfinalschwarz} has been expanded in terms of higher-dimensional spherical harmonics. As we mentioned in Section~\ref{sec:USGRcalc}, the GR calculation at small frequency can also be decomposed into spherical harmonics, though the relation between the $L,m,j,M$ eigenvalues  of the GR calculation and the $L,m_1,\ldots m_r$ eigenvalues used to label spherical harmonics in the EFT is perhaps not totally obvious. As a first step we want to establish the dictionary between the two angular sectors.

The $L$ and $m$ eigenvalues in the GR calculation simply correspond to $L$ and $m_1$ in the EFT labeling of spherical harmonics, where $m_1$ is the quantum number associated to the plane in which the black hole has nonzero spin. The quantum number $j$---which labels the angular momentum in the distinguished $n$-sphere---is related to the remaining EFT eigenvalues in a slightly more involved manner.
It obeys the condition $j =L-|m|-2k$ for $k\in \mathbb{Z}^+$. As such one can see the parity of $j$ is fixed by the values of $L$ and $m$. Let
\begin{equation}
    \Sigma_m = \sum_{i=2}^{r} |m_i| \,,
\end{equation}
where $r=\lfloor\tfrac{d}{2}\rfloor$ is the number of orthogonal planes in the spacetime (the rank of the rotation group in $d$ spatial dimensions). Then we have
\begin{equation}
    \label{eq:USEFTjformula}
    j= \begin{cases}
        \Sigma_m & \text{if } L-|m|-\Sigma_m \text{ is even} \\
        \Sigma_m+1 & \text{if } L-|m|-\Sigma_m\text{ is odd}
    \end{cases}\,.
\end{equation}
With this, we may identify the values of $L,j,m$ for any Thorne tensor $\mathcal{Y}_{L m_1 \cdots m_r}^{i_1 \cdots i_L}$ or spherical harmonic $Y_{L m_1 \cdots m_r}$. 
In the following, we match the Wilson coefficients for Thorne tensor components of the Green's function that have eigenvalues corresponding to a given assignment of $L,j,m$ in the GR calculation. This is because responses with different magnetic quantum numbers have the same coefficients if they have the same values of $L,j,m$ (much like the Schwarzschild black hole has the same response for every $m$, and only depends on the value of $L$).

\vspace{-12pt}
\paragraph{Identifying $j$:} Above we stated the rule~\eqref{eq:USEFTjformula} to infer the value of $j$ corresponding to a given assignment of magnetic quantum numbers in the EFT. Here we wish to explain this correspondence. The argument is a bit involved, and the reader willing to take~\eqref{eq:USEFTjformula} on faith can skip it.

The relationship between $(L,j,m)$ and $(L,m_1,\ldots,m_r)$ is easiest to see by constructing the Thorne tensor that corresponds to a spherical harmonic labeled by $(L,j,m)$. 

We describe the construction of Thorne tensors in Appendix~\ref{appendix:Thorne}, but here we quickly summarize the relevant points. Thorne tensors are a set of symmetric traceless tensors with the property that their contractions with unit vectors on $S^{d-1}$ are spherical harmonics:
\be
{\cal Y}_{LM}^{i_1\cdots i_L} \hat x_{i_1}\cdots \hat x_{i_L} = Y_{LM}(\theta)\,.
\ee

To construct Thorne tensors, we first define a set of basis vector pairs, $(e^A)^i_\pm$, with $A\in 1,2,\cdots, \lfloor\tfrac{d}{2}\rfloor$, where each pair spans one of the orthogonal spin planes of the spacetime. In an odd number of spatial dimensions, we also define a vector $s^i$ which is orthogonal to the rest of these basis vectors using the epsilon tensor.
A Thorne tensor is then built from any symmetric and traceless combination of the $(e^A)^i_\pm$ and $s^i$, and is labeled by eigenvalues $(L,m_1,\ldots,m_r)$. The parameter $L$ counts the total number of vectors used in the construction (or indices of the final tensor), and the $m_A$ values are the number of $(e^A)^i_+$s used minus the number of $(e^A)^i_-$s.

We can now use this technology to relate $j$ to $L,M$ in the EFT. 
From~\eqref{eq:singlespinangularLaplacianoperator} and the coordinate choice~\eqref{eq:USpolysphericalcoords}, we see that the $L$ in the GR calculation corresponds to the eigenvalue of the laplacian on $S^{d-1} = S^{n+2}$, which is then the same $L$ that appears in the EFT. Similarly, $m$ in the GR calculation is the angular momentum corresponding to the axial Killing vector $\partial_{\varphi}$. We can---without loss of generality---chose this magnetic quantum number to be $m_1$ in the EFT. We then want to build the full set of Thorne tensors that have $L$ indices and $m_1 = m$. The remaining freedom is to pick  $L-|m_1|-2k$ vectors from $(e^A)^i_\pm$, with $A$ taking values $A=2,\cdots,\lfloor\tfrac{d}{2}\rfloor$, and $s^i$. (Here $k$ counts the number of 
pairs of vectors of the form $(e_+^A)^i (e_-^A)^j$ that we include, which do not change any of the $m$ values.) Our choice of these vectors will determine the $m_2,\cdots, m_r$ quantum numbers. We identify this number as $j=L-|m_1|-2k$, which then implies $L=2k+j+|m_1|$ as desired.  If one is instead given the values $(m_2,\ldots,m_r)$, the above formula~\eqref{eq:USEFTjformula} calculates $j$. 

We can interpret $j$ as the net number of vectors that do not lie in the black hole's spin plane. If we take these vectors on their own, their traceless symmetrization would correspond to a Thorne tensor on $S^{d-3}$ with angular momentum $j$ and with maximal magnetic quantum numbers, $j=\sum_{i=2}^r m_i$.

As a final comment, we can relate $j$ to the angular sector of the $5D$ Myers--Perry black hole and the Kerr black hole. In the Myers--Perry case, $j\to m_\psi$ and $Y_j(\theta_1,\ldots,\theta_n)\to \E^{i m_\psi \psi}$. In the Kerr case, we have to set $d=3$ and we find that either $j=0$ or $j=1$ and the dependence on $j$ drops out of the equation of motion~\eqref{eq:singlespinangularLaplacianoperator}. We should interpret $j$ in this case as a harmonic on $S^0 \equiv {\mathbb Z}_2$, which is just two points. We therefore indeed expect two possible eigenvalues $\pm 1$, which is just the parity of the spherical harmonic, determined by $L$ and $m$.

\vspace{-12pt}
\paragraph{EFT calculation:}

We now have enough technology to match the ultra-spinning black hole to a point-particle EFT. In the previous discussion, we saw that we only need to specify $L, j, m$---which are related to the EFT quantum numbers via $L=L$, $m=m_1$ and~\eqref{eq:USEFTjformula}. All assignments of $m_A$ that lead to the same $j,m$ will have the same response coefficients. With this understanding, we write~\eqref{eq:smallfreq1ptfinal} as
\begin{equation}
    \label{eq:smallfreq1ptfinalUS}
    \langle \varphi_+(\vec x, t)\rangle =\E^{-i \omega t}r^L\mathcal{E}_{Ljm}\, Y_{Ljm}\left( \lambda^{(0)}_{Ljm}+(r_s\omega)\,
    \lambda^{(1)}_{Ljm}+\cdots \right)\frac{2^{2L-3}}{\pi^d} \Gamma(L+\tfrac{d}{2})\Gamma(L+\tfrac{d}{2}-1)r^{-2L+2-d}\,,
\end{equation}
and expand
\begin{equation}
    \lambda^{(0)}_{Ljm}+(r_s\omega)\,
    \lambda^{(1)}_{Ljm}= k^{(0)}_{Ljm}+i\nu^{(0)}_{Ljm} +(r_s\omega)\left(k^{(1)}_{Ljm}+i\nu^{(1)}_{Ljm}\right)\,.
\end{equation}
By comparing~\eqref{eq:smallfreq1ptfinalUS} and~\eqref{eq:USfarzoneapproximatedresponse} we find there is an overall, real prefactor of
\be
    \lambda_{Ljm}^{\rm US}= \frac{ 2^{-2L+3} \pi^{n+3}a^{2 L+n+1}\, \delta_a^{2 j+n-1} \Gamma \left(\tfrac{j+L-m+n+1}{2}\right)^2 \Gamma \left(\tfrac{j+L+m+n+1}{2}\right)^2\Gamma(\hat{j}+1)^4}{\Gamma
    \left(j+\tfrac{n-1}{2}\right) \Gamma \left(j+\tfrac{n+1}{2}\right) \Gamma \left(L+\frac{n+3}{2}\right)^2 \Gamma \left(
    L+\tfrac{n+1}{2}\right)^2\Gamma(2\hat j +1)\Gamma(2\hat j +2)}\,.
\ee
In order to facilitate comparison with previous calculations, it is convenient to note that 
the surface gravity in the ultra-spinning limit is 
\begin{equation}
    r_s \kappa=\frac{n-1}{2 r_h}+\ord{\delta_a}\,.
\end{equation}
We can now match the conservative and dissipative responses by looking at the real and imaginary parts of the response.

\subsubsection{Conservative response}
We identify the conservative response by looking at the real part of~\eqref{eq:USfarzoneapproximatedresponse}. We find
\begin{tcolorbox}[colframe=white,arc=0pt,colback=greyish2]
\vspace{-10pt}
    \begin{align}
        k^{(0)}_{Ljm}&=\lambda_{Ljm}^{\rm US} \,\frac{\tan (\pi  \hat{j})}{2 \pi }\,,\\
        k^{(1)}_{Ljm} &=\lambda_{Ljm}^{\rm US}\,\frac{ \delta_a \hat{m}}{r_s^2 \kappa}\left(\pi \cot(2\pi\hat j )+\frac{ \tan (\pi 
        \hat{j}) }{\pi }\sum_{k=1}^{\infty}\frac{1}{(k+\hat{j})^2}\right)\,.
    \end{align}
\end{tcolorbox}
\noindent
The Love number, $k^{(0)}_{Ljm}$, takes schematically the same form as the Schwarzschild result where $\hat j$ determines the zeroes and divergences, rather than $\hat L$. We see that the static Love number vanishes for integer $\hat j$ and displays classical running for half-integer $\hat j$. Similarly, the  frequency-dependent dynamical Love number diverges for integer and half integer $\hat j$, indicative of a classical running. This is an interesting case where both the static and dynamical Love numbers can run at the same angular momentum values.

An interesting curiosity is that the
conservative and dissipative sectors, up to an overall $\hat{j}$ and $\hat{L}$ dependent constant, agree with the small $U$ limit of the infinite dimension calculation discussed in Section~\ref{sec:InfDEFT}. Physically this is perhaps related to the fact that
the surface gravity becomes large in the ultra-spinning limit, driving $U=\frac{\hat{m} \Omega_H}{2\kappa}$ to $0$.

\subsubsection{Dissipative response}
We can identify the dissipative response by looking at the imaginary part of~\eqref{eq:USfarzoneapproximatedresponse}. We find
\begin{tcolorbox}[colframe=white,arc=0pt,colback=greyish2]
\vspace{-10pt}
    \begin{align}
        \nu^{(0)}_{Ljm}&=-\lambda_{Ljm}^{\rm US}\,\delta_a \hat{m}=-\lambda_{Ljm}^{\rm US}\,\frac{1}{2 r_s\kappa}\frac{m}{a}\,,\\
        \nu^{(1)}_{Ljm} &=\lambda_{Ljm}^{\rm US}\, \frac{1}{2 r_s \kappa}  \,,
    \end{align}
\end{tcolorbox}
\noindent
The dissipative response coefficients take a remarkably simple form, and notably do not diverge or vanish for any values of the angular momenta. Another interesting feature is that the conservative and dissipative responses behave somewhat like a black hole in two fewer dimensions because they depend on the angular momentum $\hat j$ rather than $\hat L$. Physically this is a consequence of the flattened geometry in the ultra-spinning limit.

\newpage
\section{Conclusions}
\label{sec:Concl}

In this paper we have studied the (scalar) tidal responses of higher-dimensional black holes. Given the large diversity of distinct solutions in generic dimension, in order to study their properties in a uniform way we have cast the discussion in the language of point-particle effective field theory.  This has the benefit of being an unambiguous characterization of responses, and allows us to encode the physical properties of objects in the coefficients of non-minimal worldline couplings to external probe fields. By allowing these couplings to involve additional degrees of freedom localized on the worldine, it is possible to capture both dissipative and conservative physical features.

We have applied the EFT formalism to a number of examples. In order to make contact with previous studies, we have considered the responses of $D$-dimensional Schwarzschild and four-dimensional Kerr black holes to external scalar profiles, reproducing their static and leading-order dynamical response (along with dissipative responses). We then turned to studying higher-dimensional black holes with spin. As a first example, we have studied the Myers--Perry black hole in five dimensions. In addition to reproducing the static conservative and dissipative responses, we have also computed their dynamical Love numbers. This is made possible by making an appropriate near zone approximation that captures the relevant physics at $\ord \omega$. In order to validate the procedure of matching the (off-shell) one-point function at this order, we have---in the case of the Kerr black hole---explicitly checked that the responses inferred in this way agree with those obtained by matching the scattering cross section of the black hole (see Appendix~\ref{app:scattering}).

In addition to reconsidering known examples, we consider the scalar responses of spinning black hole solutions in the $D\to \infty$ limit, and in the ultra-spinning regime for $D \geq 6$. In each of these cases, the equations relevant for computing scalar responses reduce to a hypergeometric equation, which allows us to analytically extract the tidal responses of these objects. We then match the responses computed by solving the wave equation in the general relativity black hole background to responses computed in point particle EFT. Since these black hole solutions have enhanced symmetry compared to a generic spinning object, there are interesting features of the EFT description, where the symmetries of the ultraviolet object manifest as symmetries of the EFT and relate different Wilson coefficients.

The scalar tidal responses that we have computed have a number of interesting features, and relations between different solutions. First, we find that for generic spin parameters, the scalar tidal Love numbers of higher-dimensional black holes do not vanish. 
This indicates that higher-dimensional black holes with general spin parameters are somewhat less special than four-dimensional black holes as objects.
However, in the special case where the spin parameters are tuned to be equal in $D=5$, Love numbers vanish when the angular momentum is two times an integer. In the ultra-spinning case, the scalar Love numbers vanish when the parameter $j$ is $D-5$ times an integer, where $j$ the angular momentum along the sphere factor of the black hole spacetime. Interestingly, the ultra-spinning black hole behaves similarly to a Schwarzschild black hole in two lower dimensions. There are some other notable features of the responses: we see that the $D=5$ Myers--Perry black hole with equal spins has similar responses to the $D=4$ Kerr black hole, both in the static and dynamical sectors. In addition we see that the large-dimension limit captures some of the features of finite-dimensional black holes.

There are a number of interesting aspects of our analysis which merit further investigation. First, there is an interesting pattern of vanishing scalar responses. In $D=4$, the vanishing of static responses can be understood in various ways from the viewpoint of symmetry~\cite{Charalambous:2021kcz,Hui:2021vcv}. In the Schwarzschild case, these structures extend to higher-dimensional black holes~\cite{Hui:2021vcv,Charalambous:2024tdj}, and some features of the symmetry structure of Myers--Perry black holes have been studied in $D=5$~\cite{Charalambous:2023jgq}. It would be interesting to more fully study the symmetries of higher-dimensional black hole solutions and their impact on tidal responses. In addition to understanding the vanishing responses at the microscopic level by studying the wave equation in black hole spacetimes, an important aspect to understand is the action of these symmetries at the level of the EFT description. In contrast to the four-dimensional case, where the symmetries of the microphysics set to zero all static tidal operators in the EFT~\cite{Hui:2021vcv}, some of the responses in higher dimensions vanish, but not all, and so the relevant EFT symmetries must force only some Wilson coefficients to vanish. It would be very interesting to understand how this works. Beyond understanding the symmetry structure, it would be interesting to study the ultra-spinning black hole in the regime where the ratio $r/a \to 0$ in the high spin limit. In this limit, the black hole flattens in the spin plane, and so looks like a membrane as viewed from infinity. This limit is not naturally matched to a point particle, but should instead be matched to the EFT of a dynamical brane, and it would be interesting to see this explicitly.

The constructions that we have studied also suggest various directions for further study. 
Perhaps the most obvious is to consider additional higher-dimensional black hole spacetimes, we have by no means exhausted the full space of known solutions. Particularly interesting would be to study black hole solutions that are supersymmetric. Static responses of some of these solutions have been studied~\cite{Cvetic:2021vxa,Cvetic:2024dvn}, and it is natural to expect that their EFT description would be particularly rich. Beyond this, it would be interesting to study the properties of black strings and branes in EFT.
In addition, it is important to generalize the study to nonlinear responses (both conservative and dissipative) of higher-dimensional black holes both in EFT and at the microscopic level. Another feature that is intriguing is that the equation of motion for a scalar field is exactly solvable as $D\to \infty$. This suggests that it should be possible to match the {\it exact} worldline two-point function in this case, to all orders in frequency. It would be interesting to more fully study the properties of this EFT, in particular it could serve as a test case for the study of the self-force problem~\cite{Cheung:2023lnj,Kosmopoulos:2023bwc}.
Finally,  none of the higher-dimensional black holes that we have studied are expected to be exactly stable (at least with trivial topology), and most likely decay through some instability similar to the Gregory--Laflamme~\cite{Gregory:1994bj} instability. It would be extremely interesting to understand the signatures of these instabilities in EFT. Some aspects of black hole stability have been considered in EFT before~\cite{Chu:2006ce,Kol:2007rx}, but we expect that the rich parameter space of higher-dimensional black hole solutions will provide additional insights.

\vspace{-12pt}
\paragraph{Acknowledgements:} Thanks to Lam Hui, Mikhail Ivanov, Leah Jenks, Donal O'Connell, Julio Parra-Martinez, Riccardo Penco, and Zihan Zhou for helpful discussions. AJ and DG are supported in part by DOE (HEP) Award DE-SC0025323. 
LS was supported by the Programme National GRAM of CNRS/INSU with INP and IN2P3 co-funded by CNES.
The work of MJR is supported through the NSF grant PHY-2309270. MJR gratefully thanks the Mitchell Family Foundation and the Centro de Ciencias de Benasque Pedro Pascual for their warm hospitality. LFT acknowledges support from USU PDRF fellowship and USU Howard L. Blood fellowship.

\newpage
\appendix

\section{Schwinger--Keldysh}
\label{app:SK}

An important feature of black holes is that things can fall in. As such, black holes need to be described as dissipative systems, and one of our goals is to match these dissipative features to a point particle effective description. In dissipative systems---or more generally out of equilibrium systems---we are unable to specify both the initial and final states of a system in order to compute quantities like transition amplitudes.
Instead we can only specify the initial state and then compute correlation functions of observables at some later time. To describe such a system we employ the so-called {\it Schwinger--Keldysh} formalism. Here we provide a brief overview of the relevant formalism, more comprehensive discussions can be found in~\cite{kamenev2011field,Akyuz:2023lsm,Haehl:2015foa,Crossley:2015evo,Liu:2018kfw,Haehl:2024pqu,Glorioso:2016gsa,Weinberg:2005vy}.

The Schwinger--Keldysh formalism~\cite{Schwinger:1960qe,kadanoff1962quantum,Keldysh:1964ud,Feynman:1963fq,Caldeira:1982iu,Calzetta:1986cq} can be used to describe closed quantum systems in nontrivial states (for example in thermal states), or with non-adiabatic time evolution, as well as open quantum systems, which exhibit dissipation into some external reservoir. The unifying feature shared by all of these situations is that the dynamics is such that one can only specify the initial state (or density matrix), and then compute correlation functions in this state at later times:
\be
\langle {\cal O}(t)\rangle = \tr\Big(\rho(t){\cal O}(t)\Big) = \tr\Big({\cal U}^\dagger (t,t_i){\cal O}(t){\cal U} (t,t_i)\rho(t_i)\Big)\,,
\label{eq:SK1}
\ee
where $\rho(t_i)$ is the initial density matrix of the system, the trace is taken over the full Hilbert space, and we are implicitly working in interaction picture so that ${\cal U}$ is the time evolution operator
\be
{\cal U}(t_2,t_1) = U^\dagger_0(t_2,t_1) U(t_2,t_1)\,,
\ee
that effectively evolves with the interaction hamiltonian. Here $U(t_2,t_1)$ is the ordinary time evolution operator
\be
U(t_1,t_2) = T \E^{-i\int_{t_1}^{t_2}\rd t H}\,,
\ee
that evolves with the full hamiltonian, and $U_0$ is the analogue with the free hamiltonian. The important feature of~\eqref{eq:SK1} is that it involves evolving the density matrix {\it forward} in time, to time $t$, where the operator is inserted, and then evolving {\it backward} in time back to the initial time $t_i$. For this reason, the formalism is also often called the in-in formalism. 

\begin{figure}[t!]
\centering
\includegraphics[scale=2.5]{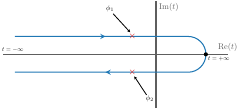}
\caption{\footnotesize Illustration of the closed-time contour used in the Schwinger--Keldysh formalism. The system is evolved from the infinite past ($t=-\infty$) to the future ($t=\infty$, here denoted by a finite point), and back along a closed contour in the complex time plane. Fields  inserted along the forward part of the contour are labeled with a $1$ subscript, e.g., $\phi_1$, while fields inserted along the reverse contour are labelled with a $2$ subscript: e.g., $\phi_2$. The Schwinger--Keldysh path integral~\eqref{eq:SKdoubledPI} generates correlators path-ordered along this contour.}
\label{fig:inincontour}
\end{figure}

It is often convenient to think of the time evolution involved in~\eqref{eq:SK1} as proceeding along a contour in the complex time plane that runs from $-\infty$ to $+\infty$ and back.\footnote{In equilibrium, the reverse part of the contour just generates a phase, and so can be removed by dividing by the sum over disconnected diagrams, which is what we do when we consider transition amplitudes in in-out quantum field theory.
In the generic situation, however, one must account for the full forward and backward evolution.} This contour is depicted in Figure~\ref{fig:inincontour}. Since it is a closed loop, it is often called a closed time path.
Then, we can consider operator insertions along either of the two branches of the contour (forward and backward). As such, it is quite natural to consider correlation functions of the following type
\be
\langle \bar T\,\big(
O_{n+m}(t_{n+m})\cdots O_{n+1}(t_{n+1})
\big) \,T\,\big(
O_n(t_{n})\cdots O_{1}(t_{1})
\big)
\rangle\,,
\ee
which consists of time-ordered operators along the forward part of the contour, and anti-time-ordered operators along the backward part. In order to construct a generating function for correlators of this type, we consider a doubled path integral, which introduces a copy of the fundamental fields on each branch of the in-in contour
\be
Z_{\rm SK}[J_1,J_2] = \int{\cal D}\phi_1{\cal D}\phi_2 \,\E^{iS[\phi_1]-iS[\phi_2]+\int\rd^Dx\left( J_1\phi_1+J_2\phi_2\right)}\,,
\label{eq:SKdoubledPI}
\ee
where, for simplicity, we have restricted ourselves to introducing sources for the fundamental fields in the path integral.

The boundary conditions for the path integral~\eqref{eq:SKdoubledPI} have been left implicit, but they are rather important. In order to specify the initial state, we require $\phi_1$ and $\phi_2$ to have some particular profile at $t_i = -\infty$, which captures the features of the initial density matrix. We then evolve these field configurations to the time $t = +\infty$ (note that for the field $\phi_2$ this is formally {\it backward} time evolution, which leads to the relative minus signs). At $t=+\infty$, we require $\phi_1(\vec x) = \phi_2(\vec x)$ because these are really the same field and have to be glued together consistently on this time slice. These combined boundary conditions lead to an effective coupling between the $\phi_1$ and $\phi_2$ fields, as can be seen, for example, by discretizing the path integral~\cite{kamenev2011field}. This coupling of the fields is reflected in the nonzero two-point function between fields on the different branches:
\be
\langle T \phi_1(x_1)\phi_2(x_2)\rangle  = \langle \phi(x_1) \phi(x_2)\rangle\,.
\ee
Since all of the fields inserted on the backward part of the contour (the ``2" fields) have later time arguments than those inserted on the forward part (the ``1" fields), the time-ordered two-point function between these fields is just the Wightman two-point function of the original theory. In general there are four different Green's functions we can define by correlating two fields, inserted on the two branches of the contour
\begin{align}
\label{eq:SKG1}
iG_F(x_1,x_2)& = \langle T \phi_1(x_1) \phi_1(x_2)\rangle\\
iG_{\bar F}(x_1,x_2) &= \langle \bar T \phi_2(x_1) \phi_2(x_2)\rangle\\
i G_<(x_1,x_2) &= \langle \phi_2(x_2)\phi_1(x_1)\rangle\\
i G_>(x_1,x_2) &= \langle \phi_2(x_1)\phi_1(x_2)\rangle\,.
\label{eq:SKG2}
\end{align}
These Green's functions capture the different possible time orderings between operators on the two branches. Since all the fields on branch $2$ of the contour are later than those on branch $1$, their two-point function is just the Wightman function (where $<$ denotes $t_1$ is earlier than $t_2$ and $>$ denotes the opposite). For fields inserted on the same branch, we have to account for their relative time-ordering, so the corresponding Green's function is either the Feynman or anti-Feynman propagator.

The correlations between the fields $\phi_1,\phi_2$ on the two parts of the contour are notably absent in~\eqref{eq:SKdoubledPI}. In order to make this feature---along with the dependence on the initial state---manifest, it is often convenient to instead work with an {\it effective} action of the form~\cite{Akyuz:2023lsm}
\be
Z_{\rm SK,eff}[J_1,J_2] = \int{\cal D}\phi_1{\cal D}\phi_2 \,\E^{iS[\phi_1]-iS[\phi_2]+iS_{\rm int}[\phi_1,\phi_2]+\int\rd^Dx\left( J_1\phi_1+J_2\phi_2\right)}\,,
\label{eq:effSK}
\ee
which explicitly couples the two fields. As a matter of practice this is what we imagine implicitly doing in Section~\ref{sec:1ptfunction}. At the practical level this is the same as keeping track of the correlations between $\phi_1$ and $\phi_2$.\footnote{There are some subtleties associated to the use of the effective action~\eqref{eq:effSK}. Most notably the fields appearing in~\eqref{eq:effSK} and those in~\eqref{eq:SKdoubledPI} are not necessarily the same. See e.g.,~\cite{Akyuz:2023lsm} for discussion of this and other subtleties in Schwinger--Keldysh effective field theory.}

The Green's functions~\eqref{eq:SKG1}--\eqref{eq:SKG2} are not all independent, they satisfy the relation\footnote{In the context of in-out quantum field theory this identity is familiar, and underlies perturbative Cutkosky rules. This identity, and generalizations to higher point correlation functions, can be derived straightforwardly from the so-called largest-time equation~\cite{Veltman:1994wz}.
}
\be
G_F(x_1,x_2)+G_{\bar F}(x_1,x_2) = G_<(x_1,x_2) +G_>(x_1,x_2)\,.
\label{eq:SKlargesttime}
\ee
It is convenient to rotate our field basis to trivialize this linear relationship by defining the Keldysh basis fields
\be
\phi_- \equiv \phi_1-\phi_2\,,  \hspace{2cm}\phi_+ \equiv \tfrac{1}{2}\left(\phi_1+\phi_2\right)\,.
\ee
With this definition, the contour-ordered Green's functions are
\begin{align}
\label{eq:skprop1}
\langle\phi_+(x_1)\phi_+(x_2)\rangle &= iG_K(x_1,x_2) = \frac{i}{2}\big(
G_<(x_1,x_2)+G_>(x_1,x_2)
\big) \\
\langle\phi_+(x_1)\phi_-(x_2)\rangle &=  iG_R(x_1,x_2)  = i\theta(t_1-t_2)\big(
G_>(x_1,x_2)-G_<(x_1,x_2)
\big)\\
\langle\phi_-(x_1)\phi_+(x_2)\rangle &=  iG_A(x_1,x_2)  = i\theta(t_2-t_1)\big(
G_<(x_1,x_2)-G_>(x_1,x_2)
\big)\\
\langle\phi_-(x_1)\phi_-(x_2)\rangle &= 0\,,
\label{eq:skprop4}
\end{align}
where we have implicitly defined the {\it Keldysh propagator} $G_K$, which is the contour-ordered two-point function of the average field $\phi_+$. It is sometimes useful to group the various propagators into a matrix
\begin{equation}
    iG^{IJ}=
    \begin{pmatrix}
        0 & G_{\text{adv}}\\
        G_R & G_K
    \end{pmatrix}\,,
    \label{eq:GABdef}
\end{equation}
where the $I,J$ indices run over $+,-$ and indices are contracted with the off-diagonal matrix 
\begin{equation}
    c^{AB}=c_{AB}=
    \begin{pmatrix}
        0 & 1 \\
        1 & 0
    \end{pmatrix}.
\end{equation}
In this basis the identity~\eqref{eq:SKlargesttime} appears as the fact that the two-point function of the difference field $\phi_-$ vanishes.\footnote{More generally, any correlation function where $\phi_-$ is the latest insertion will vanish~\cite{kamenev2011field,Haehl:2024pqu}.} 
Aside from making this property manifest, the Keldysh basis is useful because the average field has the interpretation as the {\it classical} field configuration of the system, while the difference is related to {\it noise} (sometimes called the quantum part of the system).\footnote{This can be seen, for example by taking the $\hbar\to 0$ limit, and identifying $\phi_+$ with the corresponding classical field. See, e.g., {\tt Ch.4} of~\cite{kamenev2011field}.} This is essentially why the Schwinger--Keldysh formalism is useful for describing dissipative or thermal physics, the classical degrees of freedom of the system $\phi_+$ must necessarily couple to the $\phi_-$ degrees of freedom, which can carry away energy and other conserved charges, leading to effective dissipation in the $\phi_+$ sector. We also see that the correlations between $\phi_+$ and $\phi_-$ capture the causal dynamics of the system, while the correlations of $\phi_+$ with itself contains information about the initial state~\cite{Akyuz:2023lsm}. It is convenient to introduce some diagrammatic notation that accounts for the off-diagonal nature of the correlations between $\phi_+$ and $\phi_-$~\cite{kamenev2011field}. We can denote $\phi_+$ by a solid line and $\phi_-$ by a dashed line. Then~\eqref{eq:skprop1}--\eqref{eq:skprop4} can be written as
\begin{align}
\label{eq:prop1}
\raisebox{-6.5pt}{\begin{tikzpicture}[line width=1. pt, scale=2]
\draw[fill=black,dashed,-stealth] (0,0) -- (.6,0);
\draw[fill=black] (.6,0) -- (1.2,0);
\node[scale=1] at (-.4,0) {$\phi_-(x_1)$};
\node[scale=1] at (1.6,0) {$\phi_+(x_2)$};
\end{tikzpicture}}
&= G_R(x_1,x_2)\\
\raisebox{-6.5pt}{\begin{tikzpicture}[line width=1. pt, scale=2]
\draw[fill=black] (0,0) -- (.6,0);
\draw[fill=black,dashed,-stealth] (1.2,0) -- (.6,0);
\node[scale=1] at (-.4,0) {$\phi_+(x_1)$};
\node[scale=1] at (1.6,0) {$\phi_-(x_2)$};
\end{tikzpicture}}
&= G_A(x_1,x_2)\\
\raisebox{-6.5pt}{\begin{tikzpicture}[line width=1. pt, scale=2]
\draw[fill=black] (0,0) -- (1.2,0);
\node[scale=1] at (-.4,0) {$\phi_+(x_1)$};
\node[scale=1] at (1.6,0) {$\phi_+(x_2)$};
\end{tikzpicture}}
&= G_K(x_1,x_2)\,,
\label{eq:prop2}
\end{align}
where time flows in the direction of the arrow (which points toward later times). This diagrammatic formalism makes it straightforward to, for example, compute the classical field profile in the presence of a source, which is the problem of interest in~\eqref{sec:1ptfunction}.

Though we have focused so far on propagators, we can introduce interactions straightforwardly. In a generic system, there will be all possible interactions not only in the $\phi_-$ and $\phi_+$ sectors separately, but also all possible interactions between the two sectors. In effective field theory, the interactions are constrained in an interesting and nontrivial way~\cite{kamenev2011field,Akyuz:2023lsm,Haehl:2015foa,Crossley:2015evo,Liu:2018kfw,Haehl:2024pqu,Glorioso:2016gsa}. In particular, the effective action $S_{\rm eff}[\phi_+,\phi_-]$ that describes the dynamics must satisfy
\begin{align}
\label{eq:SKreqs1}
S_{\rm eff}[\phi_+,\phi_- = 0] &= 0\\
\label{eq:SKreqs2}
S_{\rm eff}[\phi_+,\phi_- ]  +S_{\rm eff}^*[\phi_+,-\phi_-]  &= 0\\
{\rm Im}\, S_{\rm eff}[\phi_+,\phi_- ] &\geq 0\,,
\label{eq:SKreqs3}
\end{align}
as a consequence of unitarity of time evolution. Notably, the effective action is allowed to be complex, as long as~\eqref{eq:SKreqs2} is satisfied. In particular states there can be additional constraints, for example the effective action that describes a thermal state must be consistent with the Kubo--Martin--Schwinger relations~\cite{Kubo:1957mj,Martin:1959jp}.

\vspace{-12pt}
\paragraph{Dissipation:} Our motivation for working with the Schwinger--Keldysh formalism is to model the dissipative effects of a black hole in point particle EFT. There are two (related) ways that one could go about this. The first is to envision the Schwinger--Keldysh effective action as arising from starting with the Schwinger--Keldysh description of a larger (pure) system and tracing out some of the degrees of freedom. This would lead to some pattern of interactions between the low-energy fields $\phi_\pm$, that would necessarily be consistent with the constraints~\eqref{eq:SKreqs1}--\eqref{eq:SKreqs3}. We could then parameterize all possible such interactions and then match them to a general relativity calculation.

In practice, a slightly different approach is somewhat simpler to implement. It is convenient to follow~\cite{Goldberger:2005cd} and introduce in the EFT effective degrees of freedom, $X$, that capture the dissipative effects arising from tracing out some degrees of freedom in the full description. (We can think of this as purifying the EFT by introducing some additional environmental degrees of freedom.) We can then consider the combined system consisting of both $\phi$ and $X$:
\be
Z[J_\pm^\phi, J^X_\pm] = \int {\cal D}\phi_+{\cal D}\phi_-{\cal D} X_+{\cal D} X_- \E^{i S_{\rm eff}[\phi_\pm, X_\pm]+\int\rd^Dx\left(J_\pm^\phi \phi_\pm+J_\pm^X X_\pm\right)}\,,
\label{eq:pathintSKdissipation}
\ee
where we have suppressed the sum over the $+$ and $-$ fields in the sources. One of the advantages of proceeding in this way is that we can construct the effective action directly from the ordinary lagrangian for the combined $\phi,X$ system:
\be
iS_{\rm eff}[\phi_{1,2}, X_{1,2}] = iS_{\rm eff}[\phi_1, X_1] - iS_{\rm eff}[\phi_1, X_1] \,,
\ee
where $iS_{\rm eff}[\phi_1, X_1]$ contains all possible couplings between the $\phi$ and $X$ sectors that are consistent with the symmetries of the problem. We can then change variables to the Keldysh basis to obtain~\eqref{eq:pathintSKdissipation}, and parameterize the dynamics of the $X$ degrees of freedom and then integrate them out to obtain an effective action for the $\phi$ fields alone:
\be
\E^{i\Gamma_{\rm eff}[\phi_\pm]} =  \int{\cal D} X_+{\cal D} X_- \E^{i S_{\rm eff}[\phi_\pm, X_\pm]}\,.
\label{eq:appSKeff}
\ee
The advantage of proceeding in this way is that the EFT input we require is a parameterization of the interactions between the $\phi$ field and the auxiliary degrees of freedom $X$. We can construct this action using the usual rules of EFT and then upon integrating out the $X$ variables the $\phi_\pm$ action is guaranteed to be consistent with~\eqref{eq:SKreqs1}--\eqref{eq:SKreqs3}. 

In practice in Section~\ref{sec:1ptfunction} we are interested in a limited part of the effective action~\eqref{eq:appSKeff} that arises from couplings of the schematic form
\be
S[\phi, X] \supset \int\rd\tau \Big( \phi_+ \,X_-+ \phi_- \,X_+\Big)\,,
\label{eq:schematic}
\ee 
which we then use to integrate out $X$ in the presence of a classical source for $\phi$. The resulting contribution to the effective action in the form of the one-point function for $\phi_+$ has the following diagrammatic expression, in the notation of~\eqref{eq:prop1}--\eqref{eq:prop2} 
\begin{equation*}
\includegraphics[scale=.95]{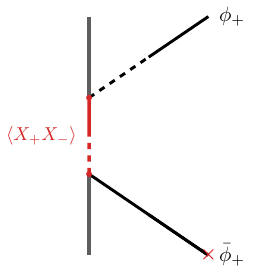}
\end{equation*}
which allows us to compute $\phi_+$ in terms of the Greens function for the $X$ degrees of freedom. Note that the propagators appearing in this expression will be the causal propagators, which is appropriate for computing the classical field profile. In the black hole case of interest, the couplings are slightly more complicated than~\eqref{eq:schematic}, but the physics is the same.

\newpage
\section{Thorne Tensors}\label{appendix:Thorne}
In several places in the main text, it is necessary to decompose symmetric traceless tensors in a standard basis. Such a basis of tensors was constructed by Thorne in~\cite{Thorne:1980ru} (and by Pirani in~\cite{trautman1965lectures}).  This basis was also used in~\cite{Charalambous:2021mea}, and we will follow them by referring to the tensors as {\it Thorne tensors}. We require the generalization to general dimension, which we describe in the following. (See also~\cite{Loganayagam:2023pfb} for some useful formulas.)

\subsection{Spherical Harmonics}

Fundamentally, the reason that symmetric traceless tensors appear recurrently is that they are closely related to spherical harmonics. (Spherical harmonics themselves appear because of the SO$(D-1)$ rotational symmetries present far from a localized object.) He we briefly review some relevant features of spherical harmonics. (See, e.g.,~\cite{Chodos:1983zi,Hui:2020xxx,Loganayagam:2023pfb} for more details.)

Recall that harmonic polynomials  in $n+1$ dimensions of the form (where $c^{(M)}_{i_1\cdots a_L}$ is both traceless and symmetric)
\be
P_{LM}(x) = c^{(M)}_{i_1\cdots a_L} x^{i_1}\cdots x^{i_L}\,,
\label{eq:harmonicpol}
\ee
when written in spherical coordinates take the form
\be
P_{LM}(r,\theta) = r^LY_{LM}(\theta)\,.
\ee
The fact that the cartesian laplacian in $n+1$ dimensions annihilates~\eqref{eq:harmonicpol} then implies that $Y_{LM}$ satisfies the eigenvalue equation
\be
\Delta_{S^n}Y_{LM}(\theta) = -L(L+n-1)Y_{LM}(\theta)\,.
\ee
and hence $Y_{LM}(\theta)$ are harmonics on the $n$-sphere. Here the multi-index $M$ catalogues the magnetic quantum numbers that label a given state in the spin-$L$ representation of SO$(n+1)$, which is carried by $Y_{LM}$. The dimension of this representation is 
\be
N_L = \frac{(L+n-2)! (2L+n+1)!}{(n-1)!L!}\,,
\ee
and correspondingly there $N_L$ different assignments of magnetic quantum numbers. There are various schemes to enumerate the possible values that 
the magnetic quantum numbers can take, which depend on the coordinate system in which we want to express the spherical harmonic. The most common choice is $n$-dimensional hyperspherical coordinates, which takes advantage of the recursive parameterization of the sphere line element
\be
\rd \Omega_{S^n}^2 = \rd\theta_n^2+\sin^2\theta_n\rd\Omega_{S^{n-1}}^2\,.
\ee
This sequence of embedded spheres indicates that an angular momentum $L$ spherical harmonic can be labeled uniquely by its eigenvalue with respect to the laplacian of each sub-sphere. In this parameterization, there are $n-1$ magnetic quantum numbers, ordered as
\be
\lvert m_1\rvert \leq m_2 \leq \cdots \leq m_{n-1} \leq L\,.
\ee

A different coordinate system that is more convenient for many computations is the $n$-dimensional generalization of the Hopf parameterization of the $3$-sphere.\footnote{These coordinates are sometimes called polyspherical coordinates, and are constructed recursively by splitting coordinates into two groups of $n/2$ coordinates and writing one group as $n^{1}_i=\cos(\theta)y_i$ and the other $n^{2}_i=\sin(\theta)y_i$, where $y_i$ are polyspherical coordinates of $S_{\frac{n}{2}}$. In odd dmensions, one eventually uses the Hopf parameterization of the $3$-sphere~\eqref{eq:n1}.}
In this parameterization, there are $r=\lfloor \frac{n+1}{2} \rfloor$ magnetic quantum numbers (the rank of the rotation group SO$(n+1)$ that acts on the $n$-sphere). These magnetic numbers can take any integer values subject to the constraint
\be
\lvert m_1\rvert +\cdots+ \lvert m_r\rvert  \leq  L\,.
\ee
The benefit of this parameterization is that in $n+1$ dimensions we can define $r$ orthogonal planes which are the analogues of axes of rotation in three spatial dimensions. The magnetic quantum numbers have the interpretation as the angular momenta in each of these planes.\footnote{More formally, so$(n+1)$ has $r$ elements of its Cartan subalgebra---those generators that can be simultaneously diagonalized. The $m_A, A\in \{1,\cdots, r\}$ are the eigenvalues of these elements of the Cartan.}

Regardless of parameterization, spherical harmonics are orthogonal in the sense
\be
\int \rd\Omega_{S^n} Y_{LM}(\theta)Y^*_{L'M'}(\theta) = \delta_{LL'}\delta_{MM'}\,,
\label{eq:sphericalharmonicortho}
\ee
where the integral is performed over the entire sphere.

We now turn to the construction of Thorne tensors, which are closely related to these spherical harmonics.

\subsection{Harmonic Tensors}

Thorne tensors are the symmetric traceless tensors ${\cal Y}_{LM}^{i_1\cdots i_L}$ that when contracted with $n+1$ dimensional unit vectors $\hat x^i \equiv x^i/\lvert \vec x\rvert$ generate spherical harmonics:
\be
{\cal Y}_{LM}^{i_1\cdots i_L} \hat x_{i_1}\cdots \hat x_{i_L} = Y_{LM}(\theta)\,.
\label{eq:thornetensordef}
\ee
That is, they are the 
concrete tensors $c^{(M)}_{i_1\cdots a_L} $ in~\eqref{eq:harmonicpol} that generate spherical harmonics with definite magnetic quantum numbers.

The Thorne tensors provide a basis for symmetric traceless tensors, meaning that it is possible to decompose any such $L$-index tensor in terms of them as
\be
T^{i_1\cdots i_L} = \sum_{M} T_{LM}\,{\cal Y}_{LM}^{i_1\cdots i_L}\,,
\ee
where the $T_{LM}$ are the coefficients of this decomposition. We can invert this relation with the help of the identity
\be
    \label{eq:Integralofsympolynomials}
    \int\rd\Omega_{S^n} \,(A_{i_1\cdots i_L}\hat x^{i_1}\cdots \hat x^{i_L})(B_{i_{L+1}\cdots i_{2L}}\hat x^{i_{L+1}}\cdots \hat x^{i_{2L}} )= A_{i_1\cdots i_L}B^{i_{1}\cdots i_{L}} A(S^n)\frac{(n-1)!!\,L!}{(n+2L-1)!!}\,,
\ee
where $A(S^n)=2 \pi^{\tfrac{n+1}{2}}/\Gamma\left(\tfrac{n+1}{2}\right)$ is the area of the $n$-sphere. Using this formula, along with~\eqref{eq:sphericalharmonicortho}, we can write
\begin{align}
    T^{LM}&= \int \rd\Omega_{S^n} \sum_{M'} T^{LM'} Y_{LM'} Y^{*}_{LM} \nonumber \\
    &= \int \rd\Omega_{S^n} T_{i_1 \cdots i_L} \hat x^{i_1}\cdots \hat x^{i_L} \mathcal{Y}^*_{LM}{}_{i_{L+1} \cdots i_{2L}} \hat x^{i_{L+1}}\cdots \hat x^{i_{2L}} \nonumber \\
    &=A(S^n)\frac{(n-1)!!\,L!}{(n+2L-1)!!} \,T_{i_1 \cdots i_L} \mathcal{Y}^*_{LM}{}^{i_1 \cdots i_L}\,.
    \label{eq:ThorneTensorcomponentNormgeneralD}
\end{align}
We can similarly derive a formula for the contraction of Thorne tensors starting from the orthogonality relation~\eqref{eq:sphericalharmonicortho}. By writing each of the spherical harmonics in this expression in terms of their corresponding Thorne tensor~\eqref{eq:thornetensordef} and using the integral formula~\eqref{eq:Integralofsympolynomials}
we find
\begin{equation}
    \label{eq:ThorneTensororthogonality}
    \mathcal{Y}_{LM}{}_{i_1\cdots i_L}\mathcal{Y}^*_{L'M'}{}^{i_1\cdots i_L}= \frac{ 2^{L-1} \pi^{-\tfrac{n+1}{2}}\Gamma(L+\tfrac{n+1}{2})}{ \Gamma (L+1)} \delta_{LL'}\delta_{MM'}\,.
\end{equation}
Finally by combining~\eqref{eq:ThorneTensororthogonality} and~\eqref{eq:sphericalharmonicortho}, we find
\begin{equation}
    \label{eq:integraloverksandarbitrarysphharmonic}
    \int d\Omega_{S^{n}}\hat{x}^{(i_1}\cdots \hat{x}^{i_L)_T} Y_{L'M'} (\hat{x})=\frac{2^{1-L'} \pi ^{\frac{n+1}{2}}  \Gamma (L'+1)}{\Gamma(L'+\tfrac{n+1}{2})} \mathcal{Y}_{L'M'}^{i_1\cdots i_{L'}} \delta_{LL'}\,.
\end{equation}
Given a function $f(x)$ expanded in spherical harmonics as
\begin{equation}
    f(x)= \sum_{L} {\cal E}_{j_1\cdots j_L} x^{j_1}\cdots x^{j_L}=\sum_{L,M} {\cal E}_{LM}\,  r^L Y_{LM}(\theta)\,,
\end{equation}
we can isolate the expansion coefficient
 ${\cal E}_{LM}$ by taking appropriate derivatives, using~\eqref{eq:ThorneTensororthogonality} and evaluating at the origin:
\begin{equation}
    \mathcal{Y}^*_{L'M'}{}^{i_1\cdots i_L}\partial_{i_1}\cdots\partial_{i_{L'}} f(x)\big\rvert_{r=0}  ={\cal E}_{LM} \,2^{L-1} \pi^{-\tfrac{n+1}{2}} \Gamma(L+\tfrac{n+1}{2})\, \delta_{LL'}\delta_{MM'}\,.
\end{equation}

For the interested reader, we present the (slightly intricate) details of the derivation of~\eqref{eq:Integralofsympolynomials} in the following inset.
\vspace{-4pt}
\begin{oframed}
{\small
\noindent
{\bf\normalsize Derivation of~\eqref{eq:Integralofsympolynomials}:} 
We start with the integral formula for integrals over $2L$ unit vectors~\cite{Thorne:1980ru}:
\begin{equation}
    \int\rd\Omega_{S^n} \,\hat x^{i_1}\cdots \hat x^{i_{2L}}= A(S^n) \frac{(2L-1)!!\,(n-1)!!}{(n+2L-1)!!}\delta^{(i_1 i_2}\cdots\delta^{i_{2L-1}i_{2L})}\,,
\end{equation}
where $A(S^n)$ is again the area of the $n$-sphere. We can write the symmetrized product of Kronecker deltas as
\begin{equation}
    \delta^{(i_1 i_2}\cdots\delta^{i_{2L-1}i_{2L})}= \frac{1}{(2L-1)!!}\sum_{j_2,j_4,\ldots j_{2L}} \delta^{i_1 i_{j_2}}\delta^{i_{j_3}i_{j_4}}\cdots \delta^{i_{j_{2L-1}}i_{j_{2L}}}\,,
\end{equation}
where the sum over the $j$ indices is slightly complicated: $j_2$ is summed from $2$ to $2L$; $j_3$ is the smallest integer not equal to $1$ or $j_2$; $j_4$ is summed over all integers $2$ to $2L$ that are not equal to $j_2$ or $j_3$; $j_5$ is the smallest integer not equal to any of the previous indices, and so on. (Note that we only have to sum over every other index, as each $\delta^{ij}$ is symmetric in its two indices.) We can see that, for a particular permutation, the choice of $j_2$ has $(2L-1)$ options, the choice of $j_4$ has $(2L-3)$ options, and so on, so the overall symmetry factor is $(2L-1)!!$. With these identities, we can write
\be
\begin{aligned}
    &\int\rd\Omega_{S^n} \,(A_{i_1\cdots i_L}\hat x^{i_1}\cdots \hat x^{i_L})(B_{i_{L+1}\cdots i_{2L}}\hat x^{i_{L+1}}\cdots \hat x^{i_{2L}} )\\
    &\hspace{2cm}=A_{i_1\cdots i_L}B_{i_{L+1}\cdots i_{2L}}A(S^n) \frac{(2L-1)!!\,(n-1)!!}{(n+2L-1)!!}\sum_{j_2,j_4,\ldots j_{2L}} \delta^{i_1 i_{j_2}}\delta^{i_{j_3}i_{j_4}}\cdots \delta^{i_{j_{2L-1}}i_{j_{2L}}}\,.
\end{aligned}
\ee
Since $A_{i_1\cdots i_L}$ and $B_{i_{L+1}\cdots i_{2L}}$ are traceless, only deltas with one index having $j_n>\,L$ and one index $j_n\leq\,L$ will survive contraction. Due to the rules of the sum, this is automatically accomplished by summing the even indices only from $L+1$ to $2L$, thus forcing the odd indices to be $\leq L$. There are then $L!$ terms that survive,
leading to~\eqref{eq:Integralofsympolynomials}.
}
\end{oframed}

It will be useful to have an explicit construction of these tensors adapted to the black hole's orientation and spin. Mathematically, the representation theory exercise that we are doing is to take a symmetric traceless tensor $c_{i_1\cdots i_L}$---which transforms in the spin-$L$ representation of SO$(n)$---and decompose it into representations of the little group that preserves an antisymmetric tensor $S_{ij}$ in the adjoint representation. Since the Lie algebra is also in the adjoint, the little group will be generated by the Cartan subalgebra. So, the branching problem we are solving is
\be
\gyoung(_4L)^{\,T}\,\xrightarrow{{\rm SO}(n)\,\rightarrow\, {\rm SO}(2)^r}
~m_1\raisebox{0ex}{\medoplus}~
\cdots
\,\raisebox{0ex}{\medoplus}~
m_r\, ,
\ee
where $m_1,\cdots, m_r$ are the magnetic quantum numbers labeling how the spin-$L$ representation transforms under the Cartan. In practice we can implement this branching by constructing a basis of intertwining vectors orthogonal to $S_{ij}$ that are charged under the various SO$(2)$ subgroups, $e_a^i$. By contracting these with $c_{i_1\cdots i_L}$ we project onto the magnetic representations of interest.

We first describe the construction for the two and three-sphere before describing the general dimensional version.

\subsubsection{Thorne Tensors on $S^2$}
\label{app:Ttensorharmonics}

We first describe Thorne tensors for ordinary two-sphere spherical harmonics. Aside from being the most physically relevant case, it also serves to illustrate many of properties of the construction.

Thorne tensors are defined in terms of a fixed vector in space, which is the axis about which we measure the angular momentum. In the black hole case, the natural vector to use is the black hole's spin, $s^i$.
It is worth mentioning that the dimension-independent data that we will have is a 2-form $S_{ij}$, which in three dimensions can be dualized to a vector using the epsilon symbol $s^i = \epsilon^{ijk}S_{jk}$, so that the two-sphere case is a bit special.

Given a vector $s^i$, we can 
construct the Thorne tensors for the spherical harmonics with angular momentum $m=-L$ to $L$ with respect to this direction.
There is one plane orthogonal to the vector $s^i$, which we denote as $P^{ij}$ (for example if we take $s^i = z^i$ then it is the $xy$ plane $P_{xy}^{ij} = x^{[i}y^{j]}$). This orthogonality can be expressed as\footnote{In terms of planes, we can phrase this as $s^i$ lying both in the $xz$ and $yz$ planes: $\epsilon_{ijk}s^i P^{jk}_{xz} =\epsilon_{ijk}s^i P^{jk}_{yz} =0$. }
\be
\epsilon_{ijk}P^{ij}s^k = s\,,
\ee
where we can always normalize the objects so that their inner product is the black hole's spin, $s$.
We can then choose a basis of vectors in the plane defined by $P_{ij}$, which we will call $e_a^i$. These can be chosen to be orthonormal 
\be
e_a^i e_b^i = \delta_{ab}\,,
\ee
but this is not required.
What is important is that they are orthogonal to the reference vector $e_a^i s_i = 0$. It is often convenient to choose a basis for the $e_a^i$ that diagonalizes the action of the rotation that keeps the plane in which the $a$ index lies invariant. In this case, we can write $e_\pm^i$, which will transform irreducibly under the SO$(2)$ of interest as $e_\pm^i\mapsto \exp(\pm i\theta)\,e_\pm^i$. Note that these vectors that carry definite helicity will no longer be orthonormal, much like circular polarizations in electromagnetism.

We can now construct Thorne tensors by building all possible combinations of $s^i$ and $e^i_a$ that have the right number of indices and are traceless. It is useful to be explicit about a few low-$L$ examples.

\vspace{-12pt}
\paragraph{$\boldsymbol{L=1}$:} The three possible tensors we can construct are
\begin{align}
m &= 0: \hspace{.5cm}s^i\,,\\
m &=\pm1:  \hspace{.18cm}e_a^i\,.
\end{align}
We can then contract these with $\hat x^i$ in order to construct spherical harmonics. 

To be totally explicit, we can enumerate a concrete basis. Write
\be
\label{eq:ThornetensorbasisS2}
\begin{aligned}
\hat x^i = 
\left(
\begin{array}{c}
\sin\theta\cos\varphi\\
\sin\theta\sin\varphi\\
\cos\theta
\end{array}
\right)
\hspace{.75cm}
s^i = 
\left(
\begin{array}{c}
0\\
0\\
1
\end{array}
\right)
\hspace{.75cm}
e_\pm^i= 
\left(
\begin{array}{c}
1\\
\pm i\\
0
\end{array}
\right)\,,
\end{aligned}
\ee
where we have normalized $s^i$ to be a unit vector.
(Note here that the $e_\pm^i$ are not actually orthonormal: $e_{\pm}\cdot e_{\pm} = 0$ and $e_{\pm}\cdot e_{\mp} =2$.)
Then we have
\begin{align}
Y_{10} &= \hat C_{10}\,s^i \hat x_i =  \frac{1}{2}\sqrt\frac{3}{\pi}\cos\theta\,,\\
Y_{11} &= \hat C_{11}\,e_+^i \hat x_i = \frac{1}{2}\sqrt\frac{3}{2\pi}\E^{i\varphi}\sin\theta\,,\\
Y_{1-1} &= \hat C_{1-1}\,e_-^i \hat x_i = -\frac{1}{2}\sqrt\frac{3}{2\pi}\E^{-i\varphi}\sin\theta\,,
\end{align}
where we have normalized with the factor
\be
\hat C_{lm} =
\begin{cases}
 (-1)^m\left(\frac{2l+1}{4\pi}\frac{(l-\lvert m\rvert)!}{(l+\lvert m\rvert)!}\right)^\frac{1}{2} \frac{(2l)!}{2^l l!(l-\lvert m\rvert)!}& m\geq 0\\
\left(\frac{2l+1}{4\pi}\frac{(l-\lvert m\rvert)!}{(l+\lvert m\rvert)!}\right)^\frac{1}{2} \frac{(2l)!}{2^l l!(l-\lvert m\rvert)!}& m<0\\
 \end{cases}\,,
\ee
which makes the spherical harmonics defined in this way orthonormal. We can therefore define the $L=1$ Thorne tensors as
\begin{align}
{\cal Y}_{10}^i &= \hat C_{10}\,s^i\,,\\
{\cal Y}_{11}^i &= \hat C_{11}\,e_+^i\,,\\
{\cal Y}_{1-1}^i &= \hat C_{1-1}\,e_-^i\,.
\end{align}

\vspace{-12pt}
\paragraph{$\boldsymbol{L=2}$:} The construction at $L=2$ is slightly more nontrivial. As before we can define
\begin{align}
m &= 0: \hspace{.5cm}{\cal Y}_{20}^{ij} = \hat C_{20}\,s^{(i}s^{j)_T}\,,\\
m &=\pm1:  \hspace{.18cm}{\cal Y}_{2\pm 1}^{ij} = \hat C_{2\pm 1} e_\pm^{(i}s^{j)_T}\,,\\
m &=\pm2:  \hspace{.18cm}{\cal Y}_{2\pm 2}^{ij} = \hat C_{2\pm 2} e_\pm^{(i}e_\pm^{j)_T}\,.
\end{align}
When we contract these with $\hat x_i\hat x_j$ we obtain the usual spherical harmonics, properly normalized. One thing worth noting is that it would seem that there are two tensors that both have little group weight zero: $s^{(i}s^{j)_T}$ and $ e_+^{(i} e_-^{j)_T}$. These two tensors are the same up to a sign, so we can use whichever one we like. It is convenient to use $s^i$ because it is necessary to construct the $m=0$ mode for odd $L$, and we will see that there is an analogue of $s^i$ for all even-dimensional spheres (odd-dimensional ambient spaces).

\vspace{-12pt}
\paragraph{General $\boldsymbol{L}$:} The construction should now be clear for general $L$. We can construct $2L+1$ Thorne tensors as
\be
{\cal Y}_{L\,\pm m}^{i_1\cdots i_L} = \hat C_{L \pm m}\,e_{\pm}^{(i_1}\cdots e_{\pm}^{i_m} s^{i_{m+1}}\cdots s^{i_L)_T}\,,
\ee
which then serve as a basis of symmetric traceless tensors. Note that the absolute value of the magnetic quantum number corresponds to the number of $e_\pm^i$ tensors that we include.\footnote{In principle it is possible to include factors of $e_+^{(i} e_-^{j)_T}$, but, as discussed previously, this is equivalent to $s^{(i} s^{j)_T}$, so without loss of generality we do not allow these factors.}
By contracting with $\hat x^i$ we obtain spherical harmonics:
\be
{\cal Y}_{Lm}^{i_1\cdots i_l}\hat x_{i_1}\cdots \hat x_{i_L} = Y_{Lm}(\theta,\varphi)\,.
\ee
Recall that spherical harmonics have the following reality property:
\be
Y^*_{Lm} = (-1)^m Y_{L\,-m}\,,
\ee
which translates into the following reality property for the Thorne tensors
\be
{\cal Y}^*{}_{L\,\pm m}^{i_1\cdots i_L} = (-1)^m {\cal Y}{}_{L\,\pm m}^{i_1\cdots i_L} \,.
\ee

It is convenient to specialize the general dimension formulas to this setting to note that we can expand a general symmetric traceless tensor in the basis of Thorne tensors as
\begin{equation}
T^{i_1\cdots i_L}=\sum_{m=-L}^{L} T^{Lm} \,\mathcal{Y}_{Lm}^{i_1\cdots i_L}\,,
\end{equation}
and that we can invert the relationship to  write\footnote{This formula can again be derived by using the integral~\cite{Thorne:1980ru}
\be
\label{eq:STFtensorcontractionintegral4D}
\int\rd\Omega_{S^2} \,(A_{i_1\cdots i_l}\hat x^{i_1}\cdots \hat x^{i_l})(B_{j_1\cdots i_p}\hat x^{i_1}\cdots \hat x^{i_p} )= 
\begin{cases}
4\pi \frac{l!}{(2l+1)!!}A_{i_1\cdots i_l}B^{i_1\cdots i_l} & p=l\\
0 & p\neq l
\end{cases}\,. \nonumber
\ee
}
\be
\label{eq:Thornetensorcomponents4D}
T^{Lm} =4\pi \frac{L!}{(2L+1)!!}{\cal Y}^*_{Lm}{}^{i_1\cdots i_L}T_{i_1\cdots i_L}\,.
\ee
It is occasionally useful to have totally explicit expressions for spherical harmonics~\cite{Thorne:1980ru}:
\begin{equation}
    Y_{Lm}(\theta,\phi)= C_{Lm}\E^{im\phi}\sin^m(\theta) \sum_{j=0}^{\lfloor \frac{L-m}{2}\rfloor} a_{Lmj} \cos^{L-m-2j}(\theta)\,,
\end{equation}
where the normalization factors are given by
\be
    C_{lm} = (-1)^m \left(\frac{2l+1}{4\pi}\frac{(l-m)!}{(l+m)!}\right)^{\frac{1}{2}}\,, \qquad\qquad    a_{lmj} = \frac{(-1)^j}{2^l j! (l-j)!} \frac{(2l-2j)!}{(l-m-2j)!}\,,
\ee
and $\theta,\phi$ are the usual angular coordinates on the two-sphere.

\vspace{-12pt}
\paragraph{Relation to spin basis:} In several places in the main text we are interested in sums of Thorne tensors of the form
\be
 \sum_{m,m'} \lambda^{ LL'mm'} \,\mathcal{Y}^{i_1\cdots i_L}_{Lm} \mathcal{Y}_{L'm'}^*{}^{j_1\cdots j_{L'}}\,,
 \label{eq:thornetensorprod}
\ee
where $ \lambda^{LL'mm'} $ parameterizes the linear combination, and is often the thing that we are trying to match between the effective description and the ultraviolet. (For example we expand the Green's function of the worldline degrees of freedom in this basis in Section~\ref{sec:1ptfunction}.)

In cases where the physics of interest preserves both angular momentum and azimuthal angular momentum, so that $L=L'$ and $m=m'$, it is often convenient to parameterize the sum
\be
 \sum_{m} \lambda^{ Lm} \,\mathcal{Y}^{i_1\cdots i_L}_{Lm} \mathcal{Y}_{Lm}^*{}^{j_1\cdots j_{L}}\,,
 \label{eq:thornetensorprod2}
\ee
in a different basis. Concretely, in~\cite{Goldberger:2020fot,Saketh:2023bul,Chia:2024bwc} responses are parameterized in terms of the basis of tensors built from $\delta_{ij},  s_i$ and $\epsilon_{ijk} s^k$ with the right index symmetries. These two bases are equivalent, but it may be useful to translate between them explicitly.

The fundamental relation needed to translate between the two bases can be understood term by term in the sum~\eqref{eq:thornetensorprod2}. Each term is of the form (restricting $m\geq0$ for simplicity, the $m<0$ case can be obtained by complex conjugation)
\be
\mathcal{Y}^{i_1\cdots i_L}_{Lm} \mathcal{Y}^*{}^{j_1\cdots j_L}_{Lm}= \lvert\hat C_{L m}\rvert^2\,e_{+}^{(i_1}\cdots e_{+}^{i_m} s^{i_{m+1}}\cdots s^{i_L)_T}\, e_{-}^{(j_1}\cdots e_{-}^{j_{m'}} s^{j_{m'+1}}\cdots s^{j_L)_T}\,.
\ee
Because both terms have the same value of $m$, this expression involves the same number of $e_+^i$ and $e_-^j$ vectors, and we can use the identity
\be
e_+^i e_-^j = \delta^{ij}- s^i  s^j - i\epsilon^{ijk} s_k\,,
\label{eq:epemidentity}
\ee
to trade the $e_\pm^i$ vectors in pairs for the basis tensors of~\cite{Goldberger:2020fot,Saketh:2023bul,Chia:2024bwc}. After this, it is just a matter of algebra to relate coefficients defined in either basis.

Let us perform this translation exercise explicitly for the $L=2$ case. Aside from illustrating the general features, this would be the case of interest for the gravitational Love numbers of an object in $D=4$. There are five terms in the sum~\eqref{eq:thornetensorprod2}
\begin{align}
\label{eq:YY1}
\mathcal{Y}^{ij}_{22} \mathcal{Y}^*_{22}\,{}_{kl} &= \lvert\hat C_{22}\rvert^2~e_{+}^{(i}e_{+}^{j)_T}\, e^{-}_{(k}e^{-}_{l)_T}\\
\mathcal{Y}^{ij}_{21} \mathcal{Y}^*_{21}\,{}_{kl} &= \lvert\hat C_{21}\rvert^2~e_{+}^{(i}s^{j)_T}\, e^{-}_{(k}s_{l)_T}\\
\mathcal{Y}^{ij}_{20} \mathcal{Y}^*_{20}\,{}_{kl} &= \lvert\hat C_{20}\rvert^2~s^{(i}s^{j)_T}\, s_{(k}s_{l)_T}\,,
\label{eq:YY2}
\end{align}
where $\mathcal{Y}^{ij}_{2-2} \mathcal{Y}^*_{2-2}\,{}_{kl}$ and $\mathcal{Y}^{ij}_{2-1} \mathcal{Y}^*_{2-1}\,{}_{kl} $ are just related by complex conjugation to the $m=2$ and $m=1$ cases. Similarly, we can construct a five-dimensional basis from the building blocks  $\delta_{ij}, s_i$ and $\epsilon_{ijk} s^k$  as~\cite{Saketh:2023bul,Chia:2024bwc}
\begin{align}
B^{(1)}{}^{ij}_{kl} &= \delta^{(i}_{(k}\delta^{j)_T}_{l)_T} & B^{(4)}{}^{ij}_{kl} &=  S^{(i}_{(k}s^{j)_T}s_{l)_T}\\
B^{(2)}{}^{ij}_{kl} &=  S^{(i}_{(k}\delta^{j)_T}_{l)_T} & B^{(5)}{}^{ij}_{kl} &=  s^{(i}s_{(k}s^{j)_T}s_{l)_T}\\
B^{(3)}{}^{ij}_{kl} &=  s^{(i}s_{(k}\delta^{j)_T}_{l)_T}\,,
\end{align}
where $S_{ij}\equiv \epsilon_{ijk}s^k$. By substituting the identity~\eqref{eq:epemidentity} into~\eqref{eq:YY1}--\eqref{eq:YY2} and decomposing the result in this basis we find
\begin{align}
\mathcal{Y}^{ij}_{22} \mathcal{Y}^*_{22}\,{}_{kl} &= \lvert\hat C_{22}\rvert^2 \left(
2B^{(1)}{}^{ij}_{kl} -2iB^{(2)}{}^{ij}_{kl} -4B^{(3)}{}^{ij}_{kl} +2iB^{(4)}{}^{ij}_{kl} +B^{(5)}{}^{ij}_{kl} 
\right)\\
\mathcal{Y}^{ij}_{21} \mathcal{Y}^*_{21}\,{}_{kl} &= \lvert\hat C_{21}\rvert^2\left(
B^{(3)}{}^{ij}_{kl} -iB^{(4)}{}^{ij}_{kl} -B^{(5)}{}^{ij}_{kl} 
\right)\\
\mathcal{Y}^{ij}_{20} \mathcal{Y}^*_{20}\,{}_{kl} &= \lvert\hat C_{20}\rvert^2\,B^{(5)}{}^{ij}_{kl} \,,
\end{align}
and where the products with $m=-2,-1$ are given by complex conjugation. We then have five different linear combinations of $B^{(n)}{}^{ij}_{kl}$ that appear in the decomposition of the five different $\mathcal{Y}^{ij}_{2m} \mathcal{Y}^*_{2m}\,{}_{kl}$, so we can invert these relationships to write the $B^{(n)}$ in terms of ${\cal Y}{\cal Y}^*$ if desired. Note that there are three real linear combinations of ${\cal Y}{\cal Y}^*$ products, and two that are purely imaginary. This corresponds to the fact that in the spin basis there are three conservative Wilson coefficients and two dissipative ones~\cite{Saketh:2023bul}.
The above translation exercise can be extended straightforwardly to higher $L$ cases as well.

\subsubsection{Thorne Tensors on $S^3$}
\label{app:s3harmonics}
Before describing the construction in general dimensions, it is useful to first consider spherical harmonics and Thorne tensors on the $3$-sphere. Not only is this case of particular interest in the context of the five-dimensional Myers--Perry black hole considered in Section~\ref{sec:MP}, but it is also the first case that displays the generic phenomenon that the angular momentum is specified by an $2$-index tensor $S^{ij}$. Unlike in the $3$-dimensional case, we cannot dualize this tensor to a vector.

We begin by specifying the two $2$-planes transverse to $S^{ij}$. (Note that this number is also the rank of the algebra so$(4)$.) We can do this by specifying all the planes in which $S^{ij}$ lies:
\be
\epsilon_{ijkl}S^{ij}P^{kl}_{(a)} = 0, \hspace{1.5cm}a=1,\cdots, 4
\ee
which are $6-2 = 4$ different planes. There are then two planes orthogonal to $S$, where we can choose normalizations so that
\be
\label{eq:spinvaluesfromspintensor}
\epsilon_{ijkl}S^{ij}P^{kl}_{(J)} = s_J\,,
\ee
which are interpretable as the angular momenta in these spin planes.

In order to construct spherical harmonics and Thorne tensors, we construct a basis for vectors in these planes. For concreteness, let us choose them to be the $xy$ and $zw$ planes (which is the choice in the Myers--Perry metric). We then have a basis for the $xy$ plane
\be
e_a^i\,,
\ee
and similarly a basis for the $zw$ plane
\be
f_{\dot a}^i\,,
\ee
where $i$ is an ambient space index (taking values in ${\mathbb R}^4$), while $a$ and $\dot a$ are indices defined in the planes. Note that the combined basis of $e, f$ vectors spans the full four-dimensional space because the $xy$ and $zw$ planes are orthogonal.
We can then build spherical harmonics by contracting these vectors with harmonic polynomials:
\be
Y_{L\, a_1\cdots a_{k_1}\,\dot a_1\cdots \dot a_{k_2}} = e_{a_1}^{i_1}\cdots e_{a_{m_1}}^{i_{k_1}}f_{\dot a_{k_1+1}}^{i_{k_1+1}}\cdots f_{\dot a_{k_1+k_2}}^{i_{k_1+k_2}}  \hat x_{(i_1}\cdots \hat x_{i_L)_T}\,.
\ee
Here the $a,\dot a$ indices carry SO$(2)$ representations, and clearly we must have $k_1+k_2 = L$ in order to contract all the indices with $L$ unit vectors $\hat x^i$.

We can organize the representations carried by the spatial indices in different ways. Two common choices are the hyperspherical representation and the Hopf representation.
For our purposes the Hopf representation is more convenient, and also more natural. 
 This parametrization is natural because the magnetic quantum numbers directly have the interpretation as the angular momentum about the spin planes.
The $a_1\cdots a_{k}$ indices are those of a symmetric traceless tensor, so we can choose the helicity basis where the helicity $\pm m$ representations decouple from each other. This corresponds to choosing the basis of circular polarizations
\be
e_\pm^i= 
\left(
\begin{array}{c}
1\\
\pm i\\
0\\
0
\end{array}
\right)
\hspace{2cm}
f_\pm^i= 
\left(
\begin{array}{c}
0\\
0\\
1\\
\pm i
\end{array}
\right)\,.
\ee
which we can then use to construct the Thorne tensors explicitly. Abstractly, they are
\be
{\cal Y}_{L\,m_1\,m_2}^{i_1\cdots i_L} = e_{+}^{(i_1}\cdots e_{+}^{i_{m_1}}f_{+}^{i_{m_1+1}}\cdots f_{+}^{i_{m_1+m_2}} \left(e_+^{i_{m_1+m_2+1}} e_-^{i_{m_1+m_2+2}} \cdots e_+^{i_{L-m_1-m_2-1}} e_-^{i_{L-m_1-m_2})_T}\right)  \,.
\ee
Here we have assumed that $m_1, m_2 \geq 0$; negative values of these magnetic quantum numbers can be obtained by substituting $+\to -$, and setting $m_a\to \lvert m_a\rvert$. Note that by construction we have $\lvert m_1\rvert+\lvert m_2\rvert \leq L$. The purpose of the part of the tensor in parenthesis is just to make up the difference between the sum of $m$s and the angular momentum $L$. (The combination $e_+^i e_-^j$ carries no helicity weight.) We could equally well replace $e_a^i$ with $f_a^i$ in this factor without changing anything.

Recall that there are 
\be
N_l = \frac{(L+1)!(L+1)}{L!}\,,
\ee
independent harmonics, and so that many Thorne tensors. For example at $L=1$, there are four harmonics: $(\pm 1,0), (0 ,\pm1)$, while for $L=2$ there are 9: $(\pm2,0), (0,\pm 2), (\pm1,\pm1),(\mp1, \pm1),(0,0)$.

\vspace{-12pt}
\paragraph{Explicit spherical harmonics:} It is sometimes useful to have explicit formulas for spherical harmonics. Here we explicitly construct harmonics in Hopf coordinates. (For expressions in hyperspherical coordinates, see~\cite{Hui:2020xxx}.) We will utilize Hopf coordinates, whose relation to a Cartesian unit vector $\hat{n}$ is as follows\footnote{Hopf coordinates have an interpretation as an isomorphism between $\mathbb{C}^2 \rightarrow \mathbb{R}^4$. If we define $z_i=x_i+ i y_i$, then this mapping is given explicitly by $(z_1,z_2) \mapsto (x_1,y_1,x_2,y_2)$. Restricting ourselves to $S^3$ implies the condition $|z_1|^2+|z_2|^2=1$, which is achieved by setting $|z_1|^2=\sin{\theta}^2,|z_2|^2=\cos^2{\theta}$. From this perspective, the coordinates $\phi$ and $\psi$ represent the phases of the two complex numbers.}
\begin{align}
\label{eq:n1}
    n_1&= \sin{\theta}\cos{\psi} & n_2 &= \sin{\theta}\sin{\psi}\\
    n_3 &= \cos{\theta}\cos{\phi} & n_4 &= \cos{\theta}\sin{\phi}
\label{eq:n4}
\end{align}
with $\phi,\psi \in [0,2\pi)$ and $\theta \in [0,\frac{\pi}{2}]$. These are called Hopf coordinates because they are the natural coordinates to describe  the Hopf fibration of the $3$-sphere. The angular laplacian in these coordinates is 
\begin{equation}
\Delta_{S^3}=\frac{1}{\cos{\theta}\sin{\theta}}\partial_{\theta}\big(\cos{\theta}\sin{\theta}\partial_{\theta}\big)+\csc^2{\theta}\partial_\phi^2+\sec^2{\theta} \partial_\psi^2\,.
\end{equation}
We are therefore looking for eigenfunctions that satisfy
\be
\Delta_{S^3}Y = -L(L+2)Y\,.
\ee
In order to solve this equation, we make the ansatz
\begin{equation}
    Y_{Lm_\phi m_\psi}(\theta,\phi,\psi)=\E^{i m_\phi \phi} \E^{i m_\psi \psi} S(\theta)\,,
\end{equation}
so that the function $S$ must satisfy the equation
\begin{equation}
    \frac{1}{\cos{\theta}\sin{\theta}}\partial_{\theta}\Big(\cos{\theta}\sin{\theta}\partial_{\theta}S(\theta)\Big)-\Big(m_\phi^2\csc^2{\theta}+m_\psi^2\sec^2{\theta} \Big)S(\theta)=-L(L+2)S(\theta)\,.
\end{equation}
This equation can be cast in a more familiar form by defining $x=\cos(2\theta)$:
\begin{equation}
    \partial_x \left[2(x^2-1) \partial_x S(x)\right]-\bigg(\tfrac{L}{2}(\tfrac{L}{2}+1)+\frac{m_\phi^2}{x-1}+\frac{m_\psi^2}{x+1}\bigg)S(x)=0\,,
\end{equation}
which is a hypergeometric equation. It is solved by~\cite{BenAchour:2015aah}
\begin{equation}
    \label{eq:5Dangularequationsolution}
    S(x)=2^{m_+}C^L_{m_+,m_-}(1-x)^{\frac{m_\psi}{2}}(1+x)^{\frac{m_\phi}{2}} P^{(m_\psi,m_\phi)}_{\frac{L}{2}-m_+}(x)\,,
\end{equation}
where $P^{(a,b)}_n$ is a Jacobi polynomial, and $m_\psi \equiv m_+ + m_-$, $m_\phi \equiv m_+ - m_-$ and the normalization factor is
\begin{equation}
    C^L_{m_+,m_-}\equiv \left(\frac{(L+1)(\tfrac{L}{2}+m_+)!(\tfrac{L}{2}-m_+)!}{2\pi^2(\tfrac{L}{2}+m_-)!(\tfrac{L}{2}-m_-)!}\right)^\frac{1}{2}\,.
\end{equation}
The solution is specified by requiring regularity at $x=\pm 1$ as well as requiring $\E^{i m_\phi \phi}$ and $\E^{i m_\psi\psi}$ to be periodic functions with period $2\pi$. These conditions, along with requiring the spherical harmonic to be nonzero, imply $|m_+| \le \frac{L}{2}$ and $\frac{L}{2}-m_\pm \in \mathbb{N}$.

We can then write the full spherical harmonic as
\be
    Y_{L m_\phi m_\psi }(\theta,\phi,\psi) =C^L_{\frac{m_\psi +m_\phi }{2},\frac{m_\psi -m_\phi }{2}}\,\E^{i m_\phi  \phi}\cos^{m_\phi} {\theta} \E^{i m_\psi  \psi}\sin^{m_\psi} {\theta}\,P^{(m_\psi ,m_\phi )}_{\frac{L-m_\psi -m_\phi }{2}}(\cos{2\theta})\,.
\ee

From~\eqref{eq:n1}--\eqref{eq:n4} we can note that $n_1+in_2=\E^{i\psi}\sin{\theta}$ and $n_3+in_4=\E^{i\phi}\cos{\theta}$. For positive values of $m_\phi$,$m_\psi$, we can use these relations and expand the Jacobi polynomial to obtain
\begin{equation}
    Y_{L m_\phi m_\psi }=\hat{C}^L_{m_\psi, m_\phi}\,(n_3+in_4)^{m_\phi}  (n_1+in_2)^{m_\psi} \sum_{j=0}^{\frac{L-m_\psi -m_\phi}{2}}a^j_{m_\psi m_\phi } (n_1+in_2)^{2j}\,,
\end{equation}
where the normalization factors are
\begin{align}
    \hat{C}^{L}_{m_\psi, m_\phi}&\equiv\left(\frac{(L+1)(\tfrac{L}{2}+\frac{m_\psi -m_\phi }{2})!}{2\pi^2(\tfrac{L}{2}-\frac{m_\psi -m_\phi }{2})!(\tfrac{L}{2}+\frac{m_\psi +m_\phi }{2})!(\tfrac{L}{2}-\frac{m_\psi +m_\phi }{2})!}\right)^\frac{1}{2}\,,\\
    a^j_{m_\psi m_\phi } &\equiv (-1)^j{\frac{L-m_\psi -m_\phi }{2}\choose j} \frac{(\frac{L+m_\psi +m_\phi }{2}+j)!}{(m_\psi +j)!}\,.
\end{align}
Upon identifying $e_+^i \hat{x}_i= n_1+in_2$, $f_+^i \hat{x}_i= n_3+in_4$, we can extract the Thorne tensor directly from this expression. This method is the analogue of the construction in~\cite{Thorne:1980ru}.

\subsubsection{General Dimension}
\label{app:sNharmonics}
Here we briefly sketch the construction of Thorne tensors in general dimension. Many of the conceptual features are the same as the cases that we have treated in detail, generalized straightforwardly.

As before, we imagine that we are given a $2$-form $S^{ij}$ that contains the information about the angular momentum of the object of interest, and our goal is to use this object to define an adapted basis of Thorne tensors for $n$-spherical harmonics. The $n$-sphere admits a natural action of SO$(n+1)$, and we can think of $S^{ij}$ as transforming in the adjoint. Writing $r=\lfloor \frac{n+1}{2}\rfloor$, we can define $r$ distinct $2$-planes that are orthogonal to $S^{ij}$ via
\be
\epsilon_{i_1\cdots i_{n+1}}S^{i_1i_2}P^{i_3\cdots i_{n+1}}_J = s_J\,.
\ee
(Recall that a $2$-plane in $d$ dimensions is specified by an antisymmetric $(d-2)$-index tensor.) The $s_J$ have the interpretation as the angular momentum of the object in the relevant planes.

For each plane, we then construct a basis of vectors in that plane
\be
\label{eq:helicityeigenvectors}
(e^J)_a^i\,.
\ee
In odd ambient dimensions, these vectors only span a $n$-dimensional subspace of the full $(n+1)$-dimensional ambient space. We therefore need one additional vector that is orthogonal to all of these basis elements. It can be defined by the condition
\be
\epsilon_{i_1\cdots i_{n+1}}(e^1)_1^{i_1}(e^1)_2^{i_2}\cdots (e^r)_1^{i_{n-1}}(e^r)_2^{i_n} S^{i_{n+1}} = 1\,.
\ee
In the two-sphere case, the role of this vector $S^i$ was played by the spin vector $s^i$ that was the dual of the original $2$-form $S^{ij}$. This is special to three ambient dimensions where we can use $\epsilon_{ijk}$.

Given these $n+1$ basis vectors, we construct the Thorne tensors and spherical harmonics as in the cases above by building traceless symmetric tensors from these basis elements. It is again natural to parameterize the basis vectors for each plane in the helicity basis, so that the spherical harmonics generated by the relevant Thorne tensors carry angular momenta in each of these $2$-planes. These are then the Thorne tensors and spherical harmonics defined in the analogue of Hopf coordinates.

\subsubsection{Thorne Tensors in Rotating Frames}

One benefit of using Thorne tensors as a basis is that they have nice transformations under a rotation of the coordinate system, which we utilized in the discussion of rotating frames in Section~\ref{sec:rotframes}. 

We would like to relate Thorne tensors defined in a rest frame to Thorne tensors defined in a time-dependent, co-rotating frame. In other words, we wish to make explicit the time dependence in  $\mathcal{Y}^{i_1\cdots i_L}_{LM}=e_{a_1}^{i_1}(T)\cdots e_{a_L}^{i_L}(T)\mathcal{Y}^{a_1\cdots a_L}_{LM}$, where the $i$ indices represent the ``lab" frame, and the $a$ indices are intrinsic ``co-rotating frame" indices.

For simplicity let us start with the $D=4$ case. In the rest frame, we have $e_0^\mu=\delta_0^\mu$. The spatial components of the vielbein are time dependent as they capture the frame of an observer stationary with respect to the horizon. Choosing to orient the object's spin axis with the $\hat z$ direction, the spatial vielbein takes the form~\cite{Goldberger:2020fot}
\be 
    e^i_a(T) =
    \begin{pmatrix}
        \cos{\Omega \, T} & -\sin{\Omega\, T} & 0 \\
        \sin{\Omega \,T} & \cos{\Omega \, T} & 0 \\
        0 & 0 & 1
    \end{pmatrix}\,,
\ee
Where $\Omega$ is the angular velocity of the black hole. Since we want to contract this vielbein with Thorne tensors, it is useful to calculate the contractions with the basis vectors defined in~\eqref{eq:ThornetensorbasisS2}.
We can just compute directly that 
\begin{equation}
    e^i_a {e}^a_\pm = \E^{\mp i \Omega s} {e}^i_\pm , \quad\qquad e^i_a s^a = s^i\,.
\end{equation}
\noindent   
As the $\mathcal{Y}^{a_1\cdots a_l}_{lm}$ have $m$ factors of ${e}^a_{\text{sgn}(m)}$, we have 
\begin{equation}
    \label{eq:vielbeinonThornetensor}
    e_{a_1}^{i_1}(T)\cdots e_{a_l}^{i_l}(T)\mathcal{Y}^{a_1\cdots a_l}_{lm}= \E^{-i m \Omega T}\, \mathcal{Y}^{i_1\cdots i_l}_{lm}\,.
\end{equation}

The higher-dimensional generalization is straightforward. If there are $r$ independent planes of rotation, denote the magnetic quantum number associated to each plane as $m_n$ and the corresponding angular frequency by $\Omega_n$. Then, we have 
\begin{equation}
    e_{a_1}^{i_1}(T)\cdots e_{a_L}^{i_L}(T)\mathcal{Y}^{a_1\cdots a_L}_{L\,m_1\cdots m_r}=\E^{-i \sum_{n=1}^{r}m_n \Omega_n T}\, \mathcal{Y}^{i_1\cdots i_L}_{L\,m_1\cdots m_r}\,,
\end{equation}
because the vielbein $e_a^i$ block diagonalizes. We therefore see that the total time dependence is just the product of the time dependences in each individual plane.

\newpage
\section{$\mathbb{C} P^N$ Eigenfunctions}
\label{app:CPNeigen}
In Section~\ref{sec:InfD} we study features of the large dimension limit of black holes with equal spin parameters. In odd spacetime dimensions, the Myers--Perry metric with equal spin parameters has components that depend only on a single radial coordinate (it is cohomogeneity-$1$)~\cite{Kunduri:2006qa}. This reflects an enhancement of symmetry for this choice of parameters. If we parameterize $D=2N+3$, then the equal-spin black hole has as its symmetry group  $\mathbb{R} \times$ U$(N+1)$ (compared to $\mathbb{R} \times$ U$(1)^{N+1}$ for a generic spinning object).
As a result of this enhanced symmetry, more of the response coefficients of the black hole will be related to each other than for a generic object. To make these symmetries more apparent, it is convenient to view the horizon--which is topologically $S^{2N+1}$---as a $S^1$ bundle over complex projective space $\mathbb{C} P^N$. Since $\mathbb{C} P^N$ has the isometry group SU$(N+1)$, this makes the symmetries of the equal-spin limit manifest.

Because of the enhanced symmetry, the wave equation in this black hole background separates if we make an ansatz of the form
\begin{equation}
    \Phi (t, r, \psi, x^a) = \E^{-i \omega t + i M \psi} R(r) \mathbb{Y}_{LM}(x^a)\, ,
    \label{eq:appcpnsplit}
\end{equation}
which decomposes the solution into plane waves in the time direction, Fourier modes around the circle factor, and $\mathbb{Y}_{LM}(x^a)$, which are eigenfunctions of the charged laplacian on $\mathbb{C} P^N$, where the charge is given by the $M$ eigenvalue. As such we will need some of the properties of these eigenfunctions.

In this Appendix, we briefly describe some of the essential features of the functions $\mathbb{Y}_{LM}(x^a)$ which appear in the decomposition~\eqref{eq:appcpnsplit}, following~\cite{Hoxha:2000jf}. Since these functions are harmonics on $\mathbb{C} P^N$, the strategy to construct them will parallel the construction of spherical harmonics in Appendix~\ref{appendix:Thorne}. That is, we will realize $\mathbb{C} P^N$ as a submanifold of a larger space, and pull back harmonic functions on this larger space to construct eigenfunctions on the $\mathbb{C} P^N$ subspace. In this case, we want to construct eigenfunctions not just of the laplacian, but of the wave operator of a charged particle. To do so, we will augment $\mathbb{C} P^N$ with a circle factor, so that the eigenvalue of the laplacian on this circle will play the role of a charge.

 Since $\mathbb{C} P^N$ is best viewed as a complex manifold, we begin by considering $(2N+2)$-dimensional flat space written in complex coordinates~$Z^A$
\begin{equation}
    \rd s^2= \rd Z^A \rd\bar{Z}_A\,,
    \label{eq:complexmetric1}
\end{equation}
The goal is to slice this space by spheres as
\begin{equation}
    \rd s^2= \rd r^2 +r^2 \rd\Omega^2_{2N+1}\,,
    \label{eq:flatspacesphereslices}
\end{equation}
so that the spheres themselves can be viewed as $S^1\hookrightarrow S^{2N+1}\to \mathbb{C} P^N$. A choice of coordinates that accomplishes this is~\cite{Hoxha:2000jf} 
\begin{equation}
    x^a \equiv \frac{Z^a}{Z^0} \,,\qquad \E^{i\psi}\equiv \frac{Z^0}{|Z^0|}\,, \qquad r \equiv\sqrt{Z^A \bar{Z}_A}\,,
    \label{eq:complexspherecoords}
\end{equation}
where we have split the $A$ coordinates into $0$ and $a\in\{1,\cdots, N\}$. In these coordinates, the sphere factor in~\eqref{eq:flatspacesphereslices} has the line element
\begin{equation}
    \label{eq:S2Np1metricasU1onCPN}
    \rd\Omega^2_{2N+1}= (\rd\psi +B)^2 +  \rd\Sigma^2_N\,,
\end{equation}
with $ \rd\Sigma^2_N$ the Fubini--Study metric on $\mathbb{C} P^N$ (which has coordinates $x^a$) and $B$ is a 1-form potential of its K\"ahler form. We will not need their explicit expressions, but they can be found in~\cite{Hoxha:2000jf}.

In order to construct eigenfunctions on $\mathbb{C} P^N$, we can follow the same strategy used to construct spherical harmonics and begin with functions annihilated by the laplacian on the full space $\mathbb{C}^{N+1}$, and then pull them back to the subspace corresponding to projective space. Consider
\begin{equation}
P_{pq}= \mathcal{E}_{ \langle A_1\cdots A_{p}\rangle_T}^{\langle B_1 \cdots B_{q}\rangle_T}Z^{A_1}\cdots Z^{A_p}\bar{Z}_{B_1}\cdots \bar{Z}_{B_q}\,,
\label{eq:harmonicfncpn}
\end{equation}
where the tensor
$\mathcal{E}_{ \langle A_1\cdots A_{p}\rangle_T}^{\langle B_1 \cdots B_{q}\rangle_T}$, is symmetric under permutations of its $A$ indices and of its $B$ indices separately, and which vanishes if we contract any $A$ index with any $B$ index, which we denote by the $\langle\cdots\rangle_T$ notation.\footnote{These tensors define a state in the representation of SU$(N+1)$ specified by the quantum numbers $(p,q)$, much like the Thorne tensors in Appendix~\ref{appendix:Thorne}.}
Since the metric~\eqref{eq:complexmetric1} is off-diagonal, raising or lowering an index in this space switches barred coordinates to unbarred and vice versa. Noting that $\frac{\partial}{\partial Z^A}\bar{Z}_B=\frac{\partial}{\partial \bar{Z}_A}Z^B=0$, it is straightforward to see that~\eqref{eq:harmonicfncpn}
is harmonic:
\begin{equation}
    \Delta_{\mathbb{C}^{N+1}} P_{pq}=\frac{\partial^2}{\partial Z^A \partial \bar{Z}_A} P_{pq} =0, 
    \label{eq:harmonicYchargedlp|}
\end{equation}
as this will necessarily involve contraction of raised with lowered indices.\footnote{This is the complexified analogue of~\eqref{eq:harmonicpol} vanishing because the tensor $c_{i_1\cdots i_L}$ is traceless.} Next, we write $P_{pq}$ in the radial coordinates~\eqref{eq:flatspacesphereslices}
\begin{equation}
    \label{eq:PhitoYtransfCPN}
    P_{pq}=r^{p+q}\E^{i(p-q)\psi} \mathbb{Y}_{pq}(x^a)\,,
\end{equation}
where we have dropped the (irrelevant) overall normalization. The fact that $P_{pq}$ satisfies~\eqref{eq:harmonicYchargedlp|} implies that $\mathbb{Y}_{pq}(x^a)$ is an eigenfunction of the charged laplacian on $\mathbb{C} P^N$. To see this, we write the laplacian in the coordinates~\eqref{eq:complexspherecoords}
\begin{equation}
    \label{eq:CPNp1laplacianinr}
    \Delta_{\mathbb{C}^{N+1}}P_{pq}=\frac{1}{r^{2N+1}}\frac{\partial}{\partial r}\left(r^{2N+1}\frac{\partial P_{pq}}{\partial r}\right) +\frac{1}{r^2}  \Delta_{S^{2N+1}}P_{pq}\,,
\end{equation}
where the laplacian on the sphere is explicitly~\cite{Hoxha:2000jf}
\begin{equation}
    \label{eq:S2nNp1laplacianinCPN}
  \Delta_{S^{2N+1}}P_{pq}= \left(\nabla_a -B_a \frac{\partial}{\partial \psi}\right)^2P_{pq}+\frac{\partial^2}{\partial \psi^2}P_{pq}\,,
\end{equation}
with $\nabla_a$ the covariant derivative of the Fubini--Study metric. 

Acting on~\eqref{eq:PhitoYtransfCPN} with~\eqref{eq:CPNp1laplacianinr}, we see that the fact that $P_{pq}$ is harmonic implies that $\mathbb{Y}_{pq}$ satisfies the eigenvalue equation
\begin{equation}
    \mathcal{D}_a \mathcal{D}^a \mathbb{Y}_{pq}= -2\big(2pq+ N(p+q)\big)\mathbb{Y}_{pq} \,,
\end{equation}
with $\mathcal{D}_a\equiv\nabla_a-i(p-q)B_a$ the covariant derivative with charge $(p-q)$, treating $B_a$ as a gauge field. 
The function $\mathbb{Y}_{pq}(x^a)$ is an eigenfunction of the charged laplacian on $\mathbb{C} P^N$ with charge $(p-q)$ and eigenvalue $-2\left(2pq+ N(p+q)\right)$. If we define $L\equiv p+q$, $M\equiv p-q$, then the field has charge $M$ and eigenvalue $L(L+2N)-M^2$, which is the parameterization used in Section~\ref{sec:InfD}. If desired, we can construct explicit harmonics and their Thorne tensors in analogy with ordinary spherical harmonics.

\newpage
\section{Gamma Function Identities}\label{appendix:Formulas}

The Gamma function (and its derivatives) appears repeatedly in the computations in the main text. Here we collect some useful identities involving the Gamma function.

Famously, the Gamma function obeys Euler's reflection formula
\begin{equation}
    \label{eq:Gammareflect}
    \Gamma(z)\Gamma(1-z)=\frac{\pi}{\sin(\pi z)}\,,
\end{equation}
when $z \notin \mathbb{Z}$. 

For $n \in \mathbb{Z}$, $b\in \mathbb{R}$, we have the following related identities which can be proven by induction
\begin{align}
    |\Gamma(1+ib)|^2 &= \frac{\pi b}{\sinh(\pi b)}\\
    |\Gamma(1+n+ib)|^2 &= \frac{\pi b}{\sinh(\pi b)} \prod_{k=1}^{n}(k^2+b^2)\\
    |\Gamma(\tfrac{1}{2} \pm n+ib)|^2 &= \frac{\pi b}{\cosh(\pi b)} \prod_{k=1}^{n}\Big((k-\tfrac{1}{2})^2+b^2\Big)^{\pm 1}\\
    |\Gamma(-n+ib)|^2 &= \frac{\pi }{b \sinh(\pi b)} \prod_{k=1}^{n}(k^2+b^2)^{-1}\,.
\end{align}

The following combination of gamma functions often appears, which we simplify using~\eqref{eq:Gammareflect} 
\begin{eqnarray}
    \frac{\Gamma(1+L+2iU)}{\Gamma(-L+2iU)}&=&\frac{\sin\big(\pi(-L+2iU)\big)}{\pi}|\Gamma(1+L+2iU)|^2\\
    \frac{\Gamma (-2 L-1)}{\Gamma (-L)}&=&-\frac{\sec (\pi  L) \Gamma (L+1)}{2 \Gamma (2 L+2)} \,.
\end{eqnarray}
Using these, we can expand the following ratio of Gamma functions that also often occurs
\be
    \frac{\Gamma(-2L-1) \Gamma(1+L+2iU)}{\Gamma(-L)\Gamma(-L+2iU)}-iU \Big(1+i\coth(2\pi U)\tan(\pi L)\Big)\frac{\Gamma (L+1)  |\Gamma (L+1+2 i U)|^2}{ \Gamma (2 L+2) |\Gamma (1+2 i U)|^2}\,,
\ee
The combination of Gamma functions appearing on the right-hand side can be simplified for integer $L$ using the following formula:
\begin{equation}
    \frac{\Gamma (L+1)  |\Gamma (L+1+2 i U)|^2}{\Gamma (2 L+2) |\Gamma (1+2 i U)|^2} \Bigg|_{L \in \mathbb{Z}}=\frac{(L!)}{(2L+1)!}\prod_{k=1}^{L} (k^2+4U^2)\,.
\end{equation}

We can similarly expand the combination
\be
\label{eq:MPresponseexpansion}
    \frac{\Gamma(-2L-1)\Gamma(L+1+2iV) \Gamma(1+L+2iU)}{\Gamma(-L+2iV)\Gamma(-L+2iU)}=
    \frac{|\Gamma(1+L+2iU)|^2|\Gamma(1+L+2iV)|^2}{\pi \Gamma(2L+2)}C_L\,,
\ee
where $C_L$ is the following combination of trigonometric functions
\be
C_L\equiv \sin\big(\pi(L-2iU)\big)\sin\big(\pi(L-2iV)\big)\csc(2L\pi)\,,
\ee
which can be split into its real and imaginary parts as
\be
\label{eq:MPtrigexpansion}
C_L=
    \frac{1}{2}\bigg[-\cot (2 \pi  L) \cosh \big(2 \pi  (U+V)\big)+\csc (2 \pi  L) \cosh \big(2 \pi  (U-V)\big)-i \sinh \big(2 \pi  (U+V)\big)\bigg]\,.
\ee

The digamma function
\be
\psi(z)  = \frac{\Gamma'(z)}{\Gamma(z)}\,,
\ee
also appears in many places. Among its many properties, one useful identity is
\begin{equation}\label{eq:digammaexpansion}
    \psi(-L+2iU)-\psi(L+1+2iU)=\pi  \cot \big(\pi  (L-2 i U)\big)+\psi (L+1-2 i U)-\psi (L+1+2 i U)\,,
\end{equation}
which splits the combination on the left into its real and imaginary parts.
The combination appearing on the right-hand side is purely imaginary, and can be written as
\begin{equation}
    \psi (L+1-2 i U)-\psi (L+1+2 i U)= i\sum_{n=1}^{\infty}\frac{-4U}{(L+n)^2+4U^2}\,.
\end{equation}
For integer $L$, we may alternatively write
\begin{equation}
    \psi(L+1-2iU)-\psi(L+1+2iU)=i \left(\sum_{n=0}^{L} \frac{4 U}{n^2+4U^2}-\frac{1}{2 U}-\pi  \coth (2 \pi  U)\right)\,,
\end{equation}
which has the benefit of being written in terms of elementary functions and finite sums.

\newpage
\section{Scattering}
\label{app:scattering}

In this Appendix, we consider the scattering of scalar waves off a particle's worldline in EFT. This slightly different physical setup can be used to cross-check the results obtained in the main text by matching the off-shell one-point function in the presence of a scalar background. This is one of the benefits of the EFT framework: once the Wilson coefficients have been matched via whatever calculation is most convenient, the resulting EFT can be used to compute any physical observable desired.
In this Appendix we describe the general formalism for computing scattering amplitudes in the EFT, and then describe the analogous scattering process for a Kerr black hole in general relativity. In Appendix~\ref{app:NZ} we use these results to validate the near zone approximation used in Section~\ref{sec:Kerr}, showing that matching the one-point function produces the same EFT coefficients as matching scattering cross sections.

\subsection{Point Particle Scattering}
\label{sec:pparticlescattering}

We begin by computing the scattering cross section for a scalar wave in the point-particle EFT constructed in Section~\ref{sec:EFT}. Essentially the same setup was considered in~\cite{Ivanov:2022qqt,Saketh:2023bul,Ivanov:2024sds}, here we want to consider the $D$-dimensional version of the problem.

We are interested in the $2\to2$ process involving a point particle and a scalar wave:
\begin{equation*}
\includegraphics[scale=1.65]{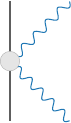}
\end{equation*}
and we would like to parameterize the cross section in terms of the Wilson coefficients of the EFT. In all cases of physical interest, the mass of the point particle is much larger than the energy carried by the incident wave, and so we can approximate the scattering as leaving the momentum of the point particle unchanged. Denoting the internal degrees of freedom of the point particle by $X$, we are therefore interested in the transition probability between the states
\be
\lvert \omega\,L\, M; X\rangle \longrightarrow \lvert \omega'\,L'\, M'; X'\rangle \,,
\ee
where we have labeled the incoming and outgoing scalar states by their frequency $\omega$ along with total angular momentum $L$ and magnetic quantum numbers, $M$ (where $M$ is a multi-index cataloging all quantum numbers).

This transition probability is captured by the $S$-matrix element
\be
{\mathds 1}+i{\cal A}_{2\to 2} = \langle \omega'\,L'\, M' \lvert \omega\,L\, M\rangle \,,
\label{eq:smatrix1}
\ee
where we have suppressed the $X$ labels because the internal state of the point particle does not change in the approximation we are making. Eventually we will be interested in the amplitude for classical waves scattering off the object, we first consider quantum-mechanical scattering, and then specialize at the end.
The states appearing in~\eqref{eq:smatrix1} are labeled by their corresponding free theory quantum numbers, in order to compute their overlap, we specify in and out states in the free theory and time evolve the in state to the far future:
\be
 \langle \omega'\,L'\, M' \lvert \omega\,L\, M\rangle =  {}_0\langle \omega'\,L'\, M' \rvert\hat S \lvert\omega\, L\, M\rangle_0\,,
\ee
where $\lvert\omega\, L\, M\rangle_0$ are asymptotic single-particle states and the operator $\hat S$ is the usual interaction picture time evolution operator:
\be
\hat S =  T\E^{-i\int_{-\infty}^\infty\rd t \hat H_{\rm int}}\,.
\label{eq:inthamil}
\ee
In the following, we will drop the $0$ subscript on the states $\lvert \cdots\rangle_0$ in a small abuse of notation, the meaning should be clear from context.

In the case of interest, the interaction Hamiltonian involves the couplings between the particle's worldline and the external scalar field constructed in Section~\ref{sec:EFT}. Since we want to compute the scattering of a single wave off of the object, we will only need the interaction
\be
   \int \rd t H_{\rm int} = -\int\rd \tau E\, \sum_L Q^{i_1\cdots i_L}(X)\,\partial_{i_1}\cdots \partial_{i_L} \phi\,,
   \label{eq:inthamil2}
\ee
which couples together the external field $\phi$ with the composite multipole operators $Q$, which are built out of the object's fundamental degrees of freedom, $X$. Here we are working in the object's rest and corotating frame, but this can easily be converted to a general frame (like the inertial lab frame) using the formulas in Section~\ref{sec:rotframes}. The main effect will be a shift of the frequency seen by the black hole degrees of freedom.
We see that the interaction~\eqref{eq:inthamil2} couples nontrivially to different multipole moments, but is time-translation invariant.
Defining $\hat S = {\mathds 1}+i\hat T$ as usual, we can relate the matrix elements of the transition matrix to connected scattering amplitudes that we compute using Feynman rules:
\be
\langle \omega'\,L'\, M' \rvert i\hat T \lvert\omega\, L\, M\rangle\ = 2\pi\,\delta(\omega-\omega')\, i{\cal A}(\omega\,L\,M;\omega'\,L'\,M')\,,
\ee
where the $\delta(\omega-\omega')$ factor is a consequence of the time-translation invariance of the interaction Hamiltonian in~\eqref{eq:inthamil} in the case of interest. 

We can compute the scattering amplitude in perturbation theory. To leading order, the amplitude is given by the diagram
\begin{equation*}
\includegraphics[scale=1.7]{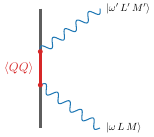}
\end{equation*}
which has the corresponding algebraic expression
\be
\label{eq:scatteringmatrixelement}
\begin{aligned}
\langle \omega'\,L'\,M'\rvert i\hat T\lvert\omega\, L\, M\rangle=
    T\bigg(-\frac{1}{2!}\sum_{\ell\,\ell'}\int\rd t&\rd t' \langle X_i\rvert Q^{i_1\cdots i_\ell }(t)Q^{j_1\cdots j_{\ell'} }(t')\lvert X_i\rangle \\[-2pt]
    &\,\times\langle \omega'\,L'\,M'\rvert\partial_{i_1}\cdots \partial_{i_{\ell}}\phi(t)\partial_{j_1}\cdots \partial_{j_{\ell'}}\phi(t')\lvert\omega\, L\, M\rangle\bigg)\,,
\end{aligned}
\ee
where the matrix element factorizes because we imagine that the Hilbert space of the $X$ degrees of freedom is tensored with that of the external scalar field. 

We first compute the matrix element involving the external field $\phi$. It can be computed by inserting a complete set of Fock states in the plane-wave basis
\be
\label{eq:phimatrix}
\begin{aligned}
\langle \omega'\,L'\,M'\rvert\partial_{i_1}\cdots \partial_{i_{\ell}}\phi(t)\partial_{j_1}\cdots \partial_{j_{\ell'}}\phi(t')\lvert\omega\, L\, M\rangle&= \\[1pt]
    \int \frac{\rd^d k}{(2\pi)^d (2\omega_k)} \frac{\rd^d k'}{(2\pi)^d (2\omega_{k'})} \langle \omega'\,L' &\,M' \rvert {\bf k}'\rangle\langle {\bf k}'\rvert\partial_{i_1}\cdots \partial_{i_{\ell}}\phi(t)\partial_{j_1}\cdots \partial_{j_{\ell'}}\phi(t')\lvert {\bf k}\rangle\langle {\bf k}\lvert\omega\, L\, M\rangle\,.
\end{aligned}
\ee
The fact that we are taking the overlap with the (single-particle) external states $\lvert\omega\, L\, M\rangle$ localizes the sum over internal states to the single-particle sector. We compute the overlap $\langle {\bf k}\lvert\omega\, L\, M\rangle$ in the following inset.

    \vspace{-4pt}
\begin{oframed}
{\small
\noindent
{\bf\normalsize Normalization of states:}
Here we want to compute the overlaps between single-particle states in different bases, following~\cite{Endlich:2016jgc}.
We normalize plane wave states with the relativistic normalization, so that two single-particle states $\lvert \mathbf{k}\rangle \equiv a_{\mathbf{k}}^\dagger |0\rangle$ with momenta ${\bf k}$ and ${\bf k}'$ have overlap
\begin{equation}
   \langle \mathbf{k}'|\mathbf{k}\rangle = 2 \omega_k (2\pi)^d \delta(\mathbf{k}-\mathbf{k}')\,,
\end{equation}
where $\omega_k = \lvert{\bf k}\rvert$ is the energy of the (massless) particle. This normalization follows from the algebra of raising operators
\begin{equation}
    \label{eq:raisingloweringcommutator}
   [a_{\mathbf{k}'},a_{\mathbf{k}}^\dagger]=\left(2\pi\right)^d 2 \omega_k \delta(\mathbf{k}-\mathbf{k}')\,.
\end{equation}
With this relativistic normalization we also have
\begin{equation}
   \int \frac{\rd^d k}{(2\pi)^d 2\omega_k}|\mathbf{k}\rangle \langle \mathbf{k}| = \mathds{1} \, .
\end{equation}
We now want to compute the overlap between the plane wave state $\lvert {\bf k}\rangle$ and the spherical wave state $\lvert \omega L M\rangle$. By symmetry, it must be of the form
\begin{equation}
   \langle \mathbf{k}| \omega,L,M\rangle= N_{\omega L M}\delta(\omega_k-\omega)Y_{LM} (\hat{k})\,,
\end{equation}
for some normalization constant $N_{\omega L M}$. We can compute this normalization by inserting a complete set of states in the matrix element $ \langle \omega'\, L'\, M'\lvert \omega\, L\, M\rangle$ to obtain
\be
\begin{aligned}
   \langle \omega'\,L'\,M'| \omega\,L\,M\rangle &= \int \frac{\rd^d k}{(2\pi)^d 2\omega_k}\langle \omega'\,L'\,M'|\mathbf{k}\rangle \langle \mathbf{k}| \omega\,L\,M\rangle \\
   &= \int \frac{\rd\Omega_{S^{d-1}} \rd k\, k^{d-2}}{2(2\pi)^d }\delta(k-\omega)\delta(k-\omega')Y_{L'M'}^* (\hat{k}) Y_{LM} (\hat{k}) \lvert N_{\omega L M} \rvert^2\\
   &=\delta(\omega-\omega')\delta_{LL'}\delta_{MM'} |N_{\omega,L,M}|^2 \frac{\omega^{d-2}}{2(2\pi)^d}\,,
\end{aligned}
\ee
where we have used~\eqref{eq:sphericalharmonicortho}. We now want to require
\be
  \langle \omega'\,L'\,M'| \omega\,L\,M\rangle=2\pi \delta(\omega-\omega')\delta_{LL'}\delta_{MM'}\,,
  \label{eq:sphericaloverlap}
\ee
so that $|N_{\omega L M}| = \left(2(2\pi)^{d+1}/\omega^{d-2}\right)^\frac{1}{2}$. With this normalization, we have, up to a phase, 
\be
\langle \mathbf{k}| \omega,L,M\rangle= \left(\frac{2(2\pi)^{d+1}}{\omega^{d-2}}\right)^\frac{1}{2}\delta(\omega_k-\omega)Y_{LM} (\hat{k})\,.
\label{eq:wavesphereoverlap}
\ee
}
\end{oframed}
    
With these normalizations, we can compute the matrix element~\eqref{eq:phimatrix} in perturbation theory by expanding in plane waves:
\be
    \phi({\bf x},t)= \int \frac{\rd^d k}{(2\pi)^d (2\omega_k)}\E^{i{\bf k} \cdot {\bf x}}\left(\E^{-i\omega_kt}a_{\mathbf{k}}+\E^{i\omega_kt}a_{\mathbf{-k}}^\dagger  \right)\,.
    \label{eq:fieldexpan}
\ee
We can then compute~\eqref{eq:phimatrix} at the location of the object (${\bf x} = 0$) by first computing
\be
\begin{aligned}
    \langle \mathbf{k}' |\partial_{i_1}\cdots \partial_{i_{\ell}}\phi(t)\partial_{j_1}\cdots \partial_{j_{\ell'}}\phi(t')|\mathbf{k}\rangle = 
    ~&i^{\ell-\ell'} k_{i_1}\cdots k_{i_{\ell}} \, k'_{j_1}\cdots k'_{j_{\ell'}} \E^{-i(kt - k' t')}\\
    &+i^{\ell'-\ell} k'_{i_1}\cdots k'_{i_{\ell}} \, k_{j_1}\cdots k_{j_{\ell'}} \E^{i(k' t-kt')}\,.
\end{aligned}
\ee
If we insert this expression into~\eqref{eq:phimatrix} and use the matrix element~\eqref{eq:wavesphereoverlap} we can then perform the integrals over the magnitude of ${\bf k}$ using the delta functions, and can 
perform the angular integrals using~\eqref{eq:integraloverksandarbitrarysphharmonic} to obtain
\begin{align}
    &\langle \omega'\,L'\,M'|\partial_{i_1}\cdots \partial_{i_{\ell}}\phi(t)\partial_{j_1}\cdots \partial_{j_{\ell'}}\phi(t')|\omega\,L\,M\rangle= \\[1pt]
    &\qquad\qquad~~~~N_{L,L'}\left(\E^{-i (\omega t- \omega' t')}\mathcal{Y}_{i_{1}\ldots i_{L}}^{L M} \mathcal{Y}^*{}_{j_{1}\ldots j_{L'}}^{L'M'} \delta_{L \ell}\delta_{L'\ell'} +\E^{i (\omega' t- \omega t')}\mathcal{Y}_{j_{1}\ldots j_{L}}^{L M}  \mathcal{Y}^*{}_{i_{1}\ldots i_{L'}}^{L'M'}  \delta_{L \ell'}\delta_{L'\ell}\right) \nonumber,
\end{align}
where we have defined the normalization factor
\begin{equation}
    N_{L,L'}=\frac{i^{L-L'}\omega^{L+\frac{d}{2}-1}\omega'^{L'+\frac{d}{2}-1}}{2(2\pi)^{d-1}}\frac{2^{1-L}\pi^{\tfrac{d}{2}}\Gamma(L+1)}{\Gamma(L+\tfrac{d}{2})}\frac{2^{1-L'}\pi^{\tfrac{d}{2}}\Gamma(L'+1)}{\Gamma(L'+\tfrac{d}{2})} .
\end{equation}

We now return to computing~\eqref{eq:scatteringmatrixelement}. For clarity, we proceed by defining the multi--indices $i_1\cdots i_L\equiv I_L$ and $j_1\cdots j_{L'}\equiv J_{L'}$.
We can now calculate
\begin{align}
    \langle \omega'\,L'\,M'\rvert i\hat T\lvert\omega\, L\, M\rangle 
    &=T\bigg(-\frac{1}{2!}\int\rd t\rd t' N_{L,L'}\mathcal{Y}_{I_{L}}^{L M} \mathcal{Y}^*{}_{J_{L'}}^{L'M'}\bigg[\E^{-i (\omega t- \omega' t')} \langle Q^{ I_L }(t)Q^{J_{L'} }(t')\rangle \\
    &~~\qquad\qquad\qquad\qquad\qquad\qquad\qquad\qquad+\E^{i (\omega' t- \omega t')}\langle Q^{ J_{L'} }(t)Q^{I_{L} }(t')\rangle\bigg]\bigg)\nonumber ,
\end{align}
where we have relabeled $I$ and $J$ in the second term while maintaining their $L,L'$ prescriptions.
We define the Fourier transform of the Wightman propagator as
\begin{equation}
    \label{eq:wightmanfourier}
    \langle Q^{ I_L }(t)Q^{J_{L'} }(t')\rangle=\int_{-\infty}^{\infty} \frac{\rd {\omega}}{2\pi} \E^{-i{\omega}(t-t')} \Delta^{I_L J_{L'}}({\omega})\,.
\end{equation}
Recall that the frequency space Wightman function is real and positive for $\omega \geq0$, and vanishes for $\omega <0$.
In terms of this propagator we can write
\begin{align}
    &\langle \omega'\,L'\,M'\rvert i\hat T\lvert\omega\, L\, M\rangle=\\
    & ~=T\bigg(\!\!-\frac{1}{2!}\int\rd t\rd t' \frac{\rd \tilde{\omega}}{2\pi} N_{L,L'}\mathcal{Y}_{I_{L}}^{L M} \mathcal{Y}^*{}_{J_{L'}}^{L'M'}\E^{-i\tilde{\omega}(t-t')}\left[\E^{-i (\omega t- \omega' t')} \Delta^{I_L J_{L'}}(\tilde{\omega}) +\E^{i (\omega' t- \omega t')}\Delta^{J_{L'}I_L}(\tilde{\omega})\right]\bigg)\,.\nonumber
\end{align}
It is convenient to change time variables to 
\begin{equation}
    t_+= \frac{t+t'}{2},\qquad\qquad t_- =t-t'\,,
\end{equation}
in terms of which we get
\begin{align}
    &\langle \omega'\,L'\,M'\rvert i\hat T\lvert\omega\, L\, M\rangle=\\
    &T\!\left(\!-\frac{N_{L,L'}}{2!}\mathcal{Y}_{I_{L}}^{L M} \mathcal{Y}^*{}_{J_{L'}}^{L'M'}\!\!\!\int_{-\infty}^{\infty} \!\!\rd t_+ \rd t_-  \frac{\rd \tilde{\omega}}{2\pi} \E^{-i(\omega-\omega')t_+-\frac{i}{2}(\omega+\omega'+2\tilde{\omega})t_-}\!\left[\Delta^{I_L J_{L'}}(\tilde{\omega})+\E^{i(\omega+\omega')t_-}\!\Delta^{J_{L'}I_L}(\tilde{\omega})\right]\right).\nonumber 
\end{align}
Time ordering acts simply in these variables (simply adding an overall factor of $ 2 \theta(t_-)$ to the expression)
We can also do the $t_+$ integral, which yields a factor of $2\pi \, \delta(\omega-\omega')$. This we pull out and set $\omega'\to \omega$. 
Putting everything together, we have
\be
    i\mathcal{A}= -N_{L,L'}\mathcal{Y}_{I_{L}}^{L M} \mathcal{Y}^*{}_{J_{L'}}^{L'M'}\int_{-\infty}^{\infty} \rd t_-  \frac{\rd \tilde{\omega}}{2\pi}\,\left(\theta(t_-) \E^{-i(\omega+\tilde{\omega})t_-}\Delta^{I_L J_{L'}}(\tilde{\omega})+\theta(t_-) \E^{i(\omega-\tilde{\omega})t_-}\Delta^{J_{L'}I_L}(\tilde{\omega})\right)\,.
\ee
If we send $\tilde{\omega}\to -\tilde{\omega}$ and $t_- \to -t_-$ in the first term we can write the amplitude in terms of the Feynman propagator of the operators $Q$
\be
i\mathcal{A}=-iN_{L,L'}\mathcal{Y}_{I_{L}}^{L M} \mathcal{Y}^*{}_{J_{L'}}^{L'M'} G^{J_{L'}I_L}_F(\omega)\,.
\label{eq:feynamp}
\ee
If we expand the Feynman propagator in the basis of Thorne tensors, 
\begin{equation}
    G^{I_L J_{L'}}_F(\omega)=\sum_{M,M'} \, \lambda_{ LML'M'}^{(F)}(\omega)\, \mathcal{Y}^{I_L}_{LM} \mathcal{Y}^*{}^{J_L'}_{L'M'}
\end{equation}
then we can write the scattering amplitude in terms of the harmonic space frequency representation
    \be
    i\mathcal{A}=-\frac{i^{L-L'+1}\omega^{L+L'+d-2}}{2(2\pi)^{d-1}}\lambda_{ L'M'LM}^{(F)}(\omega)\,.
    \label{eq:scatteringcrosssec}
    \ee

\vspace{-12pt}
\paragraph{Classical scattering:} Note that~\eqref{eq:scatteringcrosssec} is a quantum-mechanical scattering amplitude (and correspondingly is written in terms of the Feynman propagator). In order to match to the {\it classical} scattering of a wave off of a black hole, we want to extract the classical scattering amplitude from this expression. The computation of classical observables from scattering amplitudes is the study of much recent work. Roughly, there are two ways to proceed, we can assemble combinations of scattering amplitudes that reproduce classical scattering~\cite{Kosower:2018adc}, or we can define asymptotic observables that directly correspond to classical scattering~\cite{Caron-Huot:2023vxl,Biswas:2024ept}. At the level of in-in perturbation theory, these two approaches are essentially equivalent~\cite{Damgaard:2023vnx}. The final result is fairly intuitive, we effectively just replace the Feynman propagator that appears in~\eqref{eq:feynamp} with the retarded propagator, so that the classical analogue of~\eqref{eq:scatteringcrosssec} is
\begin{tcolorbox}[colframe=white,arc=0pt,colback=greyish2]
    \be
    i\mathcal{A}_{\rm cl}=-\frac{i^{L-L'+1}\omega^{L+L'+d-2}}{2(2\pi)^{d-1}}\lambda^{(R)}_{ L'M'LM}(\omega)\,,
    \label{eq:scatteringcrosssec2}
    \ee
\end{tcolorbox}
\noindent
where $\lambda^{(R)}_{ L'M'LM}$ is the frequency space representation of the retarded propagator (the same coefficients that appear for example in~\eqref{eq:freqspacegreensf} in the computation of the one-point function).

\subsection{Absorption}

An interesting physical property of an object is its absorption probability. From quite general principles we can extract this from the scattering cross section computed in Section~\ref{sec:pparticlescattering}. Writing the $S$-matrix as
\be
\hat S = {\mathds 1}+i\hat T\,,
\ee
recall that $\hat S^\dagger \hat S =  {\mathds 1}$ implies the following identity for $\hat T$:
\be
i(\hat T-\hat T^\dagger) = -\hat T^\dagger\hat T\,.
\ee
Evaluating this formula between asymptotic states, and inserting a complete set of states on the right-hand side, this becomes~\cite{Veltman:1994wz,Weinberg:1995mt}
\be
i{\cal A}(i\to f)-i{\cal A}^*(i\to f) =- \sum_X \int \rd{\rm PS}\,(2\pi)^D\delta^{(D)}(p_i-p_X){\cal A}(i\to X){\cal A}^*(X\to f)\,.
\label{eq:opticalthm}
\ee
This is the optical theorem, which relates the imaginary part of a scattering amplitude to the sum over all possible exchanged intermediate states (here denoted $X$), and integrated over the appropriate kinematic phase space.

If we consider the scattering amplitude~\eqref{eq:scatteringcrosssec2} with $L'=L$ and $M=M'$, then we can rewrite~\eqref{eq:opticalthm} as
\be
2\,{\rm Im}\,{\cal A}(i\to i) = \sum_X \int \rd{\rm PS}\,(2\pi)^D\delta^{(D)}(p_i-p_X)\lvert {\cal A}(i\to X)\rvert^2\,,
\ee
which we can represent diagrammatically as
\begin{equation*}
\raisebox{-40.5pt}{
\includegraphics[scale=1.45]{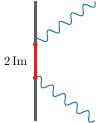}
}
~~=~~
\raisebox{-40.5pt}{
\includegraphics[scale=1.45]{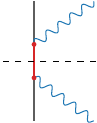}
}
~~=~~
\raisebox{-33pt}{\includegraphics[scale=1.6]{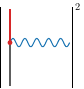}}\,.
\end{equation*}
This formula tells us that the cut of the four-point amplitude (its imaginary part) is the
square of the three-particle amplitude where the object absorbs the scalar wave, exciting some of the $X$ degrees of freedom, which is a consequence of the fluctuation dissipation theorem~\cite{1966RPPh...29..255K}. 
This (squared) amplitude is precisely the absorption probability that we would infer from infinity. In equations, this means that the absorption probability is
\be
\sigma_{\rm abs} = 2\,{\rm Im}\,{\cal A} \,.
\ee

In order to actually compute this absorption amplitude, recall that the frequency space retarded propagator can be written in terms of the Wightman function as
\be
   G_R(\omega)=-\frac{i}{2}\Big(\Delta(\omega)-\Delta(-\omega)\Big)+{\rm P}\int_{-\infty}^{\infty}  \frac{\rd \tilde{\omega}}{2\pi}\, \frac{\Delta(\tilde{\omega})-\Delta(-\tilde{\omega})}{\omega -\tilde{\omega}}\,,
\ee
where ${\rm P}$ denotes the principal value.
This formula implies that
\be
{\rm Im}\,G_R = -\frac{1}{2}\Big(\Delta(\omega)-\Delta(-\omega)\Big)\,,
\ee
where we have used that the principal part integral is purely real. We can then calculate from~\eqref{eq:scatteringcrosssec2}
\begin{tcolorbox}[colframe=white,arc=0pt,colback=greyish2]
\be
\sigma_{\rm abs}=\frac{\omega^{2L+d-2}}{2(2\pi)^{d-1}}\Big(\Delta(\omega)-\Delta(-\omega)\Big)\,.
\label{eq:absprob}
\ee
\end{tcolorbox}

We can verify that~\eqref{eq:absprob} is the correct absorption probability by direct calculation, following~\cite{Endlich:2016jgc}. Mathematically, we want to compute the probability that a black hole in some initial state absorbs a particle which excites the black hole degrees of freedom:
\begin{equation}
   |\omega\, L\,M\,; X_i \rangle \longrightarrow |0\,; X_f\rangle\,,
\end{equation}
where $|0\rangle$ is the vacuum state of the particle and $\omega$, $L$, and $M$ are the frequency, angular momentum, and azimuthal eigenvalues of the particle. Emission is simply the time-reversed process. 
This absorption process is represented by the diagram
\begin{equation*}
\includegraphics[scale=1.7]{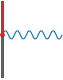}
\end{equation*}
which involves the object of interest absorbing a scalar wave and transitioning to an excited state.

The quantity we wish to calculate is 
\begin{equation}
   \label{eq:Pabs}
   P_{\rm abs}= \sum_{X_f} \frac{|\langle 0\,;X_f |\hat S|\omega\, L\,M\,; X_i\rangle |^2}{\langle \omega\, L\,M|\omega\, L\,M \rangle}\,,
\end{equation}
where we sum over all possible $|X_f\rangle$ states because we do not know the final state that the absorption will lead to. We can parameterize $\hat S$ as in~\eqref{eq:inthamil} with the interaction Hamiltonian~\eqref{eq:inthamil2}.
To first order in perturbation theory, we can then express the numerator of~\eqref{eq:Pabs} as 
\begin{align}
   \label{eq:Pabsnumerator}
   \sum_{X_f} |\langle X_f\, ; 0|&\hat S|X_i\,; \omega\, L\, M \rangle |^2  \\[-4pt]\nonumber
   &=\int \rd t \rd t' \, \langle Q^{i_1\cdots i_L}(t') Q^{j_1\cdots j_L}(t)\rangle \, \langle\omega\,L\,M| \partial_{i_1}\ldots \partial_{i_L} \phi(t') |0 \rangle \langle0|\partial_{j_1}\ldots \partial_{j_L} \phi(t)|\omega\,L\,M \rangle\,.
\end{align}
We can expand the Wightman function as
\be
\label{eq:wightmanexpan}
    \langle Q^{i_1\cdots i_L}(t) Q^{j_1\cdots j_{L'}}(t')\rangle = \sum_{M,M'}\int \frac{\rd \tilde{\omega}}{2\pi} \E^{-i{\omega}(t-t')}\Delta_{L ML'M'}({\omega}) \mathcal{Y}^{i_1 \ldots i_L}_{L M}\mathcal{Y}^*{}^{j_1\cdots j_{L'}}_{L'M'}\,,
\ee
in the basis of Thorne tensors and we compute the matrix element between the vacuum and the one-particle state $\langle0|\partial_{(j_1}\ldots \partial_{j_L)_T} \phi(t)|\omega\,L\,M \rangle$ in the following inset.

\vspace{-4pt}
\begin{oframed}
{\small
\noindent
{\bf\normalsize Matrix element:} In order to compute the absorption probability we need the matrix element $\langle0|\partial_{(j_1}\ldots \partial_{j_L)_T} \phi(t)|\omega\,L\,M \rangle$. Expanding the field in modes using~\eqref{eq:fieldexpan}, inserting a complete set of plane wave states and setting the location of the object to ${\bf x} = 0$, we have
\begin{align}
   \langle 0| \partial_{(i_1} \ldots \partial_{i_L)_T} \phi(t)|\omega\,L\,M\rangle =\int \frac{\rd^d k}{(2\pi)^d (2\omega_k)}i^L k_{(i_1}\ldots k_{i_L)_T} \langle \mathbf{k}|\omega,L,M\rangle \E^{-i\omega_kt} 
\end{align}
Plugging in~\eqref{eq:wavesphereoverlap} yields
\be
    \label{eq:kspaceintegraloverderivativesoffield}
   \langle 0| \partial_{(i_1} \ldots \partial_{i_L)_T} \phi(x,t)|\omega\,L\,M\rangle |_{x=0} =\frac{\omega^{L+\frac{d-2}{2}}}{\sqrt 2}i^L(2\pi)^{\tfrac{1-d}{2}} \E^{-i \omega t}\int \rd\Omega_{S^{d-1}}\hat{k}_{(i_1}\ldots \hat{k}_{i_L)_T} Y_{LM} (\hat{k})\,,
\ee
where we have integrated over the magnitude of ${\bf k}$ using the delta function.
Then, using~\eqref{eq:Integralofsympolynomials}, we have
\be
    \label{eq:angularintegralofkandYlm}
   \int \rd\Omega_{S^{d-1}}\hat{k}_{(i_1}\ldots \hat{k}_{i_L)_T} Y_{LM} (\hat{k})= \frac{2 \pi^{\tfrac{d}{2}}(d-2)!! L!}{\Gamma(\tfrac{d}{2})(d+2L-2)!!}\,\mathcal{Y}_{i_{1}\ldots i_{L}}^{LM}\,.
\ee
Putting everything together, we obtain, for integer $L$, 
\be
    \label{eq:wlmonderivativesoffieldfinal}
   \langle 0| \partial_{(i_1} \ldots \partial_{i_L)_T} \phi(t)|\omega,L,M\rangle =\sqrt{4\pi}i^L \omega^{L+\frac{d-2}{2}} \E^{-i \omega t}\frac{\Gamma(L+1)}{2^{L+\tfrac{d}{2}}\Gamma(L+\tfrac{d}{2})}\mathcal{Y}_{i_{1}\ldots i_{L}}^{LM}\,.
\ee
}
\end{oframed}

Now, combining this matrix element with~\eqref{eq:wightmanexpan} in~\eqref{eq:Pabsnumerator} yields
\begin{align}
\sum_{X_f} &|\langle X_f \,; 0|\hat S|X_i\,; \omega\, L\, M \rangle |^2  =\\[-6pt]\nonumber
&\int \rd t \rd t'\frac{\rd\omega'}{2\pi} \Delta_{LM}(\omega') \mathcal{Y}^{i_1 \ldots i_L}_{LM}\mathcal{Y}^*{}^{j_1 \ldots j_L}_{LM} \E^{-i(\omega-\omega')(t-t')}\omega^{2L+d-2} 4\pi \left(\frac{\Gamma(L+1)}{2^{L+\tfrac{d}{2}}\Gamma(L+\tfrac{d}{2})} \right)^2\mathcal{Y}^*{}_{i_1 \ldots i_L}^{LM}\mathcal{Y}_{j_1 \ldots j_L}^{LM}\,.
\end{align}
We can simplify the Thorne tensor contractions using~\eqref{eq:ThorneTensorcomponentNormgeneralD} and then perform the integrals to find
\be
\sum_{X_f} |\langle X_f \,; 0|\hat S|X_i\,; \omega\, L\, M \rangle |^2 =\frac{\omega^{2L+d-2}}{2(2\pi)^{d-1}} \Delta_{LM}(\omega)\, \Big(2\pi \delta(0)\Big)\,.
\ee
Then, from~\eqref{eq:sphericaloverlap} we know that $\langle \omega, L,M | \omega, L,M \rangle= 2\pi \delta(0)$, so we get
\begin{equation}
   P_{\rm abs}= \frac{\omega^{2L+d-2}}{2(2\pi)^{d-1}} \Delta_{LM}(\omega)\,.
\end{equation}

We now note that the absorption probability itself is not directly measurable from infinity. Instead we would see a mixture of the process of absorption and stimulated emission if we were to scatter a wave off of an object. The emission probability is just related by time reversal~\cite{Endlich:2016jgc}
\begin{equation}
   P_{\rm em}= \frac{\omega^{2L+d-2}}{2(2\pi)^{d-1}} \Delta_{LM}(-\omega) .
\end{equation}
The absorption cross section that we measure at infinity is the difference between absorption and stimulated emission
\be
\sigma_{\rm abs} = P_{\rm abs}-P_{\rm em} = \frac{\omega^{2L+d-2}}{2(2\pi)^{d-1}} \Big(
 \Delta_{LM}(\omega)- \Delta_{LM}(-\omega)
\Big)\,,
\ee
which agrees with~\eqref{eq:absprob}, as expected. As we saw with classical scattering, we can compute this absorption probability from
knowledge of the worldline Green's functions. Note that we have defined all quantities (including the frequency $\omega$) in the rest/corotating frame of the black hole. In a more realistic situation we will need to account for rotating frame effects that shift the frequency. This can be done straightforwardly following the discussion in Section~\ref{sec:rotframes}.

\subsection{Scattering in General Relativity}
\label{sec:MST}

The formula~\eqref{eq:scatteringcrosssec2} allows us to compute the scattering cross section of an object from knowledge of the Wightman function of the operators that an external field couples to. Once this spectral function has been matched by a microscopic calculation, the scattering cross section then becomes a prediction. We will be interested in validating our off-shell matching procedure by inferring $\Delta(\omega)$ from matching the classical field profile  response of a black hole to an external source and then using this $\Delta(\omega)$ to compute the corresponding scattering cross section and compare it to the scattering amplitude computed directly in general relativity. In order to be able to do this, we have to discuss some details about the computation of scattering amplitudes in black hole backgrounds. The MST (Mano--Suzuki--Takasugi~\cite{Mano:1996vt,Sasaki:2003xr}) formalism is a methodology to systematically compute the scattering cross section of waves off a black hole, order by order in the frequency of the waves. Here we review the aspects of the formalism that we need to compare to the EFT calculation.

\subsubsection{Mano--Suzuki--Takasugi Formalism}

We first outline the relevant features of the MST approach developed in~\cite{Mano:1996vt,Sasaki:2003xr} before applying it to the particular situation of interest. Further details can be found in~\cite{Mano:1996vt,Sasaki:2003xr}, whose notation we use. We begin by considering the Teukolsky equation~\cite{Teukolsky:1973ha}
\be
    \tilde{\Delta}^{1-s}\frac{\rd}{\rd x}\left( \tilde{\Delta}^{1+s}\frac{\rd R}{\rd x}\right)+\left[\frac{\left(2 \tilde{\Delta} \kappa  \epsilon +\tau -(2 x-1) \epsilon\right)^2}{4}- \frac{i s}{2} \left([2 \tilde{\Delta} \kappa  \epsilon -\tau ][2 x-1] +\epsilon \right)-\lambda \, \tilde{\Delta}\right]R=0\,,
\ee
where we have defined the parameters
\begin{align}
    \label{eq:MSTvariables}
\epsilon&\equiv 2M\omega, &  q& \equiv\frac{a}{M},&  \kappa &\equiv\sqrt{1-q^2}\,,\nonumber\\
\tau&\equiv\frac{\epsilon-mq}{\kappa}, &  \epsilon_\pm&\equiv\frac{\epsilon \pm \tau}{2} \,,\\
 x&\equiv\frac{\omega(r_+-r)}{\epsilon \kappa}, & \tilde{\Delta}&\equiv x(x-1)\nonumber\,.
\end{align}
The goal is to solve this equation perturbatively in the parameter $\epsilon$. We first note that we can write the solution that is ingoing at the black hole's horizon as an infinite series of hypergeometric functions:
\begin{equation}
    R_{\rm in}(x)=\E^{i \epsilon\kappa x}(-x)^{-s-i\epsilon_+}(1-x)^{i \epsilon_-} \sum_{n=-\infty}^{\infty} a_n^\nu\, 
    {}_2F_1\left[\begin{array}{c}
    n+\nu+1-i\tau\,,\,-n-\nu-i\tau\\[-3pt]
   1-s-i\epsilon-i\tau
    \end{array}\Big\rvert \,x\,\right]\,,
    \label{eq:ingpoingsolnMST}
\end{equation}
where $\nu$ is a parameter (often called ``renormalized angular momentum") that has to be solved for as part of the procedure. At leading order it is just the angular momentum $\nu = L+\ord{\omega}$.
The coefficients $a^\nu_n$ are found by substituting this expression into the Teukolsky equation and solving the resulting three-term recursion relation~\cite{Mano:1996vt,Sasaki:2003xr}. We omit the details here, as we only need the first three coefficients near $n=0$:\footnote{In the case of interest (a scalar), these formulas for $a_n^\nu$ are not valid for $L=0$ due to subtleties in solving the recursion relations~\cite{Mano:1996vt,Sasaki:2003xr}. For fields with other spin, other small values of the angular momenta also have subtleties. Since we are interested in generic $L$, this will not matter for our practical purposes.}
\be
\begin{aligned}
    \label{eq:MSTcoefficients}
    a_1^\nu&=-\frac{\kappa  \epsilon  (\tau +i [\nu +1]) (\nu -s+1)^2}{(\nu +1) (2 \nu +1) (L-\nu -1) (L+\nu +2)} \\
    a_0^\nu&= 1\\
    a_{-1}^\nu&= a_1^{-\nu-1}\,,
\end{aligned}
\ee
as all the other $a^\nu_n$ coefficients are at least $\ord{\omega^2}$, and we will ultimately only be interested in computing quantities to $\ord{\omega}$.

We can write the solution~\eqref{eq:ingpoingsolnMST} in a form amenable to expansion around $x\to \infty$ by using the hypergeometric connection formulas for each term in the sum. The result is that we can write
\begin{equation}
    R_{\rm in}=R_0^\nu+R_0^{-\nu-1},
\end{equation}
where the function $R_0^\nu$ is given by the infinite sum
\begin{align}
    R_0^{\nu}&=
    \E^{i \epsilon\kappa x}(-x)^{-s-i\epsilon_- +\nu}(1-x)^{i \epsilon_-}
    \\\nonumber
    & \,\times\!\!\! \sum_{n=-\infty}^{\infty}\!\!a_n^{\nu}\,\frac{\Gamma(1-s-i\epsilon-i\tau)\Gamma(2n+2\nu+1)}{\Gamma(n+\nu+1-i\tau)\Gamma(n+\nu+1-s-i\epsilon)} (-x)^{n}\, {}_2F_1\left[\!\begin{array}{c}
   -n-\nu-i\tau\,,\,s-n-\nu+i\epsilon\\[-3pt]
   -2n-2\nu
    \end{array}\Big\rvert \,\frac{1}{x}\,\right].
\end{align}
It is convenient to represent this instead as a series directly in $x$ as
\be
R_0^{\nu}=\E^{i \epsilon\kappa x}(-x)^{-s-i\epsilon_- +\nu}(1-x)^{i \epsilon_-} \sum_{k=-\infty}^{\infty} (-x)^k\sum_{n=k}^{\infty} C^\nu_{n,n-k} \,,
\label{eq:nearsolnseries}
\ee
where the coefficients are written in terms of sums of the expression
\begin{equation}
    C^\nu_{n,j}= a_n^{\nu}\,\frac{\Gamma(1-s-i\epsilon-i\tau)\Gamma(2n+2\nu+1)}{\Gamma(n+\nu+1-i\tau)\Gamma(n+\nu+1-s-i\epsilon)}\frac{(-n-\nu-i\tau)_j (-n-\nu+s+i\epsilon)_j}{(-2n-2\nu)_j j!}(-1)^{-j}\,.
\end{equation}
This solution, so defined, is formally valid (in the sense that it converges) for all $r<\infty$, but not literally at $r = \infty$. In order to find a solution valid at this point, we can solve the Teukolsky equation using a series of functions that are simple around $r\to \infty$, which are Coulomb wave functions. To do this, we first define the coordinate
\begin{equation}
    \hat{z}\equiv\omega(r-r_-)=\epsilon \kappa (1-x)\,.
\end{equation}
Then, we can write the solution to the Teukolsky equation as
\begin{equation}
    R_{\rm in}= K_\nu R^\nu_C+K_{-\nu-1}R^{-\nu-1}_C\,,
    \label{eq:coulombteukolsky}
\end{equation}
where at the moment $K_\nu$ are free parameters, which we want to determine, and the functions $R_C^\nu$ are defined as
\be
    R^\nu_C=\hat{z}^{-1-s}\left(1-\frac{\epsilon \kappa}{\hat{z}}\right)^{-s-i\epsilon_+} \sum_{n=-\infty}^{\infty} (-i)^n a^\nu_n \frac{(\nu+1+s-i\epsilon)_n}{(\nu+1-s+i\epsilon)_n} \, F_{n+\nu}(-is-\epsilon,\hat{z})\,,
\ee
where $F_{n+\nu}$ is an appropriately normalized confluent hypergeometric function
\be
F_L (\eta,\hat{z})\equiv  \E^{-i\hat{z}}2^L \hat{z}^{L+1} \frac{\Gamma(L+1-i\eta)}{\Gamma(2L+2)}{}_1F_1\left[\!\begin{array}{c}
   L+1-i\eta\\[-3pt]
   2L+2
    \end{array}\Big\rvert 2i\hat z\,\right].
\ee
As with~\eqref{eq:nearsolnseries} it is convenient to write the solution~\eqref{eq:coulombteukolsky} as a series in $\hat z$:
\be
    R_C^\nu = 2^{\nu } \E^{-i \hat{z}} (\epsilon \kappa   )^{-s-i\epsilon_+} \hat{z}^{\nu + i \epsilon_+} \left(\frac{\hat{z}}{\epsilon \kappa  
   }-1\right)^{-s-i \epsilon_+} \underset{k=-\infty }{\overset{\infty }{\sum }}\hat{z}^{k} \underset{n=-\infty}{\overset{k}{\sum }}D^\nu_{n,k-n}\,,
   \label{eq:coulombMST}
\ee
where the coefficients are sums over
\begin{equation}
    D^\nu_{n,j}=a_n^\nu (-1)^n  (2 i)^{j+n} \frac{\Gamma (\nu +n+1-s+i \epsilon )}{\Gamma (2 n+2 \nu +2)} \frac{(\nu +1+s-i \epsilon)_n }{(\nu +1-s+i \epsilon)_n}\frac{ (n-s+i \epsilon +\nu +1)_j}{  (2 n+2 \nu +2)_j j!}\,.
\end{equation}
The solution~\eqref{eq:coulombMST} converges for any $r>r_+$.

We now want to match~\eqref{eq:nearsolnseries} and~\eqref{eq:coulombMST} in their region of overlap  in order to extract the coefficients $K_\nu$, $K_{-\nu-1}$ that correspond to the ingoing solution~\eqref{eq:nearsolnseries}. Formally the solutions overlap for any $r_+<r<\infty$, but it is easier in practice to match if we choose $\hat{z}\gg \epsilon \kappa$. We then require that the solutions agree order-by-order in $\hat z$ so that
\be
    R_0^\nu = K_\nu R_C^\nu\,.
\ee
Equating the two series representations, we see that this implies
\be
    \E^{-i \hat{z}-i\epsilon \kappa}\left(\frac{\hat{z}}{\epsilon \kappa}\right)^{\nu-s}\sum_{k=-\infty}^{\infty} \left(\frac{\hat{z}}{\epsilon \kappa}\right)^k\sum_{n=k}^{\infty} C^\nu_{n,n-k} = K_\nu\, 2^{\nu} \E^{-i \hat{z}} \hat{z}^{\nu-s} \underset{k=-\infty }{\overset{\infty }{\sum }}\hat{z}^{k} \underset{n=-\infty}{\overset{k}{\sum }}D^\nu_{n,k-n}\,.
\ee
We require that this equality hold term-by-term in $\hat z$, which corresponds to fixing a value of $k$. Setting this value as $k = \frak{k}$, we find
\be
\begin{aligned}
    \label{eq:MSTKnu}
    K_\nu&= 2^{-\nu} \E^{i\epsilon \kappa}(\epsilon \kappa)^{s-\nu}\frac{\sum_{n={\frak k}}^{\infty}(\epsilon \kappa)^{-{\frak k}}C_{n,n-{\frak k}}}{\sum_{n=-\infty}^{{\frak k}}D^\nu_{n,{\frak k}-n}} \\
    &=2^{-\nu -{\frak k}} \E^{i \kappa  \epsilon }\frac{i^{\frak k}  \Gamma ({\frak k}+2 \nu +2) \Gamma (-s-2 i \epsilon_+ +1) (\kappa  \epsilon )^{-\nu -{\frak k}+s}}{\Gamma ({\frak k}+\nu +i \tau +1) \Gamma ({\frak k}-s+i \epsilon +\nu +1)
    \Gamma ({\frak k}+s+i \epsilon +\nu +1)}\\
    &~~~~~\times\left(\sum_{n={\frak k} }^{\infty}  a_n^\nu \,  \frac{(-1)^n\Gamma (n+{\frak k}+2 \nu +1)}{(n-{\frak k})!} \frac{\Gamma (n+\nu +i \tau +1) \Gamma (n+s+i \epsilon +\nu +1)}{\Gamma (n+\nu -i \tau +1) \Gamma (n-s-i \epsilon +\nu +1)}\right)\\
    &~~~~~~~~~\times\left(\sum_{n=-\infty}^{{\frak k}} a_n^\nu \, \frac{(-1)^n}{({\frak k}-n)! ({\frak k}+2 \nu +2)_n}\frac{(s-i \epsilon +\nu +1)_n}{(-s+i \epsilon +\nu +1)_n}\right)^{-1}\,.
\end{aligned}
\ee
Despite appearances, this expression does not depend on the value of the index
${\frak k}$. Consequently, in practice, we often just set ${\frak k}=0$. In terms of this $K^\nu$, we then have
\begin{equation}
    R_{\rm in}= K_\nu R_C^\nu+ K_{-\nu-1} R_C^{-\nu-1}\,.
    \label{eq:intocoulomb}
\end{equation}

We can now use the asymptotic expansion of the Coulomb wavefunctions in order to study the properties of the solution $R_{\rm in}$ near $r\to \infty$. It is convenient to use the identity
\begin{equation}
{}_1F_1\left[\!\begin{array}{c}
   a\\[-3pt]
   b
    \end{array}\Big\rvert x\,\right]=\frac{\E^{i \pi  a} \Gamma (b) }{\Gamma (b-a)}U(a,b,x)+\frac{\Gamma (b) \E^{x+i \pi  (a-b)} }{\Gamma (a)}U(b-a,b,-x)\,,
\end{equation}
where $U(a,b,x)$ is the irregular confluent hypergeometric function. (Note this identity holds when $\Im(x)>0$, corresponding to $r>0$ in the situation of interest.) In terms of this function, $R_C$ can be written as
\begin{equation}
    R^\nu_C=R^\nu_+ + R^\nu_-\,,
\end{equation}
where 
\begin{align}
    R^\nu_+ &=\E^{-i z} \,2^{\nu } \E^{i \pi  (\nu +1-s+ i\epsilon)}  z^{\nu +\frac{1}{2} i (\tau +\epsilon )} (z-\kappa  \epsilon )^{-s-\frac{1}{2} i (\tau +\epsilon
    )}\frac{\Gamma (-s+i \epsilon +\nu +1)}{\Gamma (s-i \epsilon +\nu +1)} \\
    &\quad~~ \times \underset{n=-\infty }{\overset{\infty }{\sum }}(2i z)^n a_n(\nu) U\big(n-s+i \epsilon +\nu +1,2 (n+\nu +1),2 i z\big) \nonumber\\
    R^\nu_- &=\E^{i z} \,2^{\nu } \E^{-i \pi  (\nu +1+s- i\epsilon)} z^{\nu +\frac{1}{2} i (\tau +\epsilon )} (z-\kappa  \epsilon )^{-s-\frac{1}{2} i (\tau +\epsilon
    )}\\
    &\quad ~~\times \underset{n=-\infty }{\overset{\infty }{\sum }} (2i z)^n a_n(\nu )\frac{ (\nu +1+s-i \epsilon )_n }{(\nu +1-s+i \epsilon )_n}U\big(n+s-i \epsilon +\nu +1,2 (n+\nu +1),2 i z\big)\,.\nonumber
\end{align}
At large $|z|$, the irregular confluent hypergeometric function has the asymptotic expansion
\begin{equation}
    U(a,b,z)=z^{-a} \sum _{d=0}^{\infty } \frac{(-z)^{-d} (a)_d (a-b+1)_d}{d!}\,,
\end{equation}
which has leading term $\sim z^{-a}$. Making this replacement, we find that
\begin{equation}
    \label{eq:RCasymptoticbehavior}
    R^\nu_C\xrightarrow{z\to\infty} A^\nu_+ \hat{z}^{-1} \E^{-i(\hat z +\epsilon \log\hat z)} +A^\nu_- \hat{z}^{-1-2s} \E^{i(\hat z +\epsilon \log\hat z)}\,,
\end{equation}
where the coefficients of the two fall-offs are
\begin{align}
    \label{eq:AplusAminus}
A^\nu_+&=2^{s-i \epsilon -1} \E^{\frac{1}{2} i \pi  (\nu -s+1)}\E^{-\frac{\pi  \epsilon }{2}}\frac{ \Gamma (-s+i \epsilon +\nu +1) }{\Gamma
    (s-i \epsilon +\nu +1)}\underset{n=-\infty }{\overset{\infty }{\sum }}a_n^\nu\,,\\
A^\nu_-&=2^{-s+i \epsilon -1} \E^{-\frac{1}{2} i \pi  (\nu +s+1)}\E^{-\frac{\pi  \epsilon }{2}} \underset{n=-\infty }{\overset{\infty }{\sum }}(-1)^n\frac{ (\nu+1+s-i\epsilon)_n }{(\nu+1-s+i\epsilon)_n}a^\nu_n\,.
\end{align}
We can obtain $A^{-\nu-1}_\pm$ by using the relation $a^{-\nu-1}_{-n}= a^\nu_n$, which leads to
\begin{align}
    A^{-\nu-1}_+&= -i \E^{-i\pi \nu}\frac{\sin(\nu-s+i\epsilon)}{\sin(\nu+s-i\epsilon)}A^\nu_+\,,\\
    A^{-\nu-1}_-&= -i \E^{i\pi \nu}A^\nu_-\,.
\end{align}

We now want to use this expansion to read off the reflection and transmission coefficients, these are defined in terms of the tortoise coordinate as the coefficients of the Coulomb fall-offs
\begin{equation}
    \label{eq:planewaveasymptoticsdefinition}
    R \to R^{\rm (in.)} \frac{1}{r}\E^{-i\omega r_\star}+ R^{\rm (ref.)} \frac{1}{r^{2s+1}}\E^{i\omega r_\star}\,,
\end{equation}
where $r_\star$ is the tortoise coordinate defined by 
\begin{equation}
    \label{eq:rstardiffeq}
    \frac{{\rm d} r_\star}{{\rm d}r}=\frac{r^2+a^2}{(r-r_+)(r-r_-)}\,,
\end{equation}
which has the solution
\begin{equation}
    r_\star=r+\frac{r_+ r_s}{r_+-r_-}\log\left(\frac{r-r_+}{r_s}\right)-\frac{r_- r_s}{r_+-r_-}\log\left(\frac{r-r_-}{r_s}\right)\,.
\end{equation}
Note, this coordinate involves a choice of an integration constant when solving~\eqref{eq:rstardiffeq}. This choice affects the phase of~\eqref{eq:MSTratioofplanewaves}.
Using the above to relate $R_C=R^\nu_C+R^{-\nu-1}_C$ to~\eqref{eq:planewaveasymptoticsdefinition}, we find
\begin{align}
    R^{\rm (in.)}&= \frac{\E^{i\epsilon \kappa}}{\omega}\left[K_\nu +i \E^{-i\pi \nu}\frac{\sin(\nu-s+i\epsilon)}{\sin(\nu+s-i\epsilon)} K_{-\nu-1}\right]A^\nu_+ \E^{-i(\epsilon \log \epsilon-\frac{1-\kappa}{2}\epsilon)}\,,\\
    R^{\rm (ref.)}&= \frac{\E^{i\epsilon \kappa}}{\omega^{2s+1}}\left[K_\nu -i \E^{i\pi \nu} K_{-\nu-1}\right]A^\nu_-\E^{i(\epsilon \log \epsilon-\frac{1-\kappa}{2}\epsilon)}\,.
\end{align}
Thus, we can read off the ratio of ingoing and outgoing waves as 
\begin{equation}
    \label{eq:MSTratioofplanewaves}
    \frac{R^{\rm (ref.)}}{R^{\rm (in.)}}=\omega^{2s}{\color{darkblue}\frac{K_\nu -i \E^{i\pi \nu} K_{-\nu-1}}{K_\nu +i \E^{-i\pi \nu}\frac{\sin(\nu-s+i\epsilon)}{\sin(\nu+s-i\epsilon)} K_{-\nu-1}}}{\color{reddish}\frac{A^\nu_-}{A^\nu_+}\E^{i\epsilon(2\log\epsilon-(1-\kappa))}}\,.
\end{equation}
Note that the two factors in~\eqref{eq:MSTratioofplanewaves} have slightly different origins in the detailed computation. The {\color{darkblue} blue} terms (first factor) arise from the expansion of the ingoing solution in terms of Coulomb wave functions~\eqref{eq:intocoulomb} and therefore reflect the physics of the matching of the near region to the far region. In contrast, the {\color{reddish} red} terms (second factor) appear from the translation between ${}_1F_1$ hypergeometric to irregular Coulomb wavefunctions, $U$. They therefore capture the physics of far region. This split into ``near" and ``far" contributions was noticed by~\cite{Ivanov:2022qqt,Saketh:2023bul}, who called it near-far factorization.\footnote{The physical origin of this factorization is still somewhat mysterious. It is possible to proceed without using this observation and instead match the full scattering amplitude as in~\cite{Ivanov:2024sds}.} This different origin suggests that the two terms capture different aspects of the physics. We should expect that the near terms will be matched by finite-size effects in the EFT, while the far terms are more universal, and should correspond to gravitational contributions.

\subsubsection{Scalar Scattering Amplitude}

We now want to specialize the MST formulas to the case of interest, a scalar scattering off a Kerr black hole. In the scalar case, the scattering amplitude is related to~\eqref{eq:MSTratioofplanewaves} via~\cite{Ivanov:2022qqt,Saketh:2023bul} 
\begin{equation}
    1-i\mathcal{A}=\eta_{LM}\E^{2i\delta_{LM}}=(-1)^{L+1}\frac{R^{\rm (ref.)}}{R^{\rm (in.)}}\,,
\end{equation}
where $\delta_{LM}$ is the phase shift, $1-|\eta_{LM}|^2$ is the absorption probability.
We would like to calculate this scattering amplitude to $\ord{\epsilon}$. Note that to this order we can set $\nu=L+\ord{\epsilon^2}$. Similarly, we can specialze to the case $s=0$ in all the above formulas.

Using MST coefficients~\eqref{eq:MSTcoefficients} in the expression for $K_\nu$~\eqref{eq:MSTKnu}, we find to leading order in $\epsilon$
\be
    \frac{K_{-\nu -1} }{K_{\nu }} = - (\kappa  \epsilon )^{2L+1}\frac{8^{-2 L-1} \Gamma \left(\frac{1}{2}-L\right)^2 \Gamma (L+1+ \frac{i m q}{\kappa})}{\Gamma \left(L+\frac{3}{2}\right)^2 \Gamma (-L+\frac{i m q}{\kappa})}\left(1+\frac{i \epsilon}{\kappa } \left[H_{-L-1+\frac{i m q}{\kappa}}-H_{L+\frac{i m q}{\kappa}}\right)\right]+\cdots\,,
\ee
where we have pulled out a kinematic factor of $(\kappa  \epsilon )^{2L+1}$. For $L>0$, this kinematic factor allows us to approximate 
\be
{\color{darkblue} \frac{K_\nu +i \E^{i\pi \nu} K_{-\nu-1}}{K_\nu -i \E^{-i\pi \nu}\frac{\sin(\nu-s+i\epsilon)}{\sin(\nu+s-i\epsilon)} K_{-\nu-1}}} =
    1-i \E^{i L \pi}\left(1+\E^{-2iL\pi}\frac{\sin (\pi  (L +i \epsilon ))}{\sin (\pi  (L -i \epsilon ))}\right) \frac{K_{-\nu -1} }{K_{\nu }}+\ord{\epsilon^{2L+3}}\,.
\ee
We can further approximate
\begin{equation}
    1+\E^{-2iL\pi}\frac{\sin (\pi  (L +i \epsilon ))}{\sin (\pi  (L -i \epsilon ))}= 2+2i\pi\cot(L\pi)\epsilon + \ord{\epsilon^2}\,.
\end{equation}
We can also expand the ``far-zone'' contribution to $\ord{\epsilon}$, yielding
\begin{align}
    (-1)^{L+1}\frac{A^\nu_-}{A^\nu_+}\E^{i\epsilon(2 \log\epsilon-(1-\kappa))} \simeq (-1)^{L+1}\left(1+i \epsilon  \big(\log 4-1+2 \log \epsilon -2 \psi ^{(0)}(L+1)\big)\right)\,.
\end{align}
Since we have neglected gravitational contributions to scattering in the EFT, we do not expect that the EFT will reproduce these far zone contributions. 
Dropping these pieces and keeping only the ``near-zone'' contributions, the MST scattering amplitude is then
\begin{tcolorbox}[colframe=white,arc=0pt,colback=greyish2]
\begin{equation}
\!\!\!\!\!\!\mathcal{A}_{\rm NZ}= \E^{i\pi L}   (\kappa  \epsilon )^{2 L+1} \frac{\Gamma \left(\frac{1}{2}-L\right)^2\Gamma \left(L+\frac{i m q}{\kappa }+1\right) }{ 4^{3 L+1} \Gamma \left(L+\frac{3}{2}\right)^2 \Gamma \left(\frac{i m
    q}{\kappa }-L\right)}\left[1 +i \pi  \epsilon  \cot (\pi  L)+\frac{i \epsilon}{\kappa}  \left(H_{\frac{i m
    q}{\kappa }-L-1}-H_{\frac{i m q}{\kappa }+L}\right)\right]\!.
    \label{eq:MSTamplitude}
\end{equation}
\end{tcolorbox}

We can make a nontrivial comparison by utilizing the EFT formula~\eqref{eq:scatteringcrosssec2} and the two-point function matched to the off-shell one-point function in Section~\ref{sec:KerrEFT}. In this context the scattering amplitude formula is
\be
\mathcal{A}_{\rm NZ}= -\frac{1}{2(2\pi)^{2}}\left(\frac{\epsilon}{r_s}\right)^{2L+1}\lambda_{LM}(\epsilon)\,,
\ee
while we can translate the two-point function from Section~\ref{sec:KerrEFT} into the MST variables as
\begin{equation}
\begin{aligned}
\lambda_{LM}^{\rm Kerr}(\epsilon)=(r_s \kappa )^{2 L+1}& \frac{2^{1-6 L}\pi^3\Gamma \left(-\frac{1}{2}-L\right)\Gamma \left(L+\tfrac{i m q}{\kappa }+1\right) }{ \Gamma \left(L+\frac{1}{2}\right)^2 \Gamma \left(L+\frac{3}{2}\right) \Gamma \left(\tfrac{i m
    q}{\kappa }-L\right)}\\
    &~~~~~~\times \left[1 +i \pi  \epsilon  \cot (\pi  L)+\frac{i \epsilon}{\kappa}  \left(H_{\frac{i m
    q}{\kappa }-L-1}-H_{\tfrac{i m q}{\kappa }+L}\right)\right]\,.
    \end{aligned}
    \label{eq:deltakerr}
\end{equation}
Combining these two formulas, we reproduce~\eqref{eq:MSTamplitude}. So, we see that matching via scattering or via the one-point function lead to the same Wilson coefficients.

\newpage
\section{Near Zone Approximation}
\label{app:NZ}

In order to solve the radial Klein--Gordon equation in Sections~\ref{sec:DdimSch}--\ref{sec:US} analytically we made some approximations to cast it in hypergeometric form.
These approximations only have a finite region of validity, and break down at sufficiently large distances from the black hole.
For this reason, this type of approximation is often called a ``near zone" approximation.
In practice the expansion parameter that will control the approximation will be the frequency of perturbations compared to all other scales in the problem---for example $\omega r_s\ll 1$.
There are many different implementations of this approximation, which simplify various features of the analysis~\cite{Page:1976df,Detweiler:1980uk,Hui:2020xxx,Hui:2022vbh,Charalambous:2021mea,Charalambous:2023jgq,Rodriguez:2023xjd,Chia:2020yla,Castro:2010fd,Maldacena:1997ih}. Specifically, by making different choices at subleading orders, one can manifest different symmetries of the dynamics (e.g.,~\cite{Castro:2010fd,Charalambous:2021kcz,Hui:2021vcv,Hui:2022vbh}), which can be used to infer properties of the dynamics at leading order in frequency.
Roughly, all approximations agree at leading order, but different choices of how to organize the expansion at subleading order leads to different effective equations of motion. In the main text, our motivation is to truncate the equation of motion in a minimal way, so that it is both analytically solvable and accurately approximates the true solution at first subleading order in the expansion parameter. Since this involves dropping some terms at sub-subleading order, the solution is not reliable at this order (which in this case corresponds to $\ord{\omega^2}$ terms in the response coefficients). 

In this Appendix, we explain the general strategy for the near zone approximations that we make in the main text. We further validate this near zone approximation in the Kerr case by comparing the response coefficients obtained by matching the near-zone one-point function to that obtained by matching to a scattering calculation, which involves solving the full equation to the relevant order in frequency.

\vspace{-12pt}
\paragraph{General Equation:} Let us first write the characteristic form of the full (unapproximated) equations that we encounter in the main text. In order to do so, we first define a small parameter $\epsilon$, which we will eventually use to control the expansion. 
In Sections~\ref{sec:DdimSch}--\ref{sec:MP} the relevant small parameter is the dimensionless frequency $\epsilon=r_s \omega$, in Section~\ref{sec:InfD} it is the inverse dimension, $\epsilon=1/N$, and in Section~\ref{sec:US} it is the dimensionless inverse spin $\epsilon=\delta_a=r_h/a$.  The general equation that we are trying to solve is of the schematic form
\be
    \label{eq:genericKGeqforNZ}
    \Delta(z) \frac{\rd}{\rd z}\left(\Delta(z) \frac{\rd R}{\rd z}\right)+\Bigg[P(\epsilon)^2+\Big(P(\epsilon)^2-\alpha(\epsilon)^2\Big)\,z-\lambda (\epsilon) \Delta(z)+\, \Big( \epsilon g_1(z)+\epsilon^2 g_2(z)+\cdots\Big)\Bigg]R(z)=0 \,,
\ee
where $z$ is a dimensionless radial coordinate which goes to $z=0$ at the horizon,
\begin{equation}
    \Delta(z)= z(z+1)+\epsilon f_1(z)+\epsilon^2 f_2(z)+\cdots\,.
\end{equation}
and  $\lambda (\epsilon),\,P(\epsilon),\, \alpha(\epsilon)$ are constants depending on $\epsilon$ and the other physical parameters of the problem. In~\eqref{eq:genericKGeqforNZ}  the functions 
$f_n(z)$, $g_n(z)$ are the $z$ dependent terms which cause the equation to differ from hypergeometric form, organized in powers of $\epsilon$.

In order to cast~\eqref{eq:genericKGeqforNZ} in hypergeometric form, we need to be able to drop the terms of the form $\epsilon^n f_n$ and $\epsilon^n g_n$. In this case, the equation will reduce to 
\begin{equation}
    \label{eq:KGformhypergeometriceq}
    z(z+1)\frac{\rd}{\rd z} \left(z(z+1)\frac{\rd R}{\rd z}\right)+\Big(P^2+(P^2-\alpha^2)z-\lambda\, z(z+1)\Big)R=0\,.
\end{equation}
We can put~\eqref{eq:KGformhypergeometriceq} in standard hypergeometric form by making the redefinition 
\begin{equation}
R(z)=z^{iP}(z+1)^{i\alpha}u(z)\,,
\end{equation}
defining the new coordinate $z=x-1$,
and the constants
\begin{equation}
    i P\equiv \frac{a+b-c}{2},\qquad\quad i \alpha\equiv \frac{c-1}{2}, \qquad\quad \lambda \equiv \frac{(a-b)^2-1}{4} \,,
\end{equation}
so that~\eqref{eq:KGformhypergeometriceq} becomes\footnote{Recall that the two solutions to the standard hypergeometric equation are ${}_2 F_1 \left[a,b,c;x\right]$ and $x^{1-c}{}_2 F_1 \left[a-c+1,b-c+1,2-c;x\right]$.}
\begin{equation}
    \label{eq:standardhypergeometricequation}
    x(1-x)\,\frac{\rd^2}{\rd x^2}u(x)+\Big(c-(a+b+1)x\Big)\,\frac{\rd}{\rd x} u(x)-ab \, u(x)=0\,.
\end{equation}
From this, we see that ultimately the validity of the near zone approximation boils down to assessing under what circumstances we can neglect terms of the form $\epsilon^n f_n$ and $\epsilon^n g_n$ in equations of the form~\eqref{eq:genericKGeqforNZ}.

\vspace{-12pt}
\paragraph{Regime of Validity:} 
In order to write~\eqref{eq:genericKGeqforNZ} as a  hypergeometric equation, we must drop the terms of the form $\epsilon^n f_n(z)$ and $\epsilon^n g_n(z)$.\footnote{In some sections in the text we make a near zone approximation with a dimensionful radial coordinate so that the horizon instead sits at $r = r_h$. There is no conceptual difference to the approximation, one must just restore dimensionful scales so that the terms we are neglecting are of the form $\epsilon^n f_n(r_h)$ rather than $\epsilon^n f_n(r)$, for example.} Though there is a formal expansion in $\epsilon$, this approximation must be done with care, because there is another dimensionless parameter present: the coordinate $z$. In principle the (dimensionless) combination $\epsilon z$ can range over all possible values from $0$ to $\infty$, and so we need to ensure that the enhancement by factors of $z$ never makes the terms that we are dropping compete with the terms that we are keeping. Specifically, we have to check the conditions
\begin{align}
    \epsilon^n f_n(z) &\ll z(z+1)  & &\hspace{-2cm}\text{for}\quad 0\leq z <\infty\,,\\[3pt]
    \epsilon^n g_n(z) &\ll \lambda (\epsilon) \Delta(z) &&\hspace{-2cm}\text{as} \quad z\to \infty\,, \\
    \frac{\epsilon^n g_n(z)}{z(z+1)} &\ll \frac{P(\epsilon)^2}{z(z+1)}& &\hspace{-2cm}\text{as}\quad z\to 0\,.
\end{align}
These conditions ensure that the terms we are neglecting do not interfere with the {\it leading-order} behavior of the solution, these terms will still contribute at $\ord{\epsilon^n}$, as expected. Thus, if we discard terms at $\ord{\epsilon^n}$, the best possible accuracy our solution could have is $\ord{\epsilon^{n-1}}$.
\

In the majority of the situations we consider, the functions $f_n(z)$, $g_n(z)$ are polynomials in $z$ with degree $k\geq 1$.\footnote{In Section~\ref{sec:US} there is a term proportional to $\frac{1}{1+z}$, but since this term does not become large at the origin it does not change our overall discussion.} Such terms are always subdominant to the terms that we keep as $z\to 0$, but are only subdominant at large $z$ when
\begin{equation}
 \epsilon^n z^k\ll 1, \quad \implies \quad z^{k} \ll \frac{1}{\epsilon^n}\,.
\end{equation}
This fact---that the approximation is only valid up to some finite distance from the black hole---is what defines the so-called ``near zone" (and what gives the approximation its name). Note that this region does not formally extend to $z=\infty$, but it can be made parametrically large by taking $\epsilon$ to be sufficiently small. Aside from Section~\ref{sec:InfD}, the deviations from hypergeometric form begin at $\ord{\epsilon^2}$, which allows us to approximate the solutions to be accurate to $\ord{\epsilon}$ within this region. 
In the case of the infinite-dimensional black hole---studied in Section~\ref{sec:InfD}---the terms we must approximate are of the form $\epsilon f_1(z)=\epsilon g_1(z)=\epsilon \log(z)$ at lowest order in $\epsilon$. These are not small near $z\to0$, so we can only proceed in the strict $\epsilon \to 0$ limit. This limits the accuracy of our final result to $\ord{\epsilon^0}$, which does not accurately capture $1/N$ corrections.

Once we have approximated~\eqref{eq:genericKGeqforNZ} to put it in the form~\eqref{eq:KGformhypergeometriceq}, we can then solve this equation to obtain a solution that approximates the solution to the full equation to some order in $\epsilon$. For example, if we had to drop terms of $\ord{\epsilon^2}$, then we should expect that we can approximate the true solution to order $\epsilon$. In doing this, we are free to approximate the parameters appearing in the equation~\eqref{eq:KGformhypergeometriceq} to this same order, if desired. For example, we in many cases will approximate
\begin{equation}
    \lambda(\epsilon)=L(L+1)+\epsilon^2\approx L(L+1)\,,
\end{equation}
since $L(L+1)$ is always $\ord{1}$.\footnote{Even at $L=0$ this is still a valid approximation, as we may treat $\epsilon^2 z(z+1)$ as subleading perturbations in the near zone.} In the main text, we often approximate $\lambda (\epsilon)$, and $\alpha(\epsilon)$ to $\ord{\epsilon}$. In some cases it is convenient to keep some terms at $\ord{\epsilon^2}$, even though these terms do not affect the actual solution at $\ord{\epsilon}$, just because they make various analytic expressions simpler. We must take special care with the parameter, $P(\epsilon)^2$. This is because this term can become large compared to other terms near $z=0$ at any finite value of $\epsilon$. This means that $P(\epsilon)^2/z$ can become $\ord{\epsilon}$ even when $P(\epsilon)^2$ is $\ord{\epsilon^2}$ by itself (see Section~\ref{sec:DdimSch}).
Relatedly, when approximating $P^2$, we need to ensure that there are no parameter choices that make the terms that we dropped of the same order or larger than the ones that we keep. So, for example we can approximate
\begin{equation}
    P(\epsilon)^2=(m\Omega-(1-\epsilon^2))^2 \approx (m\Omega-1)^2\,,
\end{equation}
but not
\begin{equation}
    P(\epsilon)^2=(m\Omega-\epsilon)^2 \not\approx m^2\Omega^2-2m\Omega\epsilon\,,
\end{equation}
as $m$ may be $0$. (Additionally, such an approximation would break continuity with the $\Omega=0$ solution.)

As a final note, one might worry that the fact that the near zone does not formally reach to $z=\infty$ will complicate matching to a point particle EFT. As a practical matter, we can imagine the EFT matching regime to be sufficiently far from the object that it is approximately a point particle, but still within the region where the near zone is valid (see also~\cite{Chia:2020yla}). Alternatively, we could imagine first matching the UV solution to a far zone and match this to scattering in the EFT. In the following we check explicitly for the Kerr case that matching the one-point function via the near zone produces the same scattering amplitude in the EFT as that matched using the MST formalism~\cite{Mano:1996vt}.

\vspace{-12pt}
\paragraph{An Example---Kerr:} It is useful to illustrate the general discussion above in the concrete example of the wave equation for a scalar field evolving in the Kerr metric. We treat this case in detail in Section~\ref{sec:Kerr}. Recall that the radial Klein--Gordon equation is~\eqref{eq:KerrradialKGequation}
\be
    \Delta \frac{\rd}{\rd r}\left(\Delta \frac{\rd R(r)}{\rd r}\right)+\Big[ \omega ^2\left(a^2+r^2\right)^2-2 a m  r_s \omega \,r+a^2 m^2-\Delta  \left(a^2 \omega ^2+\lambda \right)\Big]R(r)=0\,.
    \label{eq:KerrRdiffeq}
\ee
We can cast this in the form~\eqref{eq:genericKGeqforNZ} by first defining the dimensionless variables~\eqref{eq:kerrvars1}--\eqref{eq:kerrvars2}, so that it reads:
\be
\begin{aligned}
    z(z+1)\frac{\rd}{\rd z}\left(z(z+1)\frac{\rd R}{\rd z}\right)+ \Bigg[P^2&-\lambda z (z+1) -4r_s \kappa W U z\\
    +&W^2\bigg( \left(-4a^2\kappa^2 +8r_+\kappa \right)z + \left(8 r_+  \left(r_s+2
    r_+\right)-4a^2 \right)\kappa^2 z^2\\
    &\hspace{1cm}+ 32r_+^2r_s \kappa^3 z^3+16 r_+^2 r_s^2 \kappa^4 z^4\bigg)\Bigg]R=0\,.
\end{aligned}
\ee
This is the full, un-approximated differential equation, identical to~\eqref{eq:KerrRdiffeq}. (Recall that the separation constant $\lambda = L(L+1)+\ord{\omega^2}$.) Then, if we define the following new parameters
\begin{align}
    A&=-4 \kappa  W (r_s U+r_+ W (\kappa  r_- -2))\\
    B&= 4 \kappa ^2 r_+ W^2 \left(5 r_+ + r_s\right)\\
    L_*(L_*+1)&=\lambda-B \\
    \alpha^2&= P^2+B-A\\
  k&=\omega(r_+ - r_-)\,,
\end{align}
the equation takes the form~\eqref{eq:genericKGeqforNZ}
\begin{equation}
    z(z+1)\frac{\rd}{\rd z}\left(z(z+1)\frac{\rd R}{\rd z}\right)+\left(P^2+\left(P^2-\alpha ^2\right)z-L_*(L_*+1) z(z+1) +\frac{2 k^2 }{\kappa  r_s}z^3+k^2 z^4\right)  R=0\,.
\end{equation}
It is straightforward to define a near zone using this equation. We simply drop the $z^3$ and $z^4$ terms, whose coefficients are both $\ord{\omega^2}$.
After doing this, we are free to expand the coefficients $P^2-\alpha ^2$ and $L_*(L_*+1)$ and drop terms of $\ord{\omega^2}$. The equation then takes the form
\be
z(z+1)\frac{\rd}{\rd z} \left(z(z+1)\frac{\rd}{\rd z}R_{LM}\right)+\Big[P^2 (z+1)-(P+\omega r_s)^2z-z(z+1)L(L+1)\Big]R_{LM}=0\,.
\ee

By solving this equation and reading off the ratio of fall-offs, we can match the Wilson coefficients of the point particle EFT, which was done in Section~\ref{sec:KerrEFT}. As a practical matter we read off the low-frequency expansion of the Green's function~\eqref{eq:QQ2pt}. Once this matching calculation has been performed, we can use the EFT to compute any quantity of interest. In order to validate both the procedure of matching using a near-zone solution, and to verify the computation at $\ord{\epsilon}$, we can compare the scattering amplitude calculated in the matched EFT to the scattering amplitude computed using black hole perturbation theory for the full equation. This latter calculation employs the MST formalism, and is detailed in Section~\ref{sec:MST}.
We can first compute the scattering amplitude in the EFT in convenient variables. Translating the results of the off-shell matching using the near zone UV solution, we obtain the expansion of the retarded Green's  function:
\begin{equation}
\begin{aligned}
\lambda_{LM}^{\rm Kerr}(\epsilon)=(r_s \kappa )^{2 L+1}& \frac{2^{1-6 L}\pi^3\Gamma \left(-\frac{1}{2}-L\right)\Gamma \left(L+\tfrac{i m q}{\kappa }+1\right) }{ \Gamma \left(L+\frac{1}{2}\right)^2 \Gamma \left(L+\frac{3}{2}\right) \Gamma \left(\tfrac{i m
    q}{\kappa }-L\right)}\\
    &~~~~~~\times \left[1 +i \pi  \epsilon  \cot (\pi  L)+\frac{i \epsilon}{\kappa}  \left(H_{\frac{i m
    q}{\kappa }-L-1}-H_{\tfrac{i m q}{\kappa }+L}\right)\right]\,.
    \end{aligned}
\end{equation}
defined in terms of the parameters~\eqref{eq:MSTvariables}. The scattering amplitude in the EFT is then~\eqref{eq:scatteringcrosssec2}
\be
\mathcal{A}_{\rm NZ}= -\frac{1}{2(2\pi)^{2}}\left(\frac{\epsilon}{r_s}\right)^{2L+1}\lambda_{LM}(\epsilon)\,.
\ee
We then want to compare this to the ``near-zone" part of the MST scattering amplitude~\eqref{eq:MSTamplitude}
\begin{equation}
\!\!\!\!\!\!\mathcal{A}_{\rm NZ}= \E^{i\pi L}   (\kappa  \epsilon )^{2 L+1} \frac{\Gamma \left(\frac{1}{2}-L\right)^2\Gamma \left(L+\frac{i m q}{\kappa }+1\right) }{ 4^{3 L+1} \Gamma \left(L+\frac{3}{2}\right)^2 \Gamma \left(\frac{i m
    q}{\kappa }-L\right)}\left[1 +i \pi  \epsilon  \cot (\pi  L)+\frac{i \epsilon}{\kappa}  \left(H_{\frac{i m
    q}{\kappa }-L-1}-H_{\frac{i m q}{\kappa }+L}\right)\right]\!.
\end{equation}
Plugging in everything, we see that they agree. 
This implies both that the procedure of matching the off-shell one-point function and the scattering cross section lead to the same Wilson coefficients (as they should), but also serves as a cross check that the near-zone solution is reliable to $\ord{\epsilon}$ (also, as expected).

\newpage
\linespread{.95}
\addcontentsline{toc}{section}{References}
\bibliographystyle{utphys}
{\small
\bibliography{eftbib}
}

\end{document}